\documentclass[a4paper,12pt]{article}
\usepackage{a4wide}
\usepackage[latin1]{inputenc}
\usepackage{dsfont}
\usepackage{amsfonts}
\usepackage{amsmath}
\usepackage{amssymb}
\usepackage{amsthm}
\usepackage{graphicx}
\usepackage{booktabs}
\usepackage[numbers, sort]{natbib}
\allowdisplaybreaks[1]
\usepackage{color}
\usepackage{ulem}
\usepackage{tikz}
\usetikzlibrary{arrows}
\usepackage{floatrow}
\usepackage{array}
\usepackage{cancel}
\usepackage{pstricks}
\usepackage{axodraw4j}
\usepackage{hyperref}
\usepackage{listings}
\usepackage[para]{footmisc}

\usepackage{floatrow}
\newfloatcommand{capbtabbox}{table}[][\FBwidth]

\usepackage[font=footnotesize,bf]{caption}
\usepackage{subcaption}

\newcommand{\be}{\begin{equation}}
\newcommand{\ee}{\end{equation}}

\def\sgn{\mathop{\rm sgn}}

\def\Tr{\mathop{\rm Tr}}

\newcommand{\e}{\text{e}}

\newcommand{\GeV}{{\rm GeV}}

\newcommand{\TeV}{{\rm TeV}}

\newcommand{\SusyTC}{\texttt{SusyTC}}

\renewenvironment{abstract}
 {\small
  \begin{center}
  \bfseries \abstractname\vspace{-.5em}\vspace{0pt}
  \end{center}
  \list{}{%
    \setlength{\leftmargin}{5mm}
    \setlength{\rightmargin}{\leftmargin}%
  }%
  \item\relax}
 {\endlist}

\begin{document}

\begin{titlepage}

\begin{center}
{
\bf\Large
Predicting the Sparticle Spectrum from GUTs via\\[1mm] SUSY Threshold Corrections with \texttt{SusyTC}
}
\\[8mm]
Stefan Antusch$^{\star\dagger}$\footnote{\texttt{Stefan.Antusch@unibas.ch}},~
Constantin Sluka$^{\star}$\footnote{\texttt{Constantin.Sluka@unibas.ch}}
\end{center}
\addtocounter{footnote}{-2}

\vspace*{0.18cm}

\centerline{$^{\star}$ \it
Department of Physics, University of Basel,}
\centerline{\it
Klingelbergstr.\ 82, CH-4056 Basel, Switzerland}

\vspace*{0.4cm}

\centerline{$^{\dagger}$ \it
Max-Planck-Institut f\"ur Physik (Werner-Heisenberg-Institut),}
\centerline{\it
F\"ohringer Ring 6, D-80805 M\"unchen, Germany}

\vspace*{1.0cm}

\begin{abstract}
\noindent
Grand Unified Theories (GUTs) can feature predictions for the ratios of quark and lepton Yukawa couplings at high energy, which can be tested with the increasingly precise results for the fermion masses, given at low energies. To perform such tests, the renormalization group (RG) running has to be performed with sufficient accuracy. 
In supersymmetric (SUSY) theories, the one-loop threshold corrections (TC) are of particular importance and, since they affect the quark-lepton mass relations, link a given GUT flavour model to the sparticle spectrum. To accurately study such predictions, we extend and generalize various formulas in the literature which are needed for a precision analysis of SUSY flavour GUT models. We introduce the new software tool \SusyTC, a major extension to the Mathematica package \texttt{REAP} \cite{Antusch:2005gp}, where these formulas are implemented. \SusyTC~extends the functionality of \texttt{REAP} by a full inclusion of the (complex) MSSM SUSY sector and a careful calculation of the one-loop SUSY threshold corrections for the full down-type quark, up-type quark and charged lepton Yukawa coupling matrices in the electroweak-unbroken phase. Among other useful features, \SusyTC~calculates the one-loop corrected pole mass of the charged (or the CP-odd) Higgs boson as well as provides output in SLHA conventions, i.e.\ the necessary input for external software, e.g.\ for performing a two-loop Higgs mass calculation. We apply \SusyTC~to study the predictions for the parameters of the CMSSM (mSUGRA) SUSY scenario from the set of GUT scale Yukawa relations $\frac{y_e}{y_d}=-\frac{1}{2}$, $\frac{y_\mu}{y_s}=6$, and $\frac{y_\tau}{y_b}=-\frac{3}{2}$, which has been proposed recently in the context of SUSY GUT flavour models. 
\end{abstract}

\end{titlepage}

\setcounter{footnote}{0}

\section{Introduction}

Supersymmetric (SUSY) Grand Unified Theory (GUT) models of flavour are promising candidates towards solving the open questions of the Standard Model (SM) of Particle Physics. They embrace the unification of the SM gauge couplings, a dark matter candidate, a solution to the gauge hierarchy problem and an explanation of the hierarchies of masses and mixing angles in the flavour sector. Whether a flavour GUT model can successfully explain the observations in the flavour sector, depends on the renormalization group (RG) evolution of the Yukawa matrices from the GUT scale to lower energies. Furthermore, it is known that $\tan\beta$ enhanced supersymmetric threshold corrections (see e.g.\ \cite{threshold}) are essential in the investigation of mass (or Yukawa coupling) ratios predicted at the GUT scale. Interesting well-known GUT predictions in this context are $b-\tau$ and $b-\tau-t$ unification (for early work see e.g.\cite{Ibanez:1981yh}) and $y_\mu =  3 y_s$ \cite{GJ}.\footnote{$y_i$ are the Yukawa couplings in the diagonal basis.} Other promising quark-lepton mass relations at the GUT scale have been discussed in \cite{Antusch:2009gu,Antusch:2013rxa}, e.g.\ $y_\tau=\pm \frac{3}{2} y_b$, $y_\mu = 6 y_s$ or $y_\mu = \frac{9}{2} y_s$. Various aspects regarding the impact of such GUT relations for phenomenology have been studied in the literature, see e.g.\ \cite{Ross:2007az} for recent works. 

With the discovery of the Higgs boson at the LHC \cite{Aad:2012tfa} and the possible discovery of sparticles in the near future, the question whether a set of SUSY soft-breaking parameters can be in agreement with both, specific SUSY threshold corrections as required for realizing the flavour structure of a GUT model, and constraints from the Higgs boson mass and results on the sparticle spectrum, gains importance. To accurately study this question, we introduce the new software tool \SusyTC, a major extension to the Mathematica package \texttt{REAP} \cite{Antusch:2005gp}. 

\texttt{REAP}, which is designed to run the SM and neutrino parameters in seesaw scenarios with a proper treatment of the right-handed neutrino thresholds, is a convenient tool for top-down analyses of flavour (GUT) models, with the advantage of a user-friendly Wolfram Mathematica front-end. However, the SUSY sector is not included. In order to take supersymmetric threshold corrections into account in the analyses, for example of the $A_4$ flavour GUT models in \cite{Antusch:2013kna,Antusch:2013tta}, the following procedure was undertaken: First \texttt{REAP} was used to run the Yukawa matrices in the MSSM from the GUT-scale to a user-defined ``SUSY''-scale. At this scale, the SUSY threshold corrections were incorporated as mere model parameters, in a simplified treatment assuming, e.g., degenerate first and second generation sparticle masses (cf.\ \cite{Antusch:2013jca}), without specializing any details on the SUSY sector. Finally, the Yukawa matrices, corrected by these $\tan\beta$-enhanced thresholds, were taken as input for a second run of \texttt{REAP}, evolving the parameters in the SM from the ``SUSY''-scale to the top-mass scale.\footnote{Such a treatment of threshold corrections as additional model parameters is now implemented in the latest version of \texttt{RGEMSSM0N.m} of \texttt{REAP} 11.1.2.}
Although this procedure is quite SUSY model-independent, it only allows to study the constraints on the SUSY sector indirectly (i.e.\ via the introduced additional parameters and with simplifying assumptions), and it is unclear whether an explicit SUSY scenario with these assumptions and requirements can be realised.

The aim of this work is to make full use of the SUSY threshold corrections to gain information on the SUSY model parameters from GUTs. Towards this goal we extend and generalize various formulas in the literature which are needed for a precision analysis of SUSY flavour GUT models and implement them in \SusyTC. For example, \SusyTC~includes the full (CP violating) MSSM SUSY sector, sparticle spectrum calculation, a careful calculation of the one-loop SUSY threshold corrections for the full down-type quark, up-type quark and charged lepton Yukawa coupling matrices in the electroweak-unbroken phase, and automatically performs the matching of the MSSM to the SM, including $\overline{\text{DR}}$ to $\overline{\text{MS}}$ conversion. Among other useful features, \SusyTC~calculates the one-loop corrected pole mass of the charged (or the CP-odd) Higgs boson as well as provides output in SLHA conventions, i.e.\ the necessary input for external software, e.g.\ for performing a two-loop Higgs mass calculation. 

\SusyTC~is specifically developed to perform top-down analyses of SUSY flavour GUT models. This is a major difference to other well-known SUSY spectrum generators (e.g.\ \cite{Baer:1993ae,Allanach:2001kg,Djouadi:2002ze,Porod:2003um}, see e.g.\ \cite{Djouadi:2002nh} for a comparison), which run experimental constraints from low energies to high energies, apply GUT-scale boundary conditions, run back to low energies and repeat this procedure iteratively. \SusyTC~instead starts directly from the GUT-scale, allowing the user to define general (complex) Yukawa, trilinear, and soft-breaking matrices, as well as non-universal gaugino masses, as input. These parameters are then run to low energies, thereby enabling an investigation whether the GUT-scale Yukawa matrix structures of a given SUSY flavour GUT model are in agreement with experimental data.

We apply \SusyTC~to study the predictions for the parameters of the Constrained MSSM (mSUGRA) SUSY scenario from the GUT-scale Yukawa relations $\frac{y_e}{y_d}=-\frac{1}{2}$, $\frac{y_\mu}{y_s}=6$, and $\frac{y_\tau}{y_b}=-\frac{3}{2}$, which have been proposed recently in the context of SUSY GUT flavour models. With a Markov Chain Monte Carlo analysis we find a ``best-fit'' benchmark point as well as the 1$\sigma$ ranges for the sparticle masses and the correlations between the SUSY parameters. Without applying any constraints from LHC SUSY searches or dark matter, we find that the considered GUT scenario predicts a CMSSM sparticle spectrum above past LHC sensitivities, but within reach of the current LHC run or a future high-luminosity upgrade. Furthermore, the scenario generically features a bino-like neutralino LSP and a stop NLSP with a mass that can be close to the present bounds.    

This paper is organized as follows: In section 2 we review GUT predictions for Yukawa coupling ratios. In section 3, we describe the numerical procedure in \SusyTC~and present the main used formulas. We give a short introduction to the new features \SusyTC~adds to \texttt{REAP} in section 4. In section 5 we study the predictions for the parameters of the CMSSM SUSY scenario from the above mentioned GUT-scale Yukawa relations with \SusyTC. In the appendices we present other relevant formulas and a detailed documentation of \SusyTC.

\section{Predictions for Yukawa Coupling ratios from GUTs}
\label{sec:Clebsch}
GUTs not only contain a unification of the SM forces, they also unify fermions into joint representations. After the GUT gauge group is broken to the SM gauge group, this can lead to predictions for the ratios of down-type quark and charged lepton Yukawa couplings which result from group theoretical Clebsch Gordan (CG) factors. To confront such predictions of GUT models with the experimental data, the RG evolution of the Yukawa couplings from high to low energies has to be performed, including (SUSY) threshold corrections.

In $SU(5)$ GUTs, for example, the right-handed down-type quarks and the lepton doublets are unified in five-dimensional representations of $SU(5)$ and the quark doublets plus right-handed up-type quarks and right-handed charged leptons are unified in a ten-dimensional $SU(5)$ representation. The Higgs doublets are supplemented by $SU(3)_c$ triplets and embedded into five-dimensional representations of $SU(5)$. Using only these fields and a single renormalizable operator to generate the Yukawa couplings for the down-type quarks and charged leptons, so-called minimal $SU(5)$ predicts $Y_e=Y_d^T$ for the Yukawa matrices at the GUT scale. To correct this experimentally challenged scenario, $SU(5)$ GUT flavour models often introduce a 45-dimensional Higgs representation, which can lead to the Georgi-Jarlskog relations $y_\mu = - 3 y_s$ and $y_e = \frac{1}{3} y_d$ \cite{GJ}. 

It was pointed out in \cite{Antusch:2009gu} (see also \cite{Antusch:2013rxa}) that other promising Yukawa coupling GUT ratios can emerge in $SU(5)$, e.g.~$y_\tau = \pm \frac{3}{2} y_b$, $y_\mu = 6 y_s$ or $y_\mu = \frac{9}{2} y_s$, and  $y_e = - \frac{1}{2} y_d$ from higher dimensional GUT operators containing for instance a GUT breaking 24-dimensional Higgs representation. A convenient test whether GUT predictions for the first two families can be consistent with the experimental data is provided by the -- RG invariant and SUSY threshold correction invariant\footnote{The invariance under SUSY threhold corrections holds under some generic conditions, cf.\ \cite{Antusch:2013jca}.} -- double ratio \cite{Antusch:2013jca}
\be
\left|\frac{y_\mu}{y_s}\frac{y_d}{y_e}\right|=10.7^{+1.8}_{-0.8}\;.
\ee
While the Georgi-Jarlskog relations \cite{GJ} imply a double ratio of $9$, disfavoured by more than 2$\sigma$, other combinations of CG factors \cite{Antusch:2009gu,Antusch:2013rxa}, e.g.\ $y_\mu = 6 y_s$ and  $y_e = - \frac{1}{2} y_d$ can be in better agreement (here with a double ratio of $12$, within the $1\sigma$ experimental range).

\begin{figure}
    \begin{floatrow}
    \capbtabbox{
        \caption{Dimension 5 effective operators $(A B)_{R} (C D)_{\bar{R}}$ and CG factors emerging from the supergraphs of Figure~\ref{fig:messengerdiagram1} when $R$ and $\bar{R}$ are integrated out \cite{Antusch:2009gu}.
       } \label{tab:effectiveFToperators11}
    }
    {
        \rule{7mm}{0pt}\newline
        \begin{tabular}{cccc}
        \toprule
         $A \, B$ & $C \, D$ & $R$ & $(Y_e)_{ji} / (Y_d)_{ij}$  \\
        \midrule
         $H_{24} \, F$ & $T \, \bar H_{45}$ & $\overline{\mathbf{45}}$  &  $-\tfrac{1}{2}$  \\[0.3ex]
         $H_{24} \, F$ & $T \, \bar H_5$ & $\bar{\mathbf{5}}$          & $-\tfrac{3}{2}$  \\[0.3ex]
         $H_{24} \, T$ & $F \, \bar H_5$ & $\mathbf{10}$     &            $6$  \\
        \bottomrule
        \end{tabular}
    }
    \ffigbox{
        \includegraphics[scale=0.55,page=1]{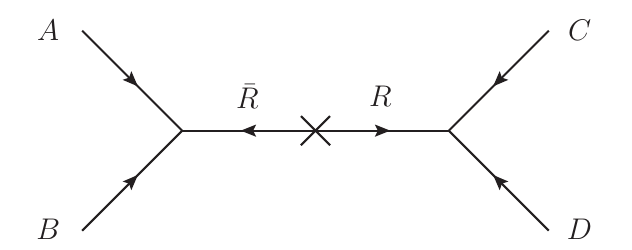}
    }{
        \caption{Supergraphs corresponding to the operators of Table~\ref{tab:effectiveFToperators11} generating effectively Yukawa couplings when the pair of messengers fields $R$
         and $\bar{R}$ is integrated out.}
         \label{fig:messengerdiagram1}
    }
\end{floatrow}
\end{figure}

The combination of GUT-scale Yukawa relations $y_e = - \frac{1}{2} y_d$, $y_\mu = 6 y_s$, and $y_\tau = -\frac{3}{2} y_b$ (as direct result of CG factors, cf.\ Table~\ref{tab:effectiveFToperators11} and Figure~\ref{fig:messengerdiagram1}, or as approximate relation after diagonalization of the GUT-scale Yukawa matrices $Y_d$ and $Y_e$) has been used to construct $SU(5)$ SUSY GUT flavour models in Refs.\ \cite{Antusch:2013kna,Antusch:2013tta,Antusch:2014poa,Gehrlein:2014wda}. A subset of these relation, $y_\mu = 6 y_s$, and $y_\tau = -\frac{3}{2} y_b$, has been used in \cite{Meroni:2012ty}. In addition to providing viable quark and charged lepton masses, the GUT CG factors $(Y_e)_{ji} / (Y_d)_{ij} = -\tfrac{1}{2}$ and $6$ can also be applied to realize the promising relation between the lepton mixing angle $\theta_{13}^\text{PMNS}$ and the Cabibbo angle, $\theta_{13}^\text{PMNS} \simeq \theta_C \sin\theta_{23}^\text{PMNS}$,  in flavour models, as discussed in \cite{Antusch:2012fb}. 

As mentioned above, in supersymmetric GUT models the SUSY threshold corrections can have an important influence on the Yukawa coupling ratios. When the MSSM is matched to the SM, integrating out the sparticles at loop-level leads to the emergence of effective operators, which can contribute sizeably to the Yukawa couplings, depending on the values of the sparticle masses, $\tan\beta$, and the soft-breaking trilinear couplings. Thereby, via the SUSY threshold corrections, a given set of GUT predictions for the ratios $\frac{y_\tau}{y_b}$, $\frac{y_\mu}{y_s}$ and $\frac{y_e}{y_d}$ imposes important constraints on the SUSY spectrum.

The SUSY threshold corrections can be subdivided into two classes: While at tree-level the down-type quarks only couple to the Higgs field $H_d$, via exchange of sparticles at one-loop level they can also couple to $H_u$, as shown in Figure \ref{fig:loops1} in section \ref{sec:thresh}. When the sparticles are integrated out the emerging effective operator is enhanced for large $\tan\beta$ (i.e.\ ``$\tan\beta$-enhanced''). Analogously, there are also $\tan\beta$-enhanced threshold corrections to the charged lepton Yukawa couplings. For $Y_u$, however, the threshold effects emerging from effective couplings to $H_d$ are $\tan\beta$-suppressed. The second class of threshold corrections emerges from the supersymmetric loops shown in Figure \ref{fig:loops2}. While some of them are strongly suppressed, others lead to the emergence of effective operators proportional to the soft SUSY-breaking trilinear couplings. For large trilinear couplings, they too can become important. Given the importance of the SUSY threshold corrections, we will discuss them and their implementation in \SusyTC~in detail in the next section.

\section{SUSY threshold corrections \& numerical procedure}
\label{sec:thresh}

We follow the notation of \texttt{REAP} \cite{Antusch:2005gp} (see also \cite{Antusch:2002ek}) and use a RL convention for the Yukawa matrices. The MSSM superpotential extended by a type-I seesaw mechanism \cite{seesaw} is thus given by
\begin{align}
\label{eq:W}
W_\text{MSSM} & = Y_{e_{ij}} E^c_i H_d\cdot L_j + Y_{\nu_{ij}} N^c_i H_u\cdot L_j \nonumber \\
& + Y_{d_{ij}} D^c_i H_d\cdot Q_j - Y_{u_{ij}}U^c_i H_u\cdot Q_j + \frac{1}{2} M_{n_{ij}}N_i^c N_j^c + \mu H_u \cdot H_d\;,
\end{align}
where the left-chiral superfields $\Phi^c$ contain the charge conjugated fields $\psi^\dagger$ and $\tilde\phi_R^*$. We use the totally antisymmetric $SU(2)$ tensor $\epsilon_{12} = -\epsilon_{21}=1$ for the  product $\Phi \cdot \Psi \equiv \epsilon_{ab} \Phi^a \Psi^b$. The soft-breaking Lagrangian is given by
\begin{align}
\label{eq:Lsoft}
-\mathcal{L}_\text{soft} &= -\frac{1}{2} M_a \lambda^a\lambda^a + h.c. \nonumber \\
& + T_{e_{ij}} \tilde e_{R_i}^* H_d \cdot \tilde L_j + T_{\nu_{ij}} \tilde \nu_{R_i}^* H_u \cdot \tilde L_j +  T_{d_{ij}} \tilde d_{R_i}^* H_d \cdot \tilde Q_j - T_{u_{ij}} \tilde u_{R_i}^* H_u \cdot \tilde Q_j + h.c. \nonumber \\
&+ \tilde Q_i^\dagger m_{\tilde Q_{ij}}^2 \tilde Q_j + \tilde L_i^\dagger m_{\tilde L_{ij}}^2 \tilde L_j + \tilde u_{R_i}^* m_{\tilde u_{ij}}^2 \tilde u_{R_j} + \tilde d_{R_i}^* m_{\tilde d_{ij}}^2 \tilde d_{R_j} + \tilde e_{R_i}^* m_{\tilde e_{ij}}^2 \tilde e_{R_j} + \tilde \nu_{R_i}^* m_{\tilde \nu_{ij}}^2 \tilde \nu_{R_j} \nonumber \\
& + m^2_{H_u} |H_u|^2 + m^2_{H_d} |H_d|^2 + (m_3^2 H_u \cdot H_d + h.c.)\;.
\end{align}
Note that these conventions differ from SUSY Les Houches Accord 2 \cite{Allanach:2008qq}. They can easily be translated by
\begin{table}[h]
\begin{tabular}{c|ccccccccc}
\texttt{REAP} \& \SusyTC & $Y_u$ & $Y_d$ & $Y_e$ & $T_u$ & $T_d$ & $T_e$ & $m_{\tilde u}^2$ & $m_{\tilde d}^2$ & $m_{\tilde e}^2$\\
SLHA 2 & $Y_u^T$ & $Y_d^T$ & $Y_e^T$ & $T_u^T$ & $T_d^T$ & $T_e^T$ & $(m_{\tilde u}^2)^T$ & $(m_{\tilde d}^2)^T$ & $(m_{\tilde e}^2)^T$
\end{tabular}
\end{table}

Since \texttt{REAP} includes the RG running in the type-I seesaw extension of the MSSM (with the $\overline{\text{DR}}$  two-loop $\beta$-functions for the MSSM parameters and the neutrino mass operator given in \cite{Antusch:2002ek}), we have calculated the $\overline{\text{DR}}$ two-loop $\beta$-functions of the gaugino mass parameters $M_a$, the trilinear couplings $T_f$, the sfermion squared mass matrices $m^2_{\tilde f}$, and soft-breaking Higgs mass parameters $m_{H_u}^2$ and $m^2_{H_d}$ in the presence of $Y_\nu$, $M_n$ and $m_{\tilde\nu}^2$ (using the general formulas of \cite{Martin:1993zk}). We list these $\beta$-functions in appendix \ref{app:RGE}.

The Yukawa matrices and soft-breaking parameters are evolved to the SUSY scale 
\be
Q = \sqrt{m_{\tilde t_1}m_{\tilde t_2}}\;,
\ee
where the stop masses are defined by the up-type squark mass eigenstates $\tilde u_i$ with the largest mixing to $\tilde t_1$ and $\tilde t_2$.\footnote{\SusyTC~can also be set to use the convention $Q=\sqrt{m_{\tilde u_1}m_{\tilde u_6}}$ or a user-defined SUSY scale, as described in appendix \ref{app:DOC}.} \texttt{REAP} automatically integrates out the right-handed neutrinos, as described in \cite{Antusch:2005gp}. We assume that $M_n$ is much larger than the SUSY scale $Q$. \texttt{REAP} also features the possibility to add one-loop right-handed neutrino thresholds for the SM parameters, following \cite{Antusch:2015pda}.

\begin{figure}[h]
\begin{subfigure}{0.45\textwidth}
\includegraphics[scale=.8]{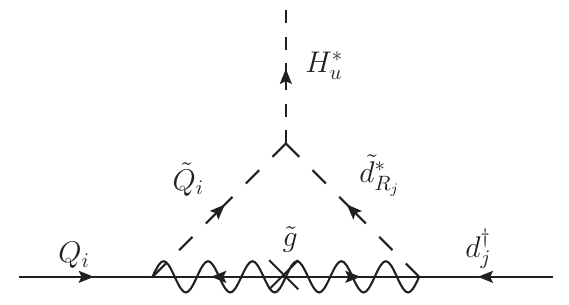}
\end{subfigure}\qquad
\begin{subfigure}{0.45\textwidth}
\includegraphics[scale=.8]{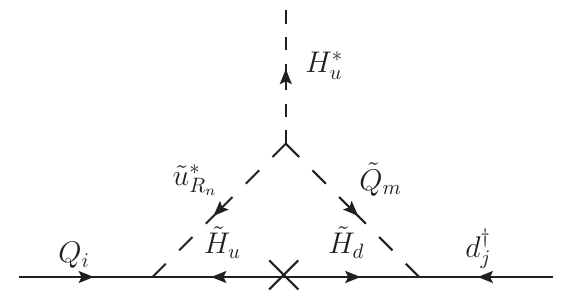}
\end{subfigure}
\begin{subfigure}{0.45\textwidth}
\includegraphics[scale=.8]{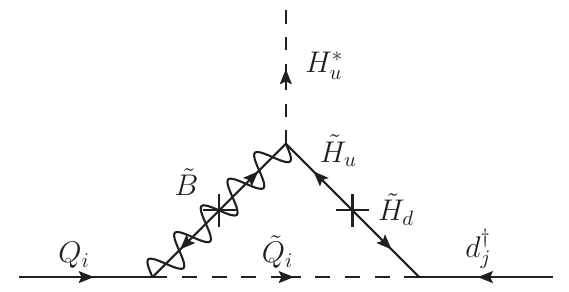}
\end{subfigure}\qquad
\begin{subfigure}{0.45\textwidth}
\includegraphics[scale=.8]{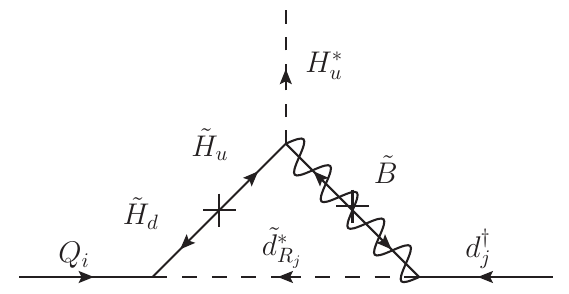}
\end{subfigure}
\begin{subfigure}{0.45\textwidth}
\includegraphics[scale=.8]{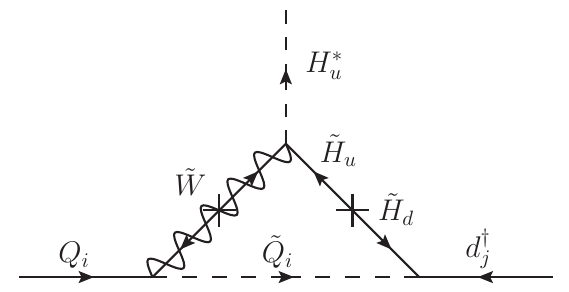}
\end{subfigure}\qquad
\begin{subfigure}{0.45\textwidth}
\includegraphics[scale=.8]{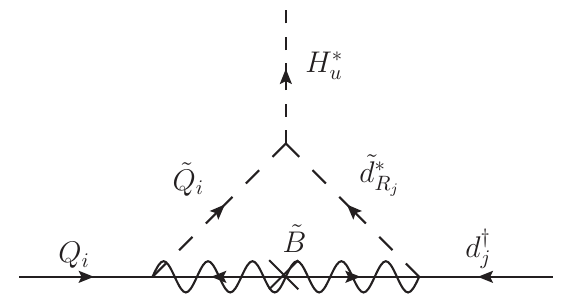}
\end{subfigure}
\caption{$\tan\beta$ - enhanced SUSY threshold corrections to $Y_d$.}
\label{fig:loops1}
\end{figure}

\begin{figure}[h]
\begin{subfigure}{0.45\textwidth}
\includegraphics[scale=.8]{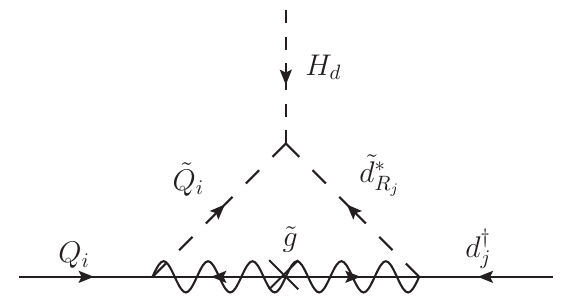}
\end{subfigure}\qquad
\begin{subfigure}{0.45\textwidth}
\includegraphics[scale=.8]{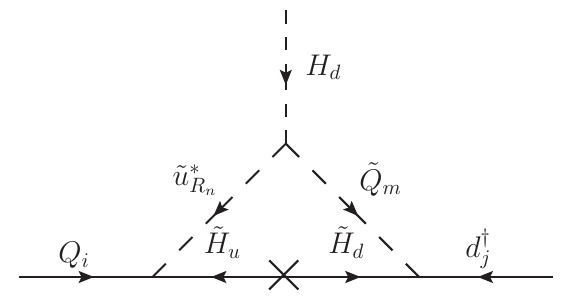}
\end{subfigure}
\begin{subfigure}{0.45\textwidth}
\includegraphics[scale=.8]{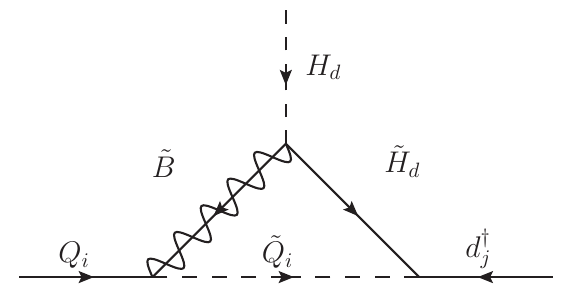}
\end{subfigure}\qquad
\begin{subfigure}{0.45\textwidth}
\includegraphics[scale=.8]{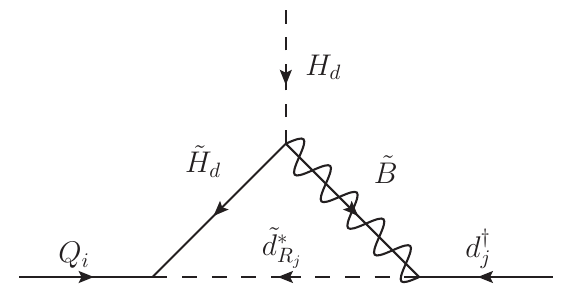}
\end{subfigure}
\begin{subfigure}{0.45\textwidth}
\includegraphics[scale=.8]{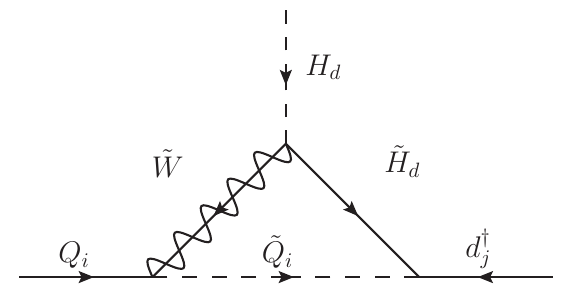}
\end{subfigure}\qquad
\begin{subfigure}{0.45\textwidth}
\includegraphics[scale=.8]{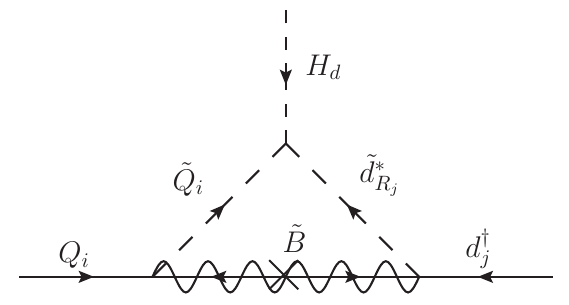}
\end{subfigure}
\caption{None $\tan\beta$ - enhanced SUSY threshold corrections to $Y_d$.}
\label{fig:loops2}
\end{figure}

At the SUSY scale $Q$ the tree-level sparticle masses and mixings are calculated. Considering heavy sparticles and large $Q\gtrsim\TeV$ the SUSY threshold corrections are calculated in the electroweak (EW) unbroken phase. In the EW unbroken phase there are in total twelve types of loop diagrams contributing to the SUSY threshold corrections for $Y_d$ (cf.\ Figures 2 and 3). The SUSY threshold corrections to $Y_d$ are calculated in the basis of diagonal squark masses and are given by
\begin{align}
\label{eq:Ydthreshold}
\tilde Y_{d_{ij}}^\text{SM} & = 
      	\tilde Y_{d_{ij}}^\text{MSSM} \cos\beta \left(1 + \frac{1}{16\pi^2}\tan\beta \left(\eta_{ij}^G + \eta_{ij}^W + \eta_{ij}^B + \eta_{ij}^T \right)+ \frac{1}{16\pi^2}\left(\epsilon_{ij}^W+\epsilon_{ij}^B+\epsilon_{ij}^T\right)\right)\nonumber \\
      	& + \tilde T_{d_{ij}} ^\text{MSSM} \cos\beta\frac{1}{16\pi^2}\left(\vphantom{\frac{1}{16\pi^2}}\zeta_{ij}^G+\zeta_{ij}^B\right)\;,\\
      	\intertext{where}
      	\label{eq:Ydthreshold2}
      	\eta^G_{ij} & = - \frac{8}{3} g_3^2 \frac{\mu^*}{M_3} H_2\left(\frac{m^2_{\tilde d_i}}{M_3^2}, \frac{m^2_{\tilde Q_j}}{M_3^2}\right)\;,\nonumber \\   
      	\eta_{ij}^W & = \frac{3}{2} g_2^2 \frac{M_2^*}{\mu} H_2 \left(\frac{M_2^2}{\mu^2}, \frac{m^2_{\tilde Q_j}}{\mu^2}\right)\;,\nonumber \\
\eta_{ij}^B & = \frac{3}{5} g_1^2 \left(
        	  \frac{1}{9} \frac{\mu^*}{M_1} H_2\left(\frac{m^2_{\tilde d_i}}{M_1^2}, \frac{m^2_{\tilde Q_j}}{M^2_1}\right) + \frac{1}{3} \frac{M_1^*}{\mu} H_2\left(\frac{m^2_{\tilde d_i}}{\mu^2},\frac{M^2_1}{\mu^2}\right) + \frac{1}{6} \frac{M_1^*}{\mu} H_2\left(\frac{M^2_1}{\mu^2}, \frac{m^2_{\tilde Q_j}}{\mu^2}\right) \right)\;,\nonumber\\
\eta_{ij}^T & = -\frac{1}{\mu \tilde Y_{d_{ij}}} \sum_{n,m}  \tilde Y_{d_{im}} \tilde T_{u_{mn}}^\dagger \tilde Y_{u_{nj}} H_2\left(\frac{m^2_{\tilde u_n}}{\mu^2}, \frac{m^2_{\tilde Q_m}}{\mu^2}\right)\;, \\
        	  \intertext{correspond to the $\tan\beta$-enhanced loops of Figure \ref{fig:loops1}, and}
        	  \label{eq:Ydthreshold3}
\epsilon^W_{ij} & = -12 g_2^2 C_{00}\left(\frac{\mu^2}{Q^2},\frac{M_2^2}{\mu^2}, \frac{m^2_{\tilde Q_j}}{\mu^2}\right)\;,\nonumber\\
\epsilon^B_{ij} & = -\frac{3}{5} g_1^2 \left(\frac{4}{3}C_{00}\left(\frac{\mu^2}{Q^2},\frac{M_1^2}{\mu^2}, \frac{m^2_{\tilde d_i}}{\mu^2}\right)+\frac{2}{3}C_{00}\left(\frac{\mu^2}{Q^2},\frac{M_1^2}{\mu^2}, \frac{m^2_{\tilde Q_j}}{\mu^2}\right)\right)\;,\nonumber\\
\epsilon_{ij}^T & = \frac{1}{\tilde Y_{d_{ij}}} \sum_{n,m}  \tilde Y_{d_{im}} \tilde Y_{u_{mn}}^\dagger \tilde Y_{u_{nj}} H_2\left(\frac{m^2_{\tilde u_n}}{\mu^2}, \frac{m^2_{\tilde Q_m}}{\mu^2}\right)\;,\nonumber \\
\zeta^G_{ij} & = \frac{8}{3} g_3^2 \frac{1}{M_3} H_2\left(\frac{m^2_{\tilde d_i}}{M_3^2},\frac{m^2_{\tilde Q_j}}{M_3^2}\right)\;,\nonumber \\
\zeta_{ij}^B & = -\frac{3}{5} g_1^2 \frac{1}{9} \frac{1}{M_1} H_2\left(\frac{m^2_{\tilde d_i}}{M_1^2},\frac{m^2_{\tilde Q_j}}{M_1^2}\right)\;,
\end{align}
correspond to the loops in Figure \ref{fig:loops2}, respectively, where the contributions $\zeta^G_{ij}$ and $\zeta^B_{ij}$ can become important in cases of small $\tan\beta$ and large trilinear couplings. The loop functions $H_2$ and $C_{00}$ are defined as
\begin{align}
H_2(x,y) := & \frac{x \log(x)}{(1 - x) (x - y)} + \frac{y \log(y)}{(1 - y)(y - x)}\;, \\
C_{00}(q,x,y) := & \frac{1}{4}\left(\frac{3}{2}-\log(q)+\frac{x^2 \log(x)}{(1 - x) (x - y)} + \frac{y^2 \log(y)}{(1 - y)(y - x)}\right)\;.
\end{align}
$\tilde Y$, $\tilde T$ are the Yukawa- and trilinear coupling matrices rotated into the basis where the squark mass matrices are diagonal, using the transformations
\begin{align}
\label{eq:rotY}
Y_u & = \tilde W_{\tilde u}\tilde Y_u \tilde W_{\tilde Q}^\dagger\;,\nonumber \\
T_u & = \tilde W_{\tilde u}\tilde T_u \tilde W_{\tilde Q}^\dagger\;,\nonumber \\
m^2_{\tilde Q} & = \tilde W_{\tilde Q} m^{2~diag}_{\tilde Q} \tilde W_{\tilde Q}^\dagger\;, \nonumber \\
m^2_{\tilde u} & = \tilde W_{\tilde u} m^{2~diag}_{\tilde u} \tilde W_{\tilde u}^\dagger \;,
\end{align}
and analogously for down-type (s)quarks and charged (s)leptons.

The SUSY threshold corrections to $Y_e$ are given by
\begin{align}
\tilde Y_{e_{ij}}^\text{SM} & =     	\tilde Y_{e_{ij}}^\text{MSSM} \cos\beta \left(1 + \frac{1}{16\pi^2}\tan\beta \left(\tau_{ij}^W + \tau_{ij}^B \right)+\frac{1}{16\pi^2}\left(\delta_{ij}^W+\delta_{ij}^B\right)\right) \nonumber\\
& + \tilde T_{e_{ij}}^\text{MSSM} \cos\beta \frac{1}{16\pi^2} \xi_{ij}^B \;,\\
      	\intertext{with the $\tan\beta$-enhanced contributions}
      	\tau_{ij}^W & = \frac{3}{2} g_2^2 \frac{M_2^*}{\mu} H_2\left(\frac{M_2^2}{\mu^2}, \frac{m^2_{\tilde L_j}}{\mu^2}\right)\;, \nonumber\\ 
\tau_{ij}^B & = \frac{3}{5} g_1^2 \left(- \frac{\mu^*}{M_1} H_2\left(\frac{m^2_{\tilde e_i}}{M_1^2},\frac{m^2_{\tilde L_j}}{M_1^2}\right) + \frac{M_1^*}{\mu} H_2\left(\frac{ m^2_{\tilde e_i}}{\mu^2},\frac{M_1^2}{\mu^2}\right) - \frac{1}{2} \frac{M^*_1}{\mu} H_2\left(\frac{M_1^2}{\mu^2}, \frac{m^2_{\tilde L_j}}{\mu^2}\right) \right)\;,\\
\intertext{and}
\delta^W_{ij} & = - 12 g_2^2 C_{00}\left(\frac{\mu^2}{Q^2},\frac{M_2^2}{\mu^2}, \frac{m^2_{\tilde L_j}}{\mu^2}\right)\;,\nonumber\\
\delta^B_{ij} & = -\frac{3}{5} g_1^2 \left(-2C_{00}\left(\frac{\mu^2}{Q^2},\frac{M_1^2}{\mu^2}, \frac{m^2_{\tilde L_j}}{\mu^2}\right)+4C_{00}\left(\frac{\mu^2}{Q^2},\frac{M_1^2}{\mu^2}, \frac{m^2_{\tilde e_i}}{\mu^2}\right)\right)\;,\nonumber\\
\xi_{ij}^B & = \frac{3}{5} g_1^2 \frac{1}{M_1} H_2\left(\frac{m^2_{\tilde e_i}}{M_1^2},\frac{m^2_{\tilde L_j}}{M_1^2}\right)\;.
\end{align}
The diagrams for the $Y_e$ SUSY threshold corrections are analogous to the ones in Figures \ref{fig:loops1} and \ref{fig:loops2}, with the exception that the loop diagrams shown in the top rows do not exist.

Turning to $Y_u$, the types of diagrams which were $\tan\beta$-enhanced for $Y_d$ and $Y_e$ are now $\tan\beta$-suppressed. However, there also exist SUSY threshold corrections which are independent of $\tan\beta$ and enhanced by large trilinear couplings. These SUSY threshold corrections to $Y_u$ can have  important effects. For example the SUSY threshold corrections to the top Yukawa coupling $y_t$ can be of significance in analyses of the Higgs mass and vacuum stability. The expression for the $Y_u$ SUSY threshold corrections can be readily obtained from the SUSY threshold corrections to $Y_d$ \eqref{eq:Ydthreshold}--\eqref{eq:Ydthreshold3} by the replacement
\begin{align}
d & \rightarrow u\;, \nonumber \\
\cos\beta & \rightarrow \sin\beta\;,
\end{align}
with the exception of the bino-loops, whose contribution become
\begin{align}
\eta_{ij}^B & = \frac{3}{5} g_1^2 \left(
        	  -\frac{2}{9} \frac{\mu^*}{M_1} H_2\left(\frac{m^2_{\tilde d_i}}{M_1^2}, \frac{m^2_{\tilde Q_j}}{M^2_1}\right) + \frac{2}{3} \frac{M_1^*}{\mu} H_2\left(\frac{m^2_{\tilde u_i}}{\mu^2},\frac{M^2_1}{\mu^2}\right) - \frac{1}{6} \frac{M_1^*}{\mu} H_2\left(\frac{M^2_1}{\mu^2}, \frac{m^2_{\tilde Q_j}}{\mu^2}\right) \right)\;,\nonumber\\
\epsilon^B_{ij} & = -\frac{3}{5} g_1^2 \left(\frac{8}{3}C_{00}\left(\frac{\mu^2}{Q^2},\frac{M_1^2}{\mu^2}, \frac{m^2_{\tilde u_i}}{\mu^2}\right)-\frac{2}{3}C_{00}\left(\frac{\mu^2}{Q^2},\frac{M_1^2}{\mu^2}, \frac{m^2_{\tilde Q_j}}{\mu^2}\right)\right)\;,\nonumber \\
\zeta_{ij}^B & = \frac{3}{5} g_1^2 \frac{2}{9} \frac{1}{M_1} H_2\left(\frac{m^2_{\tilde u_i}}{M_1^2},\frac{m^2_{\tilde Q_j}}{M_1^2}\right)\;,
\end{align}
due to the different $U(1)$ hypercharges of the (s)particles in the loop. The loop diagrams are identical to the ones of Figure \ref{fig:loops1} and \ref{fig:loops2}, with  $u$ exchanged by $d$.

Finally \SusyTC~calculates the value of $|\mu|$ and $m_3$ from $m^2_{H_u}$, $m^2_{H_d}$, $\tan\beta$ and $M_Z$ by requiring the existence of spontaneously broken EW vacuum, which is equivalent to vanishing one-loop corrected tad-pole equations of $H_u$ and $H_d$
\be
\label{eq:mu}
\mu = \e^{i\phi_\mu} \sqrt{\frac{1}{2} \left(\tan(2 \beta) \left(\bar m^2_{H_u} \tan\beta
      - \bar m^2_{H_d}\cot\beta \right)- M_Z^2 - Re\left(\Pi_{ZZ}^T\left(M_Z^2\right)\right)\right)}\;,
\ee
with $\bar m^2_{H_u} \equiv m^2_{H_u} - t_u $ and $\bar m^2_{H_d} \equiv m^2_{H_d} - t_d$. In the real (CP conserving) MSSM the phase $\phi_\mu$ is restricted to $0$ and $\pi$. The expressions for the one-loop tadpoles $t_u$, $t_d$ and the transverse Z-boson self energy $\Pi_{ZZ}^T$ are based on \cite{Pierce:1996zz}, but extended to include inter-generational mixing, and are presented in appendix \ref{app:FORM}. Because $\mu$ enters the one-loop formulas for the threshold corrections, treating $t_u$, $t_d$ and $\Pi_{ZZ}^T$ as functions of tree-level parameters is sufficiently accurate. The one-loop expression of the soft-breaking mass $m_3$ is calculated as
\be
\label{eq:m3}
m_3 =  \sqrt{\frac{1}{2} \left(\tan(2 \beta) \left(\bar m^2_{H_u} - \bar m^2_{H_d} \right)- \left(M_Z^2 + Re\left(\Pi_{ZZ}^T\left(M_Z^2\right)\right)\right)\sin(2 \beta)\right)}\;.
\ee

If desired, \texttt{SusyTC} allows to outsource a two-loop Higgs mass calculation to external software, e.g.~\texttt{FeynHiggs} \cite{Heinemeyer:1998yj,Borowka:2014wla,Hollik:2014bua}, by calculating the pole mass $m_{H^+}$ ($m_A$) as input for the complex (real) MSSM

\begin{align}
m^2_{H^+}  = \frac{1}{\cos(2 \beta)} & \left(\bar m^2_{H_u}-\bar m^2_{H_d} -M_Z^2 -Re\left(\Pi_{ZZ}^T\left(M_Z^2\right)\right)+\hat M_W^2 \vphantom{\frac{1}{2}}\right. \nonumber \\
& \left.\vphantom{\frac{1}{2}} - Re\left(\Pi_{H^+H^-}\left(m_{H^+}\right)\right)+ t_d\sin^2\beta +t_u\cos^2\beta \right)\;, \\
m^2_A = \frac{1}{\cos(2 \beta)} & \left(\bar m^2_{H_u}-\bar m^2_{H_d} -M_Z^2 -Re\left(\Pi_{ZZ}^T\left(M_Z^2\right)\right)\vphantom{\frac{1}{2}}\right. \nonumber \\
& \left.\vphantom{\frac{1}{2}} Re\left(\Pi_{AA}\left(m_A\right)\right)+ t_d\sin^2\beta +t_u\cos^2\beta \right)\;,
\end{align}
where $\hat M_W$ is the $\overline{\text{DR}}$ W-boson mass given as
\be
\hat M^2_W(Q) = M_W^2 + Re\left(\Pi_{WW}^T\left(M_W^2\right)\right) = g_2 \frac{\hat v(Q)}{2}\;,
\ee
with $M_Z$ and $M_W$ pole masses and the $\overline{\text{DR}}$ vacuum expectation value $\hat v(Q)$ given by
\be
\label{eq:vev}
\hat v^2(Q) =4 \frac{M_Z^2+Re\left(\Pi_{ZZ}^T\left(M_Z^2\right)\right)}{\frac{3}{5}g_1^2(Q) + g_2^2(Q)}\;.
\ee
As in the previous formulas, the self energies $\Pi_{H^+H^-}$ and $\Pi_{AA}$ are based on \cite{Pierce:1996zz}, but extended to include inter-generational mixing, and are understood as functions of tree-level parameters. They are given in appendix \ref{app:FORM}.

\section{The \texttt{REAP} extension \SusyTC}
\label{sec:doc}
In this section we provide a ``Getting Started'' calculation for \SusyTC. A full documentation of all features is included in appendix \ref{app:DOC}. Since \SusyTC~is an extension to \texttt{REAP}, an up-to-date version of \texttt{REAP-MPT} \cite{Antusch:2005gp} (available at \url{http://reapmpt.hepforge.org}) needs to be installed on your system. \SusyTC~consists out of the \texttt{REAP} model file \texttt{RGEMSSMsoftbroken.m}, which is based on the model file \texttt{RGEMSSM.m} of \texttt{REAP 1.11.2} and additionally contains, among other things, the RGEs of the MSSM soft-breaking parameters and the matching to the SM, and the file \texttt{SusyTC.m}, which includes the formulas for the sparticle spectrum and SUSY threshold correction calculations. Both files can be downloaded from \url{http://particlesandcosmology.unibas.ch/pages/SusyTC.htm} and have to be copied into the \texttt{REAP} directory.

To begin a calculation with \texttt{SusyTC}, one first needs to import \texttt{RGEMSSMsoftbroken.m}:\\

\hspace{1cm}\textbf{	Needs[}\texttt{"REAP`RGEMSSMsoftbroken`"}\textbf{];}\\

\noindent
The model \texttt{MSSMsoftbroken} is then defined by \texttt{RGEAdd}, including additional options such as \texttt{RGEtan$\beta$}:\\

\hspace{1cm}\textbf{	RGEAdd[}\texttt{"MSSMsoftbroken"},RGEtan$\beta\rightarrow$ 30\textbf{];}\\

\noindent
In \texttt{MSSMsoftbroken} all \texttt{REAP} options of the model \texttt{MSSM} are available. The options additionally available in \SusyTC~are given in appendix \ref{app:DOC}. The input is given by \texttt{RGESetInitial}. Let us illustrate some features of \SusyTC: To test for example the GUT scale prediction for the Yukawa coupling ratio $\frac{y_\mu}{y_s}=6$, considering a given example parameter point in the Constrained MSSM, one can type:\\

\hspace{1cm}\textbf{RGESetInitial[}\texttt{2$\cdot$10$^\wedge$16,}\\
\hspace*{2cm}\texttt{RGEYd $\rightarrow$}
\textbf{DiagonalMatrix[}\texttt{\{1.2$\cdot$10$^\wedge$-3, 2.2$\cdot$10$^\wedge$-3, 0.16\}}\textbf{]}\texttt{,}\\
\hspace*{2cm}\texttt{RGEYe $\rightarrow$}
\textbf{DiagonalMatrix[}\texttt{\{1.2$\cdot$10$^\wedge$-3, 6$\cdot$2.2$\cdot$10$^\wedge$-3, 0.16\}}\textbf{]}\texttt{,}\\
\hspace*{2cm}\texttt{RGEM12 $\rightarrow$ 2000, RGEA0 $\rightarrow$ 1000, RGEm0 $\rightarrow$ 2500}\textbf{];}\\

\noindent Of course, any general matrices can be used as input for the Yukawa, trilinear, and soft-breaking matrices, as given by the specific SUSY flavour GUT model under consideration. Also, non-universal gaugino masses can be specified.  
The RGEs are then solved from the GUT scale to the Z-boson mass scale by\\

\hspace{1cm}\textbf{RGESsolve[}\texttt{91,2$\cdot$10$^\wedge$16}\textbf{];}\\

\noindent
The ratio of the $\mu$ and strange quark Yukawa couplings at the Z-boson mass scale can now be obtained with \texttt{RGEGetSolution, CKMParameters,} and \texttt{MNSParameters}:\\

\hspace{1cm}\texttt{Yu = }\textbf{RGEGetSolution[}\texttt{91, RGEYu}\textbf{];}\\
\hspace*{1.6cm}\texttt{Yd = }\textbf{RGEGetSolution[}\texttt{91, RGEYd}\textbf{];}\\
\hspace*{1.6cm}\texttt{Ye = }\textbf{RGEGetSolution[}\texttt{91, RGEYe}\textbf{];}\\
\hspace*{1.6cm}\texttt{M$\nu$ = }\textbf{RGEGetSolution[}\texttt{91, RGEM$\nu$}\textbf{];}\\
\hspace*{1.6cm}\textbf{MNSParameters[}\texttt{M$\nu$,Ye}\textbf{]}\texttt{[[3, 2]]/}\textbf{CKMParameters[}\texttt{Yu, Yd}\textbf{]}\texttt{[[3, 2]]}\\

\begin{figure}[H]
\includegraphics[scale=.7]{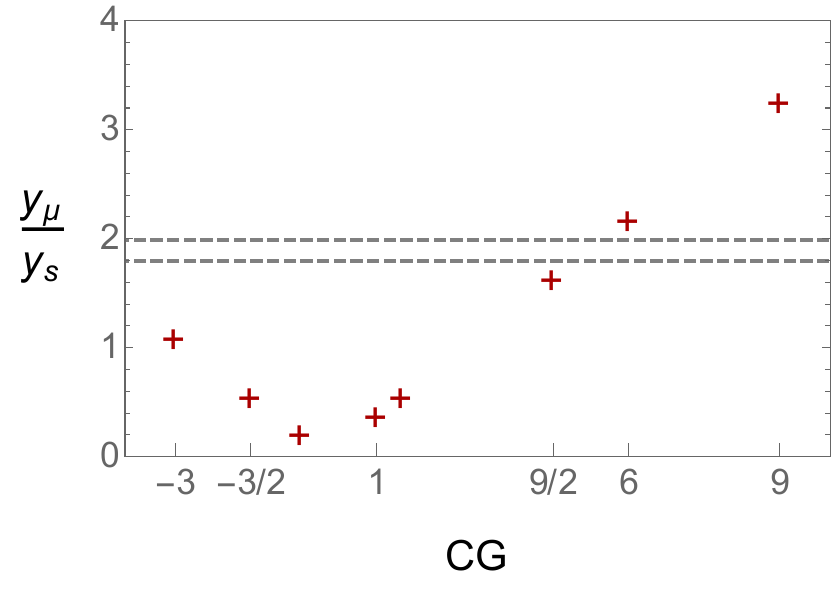}
\caption{Example results for $\frac{y_\mu}{y_s}$ at the electroweak scale, considering the $SU(5)$ GUT-scale CG factors from Table 2 of \cite{Antusch:2009gu}, i.e.\ the GUT predictions $\frac{y_\mu}{y_s} = \text{CG}$, for a given example Constrained MSSM parameter point with $\tan\beta=30$, $m_{1/2}=2000~\GeV$, $A_0=1000~\GeV$, and $m_0=2500~\GeV$. The area between the dashed gray lines corresponds to the experimental $1\sigma$ range \cite{Antusch:2013jca}.}
\label{fig:clebsch}
\end{figure}

\noindent
Repeating this calculation with all $SU(5)$ CG factors listed in Table 2 of \cite{Antusch:2009gu}, one obtains the results shown in Figure \ref{fig:clebsch}.

As described in appendix \ref{app:DOC}, \SusyTC~can also read and write ``Les Houches'' files \cite{Allanach:2008qq,Skands:2003cj} as input and output.

\section{The Sparticle Spectrum predicted from CG factors}

In this section we apply \SusyTC~to investigate the constraints on the sparticle spectrum which arise from a set of GUT scale predictions for the quark-lepton Yukawa coupling ratios $\frac{y_e}{y_d}$, $\frac{y_\mu}{y_s}$, and $\frac{y_\tau}{y_b}$. As GUT scale boundary conditions for the SUSY-breaking terms we take the Constrained MSSM. The experimental constraints are given by the Higgs boson mass $m_H=125.7\pm 0.4~\text{GeV}$ \cite{Agashe:2014kda} as well as the charged fermion masses (and the quark mixing matrix). 
We use the experimental constraints for the running $\overline{\text{MS}}$ Yukawa couplings at the Z-boson mass scale calculated in \cite{Antusch:2013jca}, where we set the uncertainty of the charged lepton Yukawa couplings to one percent to account for the estimated theoretical uncertainty (which here exceeds the experimental uncertainty). When applying the measured Higgs mass as constraint, we use a 1$\sigma$ interval of $\pm 3~\GeV$, including the estimated theoretical uncertainty. 

For our study, we consider GUT scale Yukawa coupling matrices which feature the GUT-scale Yukawa relations $\frac{y_e}{y_d}=-\frac{1}{2}$, $\frac{y_\mu}{y_s}=6$, and $\frac{y_\tau}{y_b}=-\frac{3}{2}$ (cf.\ \cite{Antusch:2009gu}):
\begin{align}
\label{eq:model}
Y_d & = \begin{pmatrix}
y_d & 0 & 0\\
0 & y_s & 0\\
0 & 0 & y_b
\end{pmatrix},  \quad
Y_e=\begin{pmatrix}
-\frac{1}{2}y_d & 0& 0\\
0 & 6 y_s & 0\\
0 & 0 & -\frac{3}{2}y_b
\end{pmatrix},\nonumber \\
Y_u & =\begin{pmatrix}
y_u & 0 & 0\\
0 & y_c & 0\\
0 & 0 & y_t
\end{pmatrix} U_\text{CKM}(\theta_{12},\theta_{13},\theta_{23},\delta)\;,
\end{align}
These GUT relations can emerge as direct result of CG factors in SU(5) GUTs or as approximate relation after diagonalization of the GUT-scale Yukawa matrices $Y_d$ and $Y_e$ (cf.\ \cite{Antusch:2013kna,Antusch:2013tta,Antusch:2014poa,Gehrlein:2014wda}). For the soft-breaking parameters we restrict our analysis to the Constrained MSSM parameters $m_0$, $m_{1/2}$, and $A_0$, with $\mu$ determined from requiring the breaking of electroweak symmetry as in \eqref{eq:mu} and set $\sgn(\mu)=+1$. We note that in specific models for the GUT Higgs potential, for instance in \cite{Antusch:2014poa}, $\mu$ can be realized as an effective parameter of the superpotential with a fixed phase, including the case that $\mu$ is real. The value of $\tan\beta$ is fixed at the SUSY scale $Q$ to $\tan\beta=30$.\footnote{We have not included $\tan\beta$ in the fit, however we tried different values and $\tan\beta=30$ turned out to be the best choice for obtaining a good fit.}

We note that we have also added a neutrino sector, i.e.\ a neutrino Yukawa matrix $Y_\nu$ and and a mass matrix $M_n$ of the right-handed neutrinos, but we have set the entries of $Y_\nu$ to very small values below $\mathcal{O}(10^{-3})$, such that their effects on the RG evolution can be safely neglected, and the masses of the right-handed neutrinos to values many orders of magnitude higher than the expected SUSY scale. With these parameters, the neutrino sector is decoupled from the main analysis. Such small values of the neutrino Yukawa couplings are e.g.\ expected in the models \cite{Antusch:2013kna,Antusch:2013tta, Gehrlein:2014wda}, where they arise as effective operators.

Using one-loop RGEs, \texttt{REAP} 1.11.2 and \SusyTC~we determine the soft-breaking parameters and $\mu$ at the SUSY scale, as well as the pole mass $m_{H^+}$. This output is then passed to \texttt{FeynHiggs} 2.11.2 \cite{Heinemeyer:1998yj,Borowka:2014wla,Hollik:2014bua} in order to calculate the two-loop corrected pole masses of the Higgs bosons in the complex MSSM. The MSSM is automatically matched to the SM and we compare the results for the Yukawa couplings at the Z-boson mass scale with the experimental values reported in \cite{Antusch:2013jca}. 

When fitting the GUT-scale parameters to the experimental data, we found that our results for the up-type quark Yukawa couplings and CKM angles and CP-phase could be fitted to agree with observations to at least $10^{-3}$ relative precision, by adjusting the parameters of $Y_u$. The remaining six parameters are used to fit the Yukawa couplings of down-type quarks and charged leptons, as well as the mass of the SM-like Higgs boson. We find a benchmark point with a $\chi^2=3.75$:
\begin{table}[H]
\begin{tabular}{ccc}
\toprule
\multicolumn{3}{c}{input GUT-scale parameters}\\
\midrule
 $y_d$ & $y_s$ & $y_b$ \\
 $1.27\cdot 10^{-4}$ & $2.25\cdot 10^{-3}$ & $0.160$ \\[1ex]
 $m_0$ & $A_0$ & $m_{1/2}$ \\
 $2129.3$ & $-5822.27$ & $1007.94$ \\
 \bottomrule
\end{tabular}
\end{table}
\noindent

\begin{figure}[h]
\includegraphics[scale=0.5]{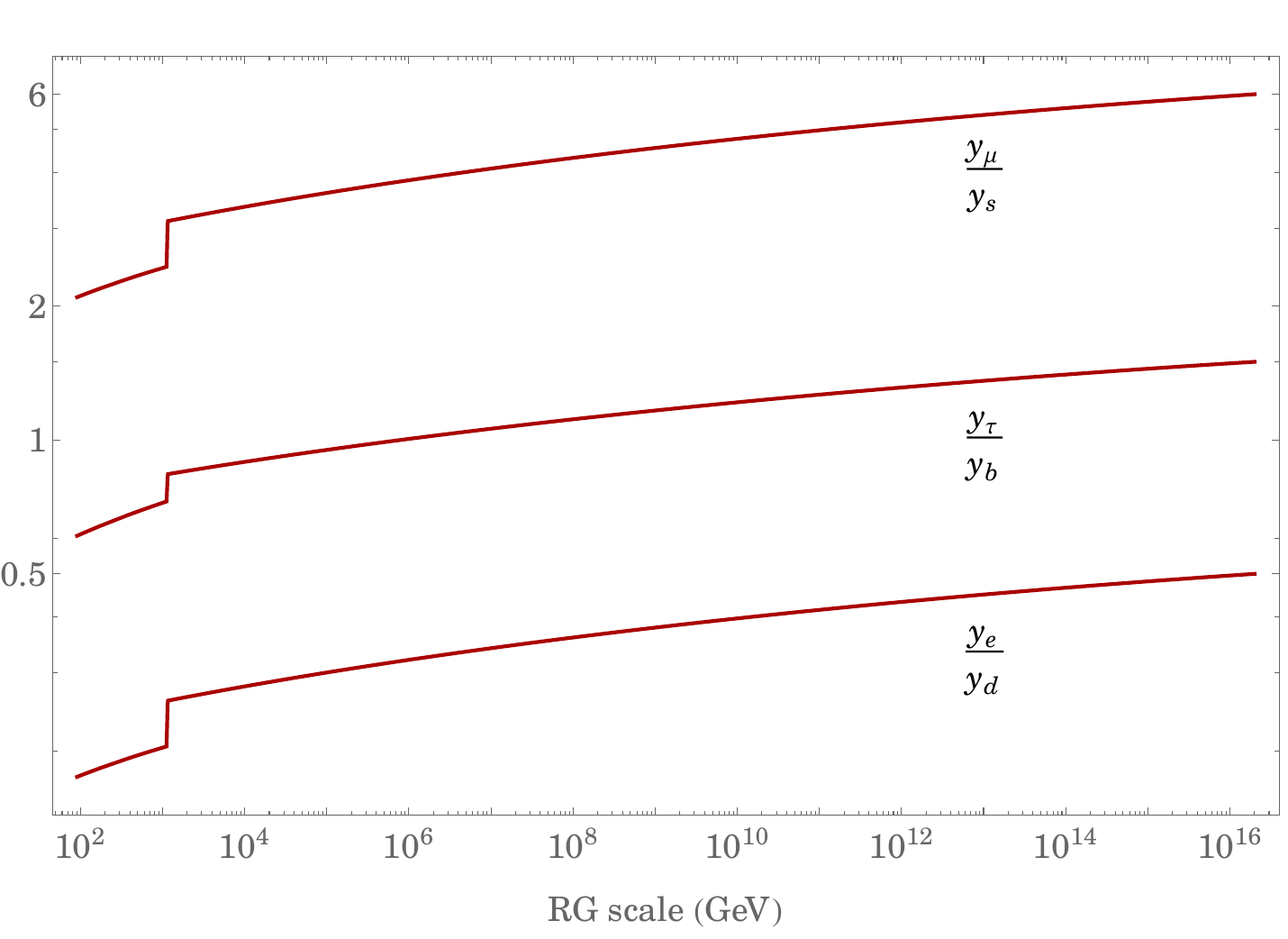}
\caption{RG evolution of the Yukawa coupling ratios of the first, second, and third family from the GUT-scale to the mass scale of the Z-boson. The GUT scale parameters correspond to our benchmark point from Section 5. The effects of the threshold corrections are clearly visible at the SUSY scale $Q=1128~\GeV$.}
\label{fig:runningratio}
\end{figure}

\begin{figure}
\includegraphics[scale=0.7]{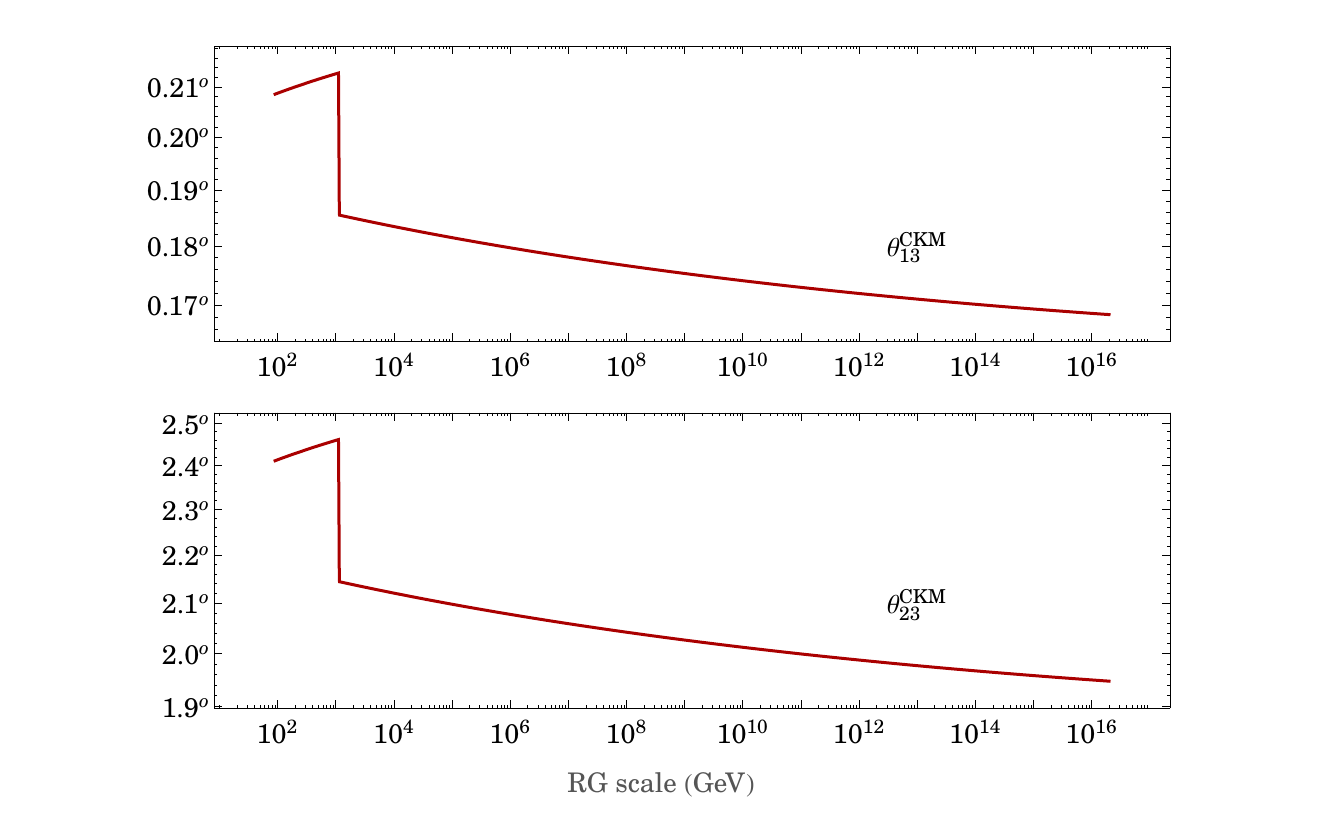}
\caption{RG evolution of the CKM mixing angles $\theta_{13}^\text{CKM}$ and $\theta_{23}^\text{CKM}$ from the GUT-scale to the mass scale of the Z-boson. The GUT scale parameters correspond to our benchmark point from Section 5. The effects of the threshold corrections are clearly visible at the SUSY scale $Q=1128~\GeV$.}
\label{fig:runningangles}
\end{figure}

\begin{figure}
\includegraphics[scale=0.6]{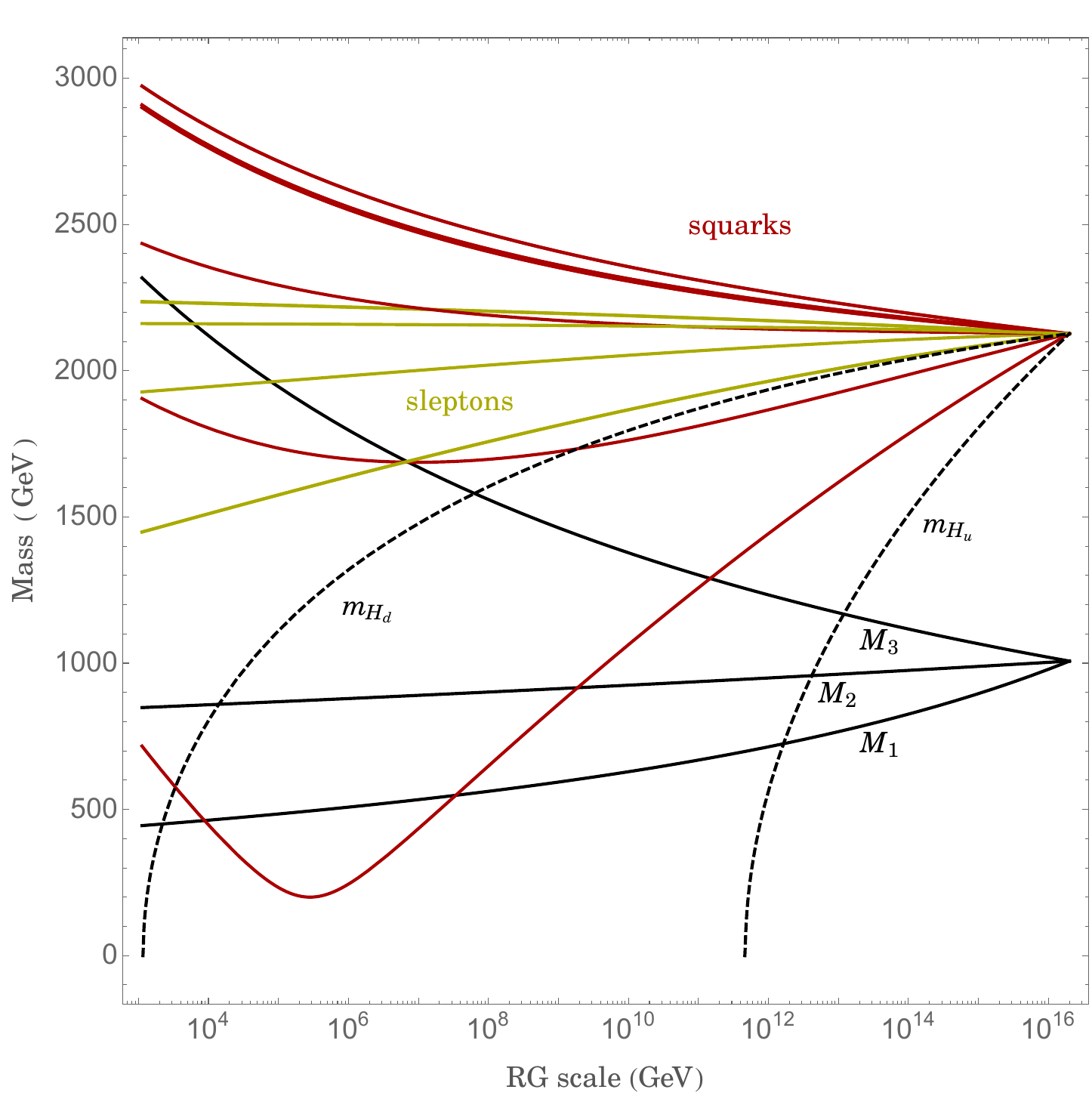}
\caption{RG evolution of the soft-breaking gaugino masses (solid black), squark masses (red), slepton masses (yellow), and $m_{H_u}$ and $m_{H_d}$ (dashed black), from the GUT-scale to the SUSY scale $Q=1128~\GeV$. The GUT scale parameters correspond to our benchmark point from Section 5. }
\label{fig:runningsoftterms}
\end{figure}

\begin{table}[H]
\begin{tabular}{cccc}
\toprule
\multicolumn{4}{c}{low energy results}\\
\midrule
 $y_e$ & $y_\mu$  &  $y_\tau$ & \\
 $2.79\cdot 10^{-6}$ & $5.92\cdot 10^{-4}$  & $1.00\cdot 10^{-2}$ & \\[1ex]
 $m_H$ & $y_d$ & $y_s$ & $y_b$  \\
 $125.0$  & $1.60\cdot 10^{-5}$ & $2.82\cdot 10^{-4}$ & $1.64\cdot 10^{-2}$\\
 \bottomrule
\end{tabular}
\end{table}
\noindent
Looking at our results for the low-energy Yukawa coupling ratios, $\frac{y_e}{y_d}=0.17$, $\frac{y_\mu}{y_s}=2.10$, and $\frac{y_\tau}{y_b}=0.61$, the importance of SUSY threshold corrections in evaluating the GUT-scale Yukawa ratios becomes evident. This can also be seen in Figure \ref{fig:runningratio}. Additionally, as shown in Figure \ref{fig:runningangles}, SUSY threshold corrections also affect the CKM mixing angles. Finally, in Figure \ref{fig:runningsoftterms} we show the results for the RG running of the soft-breaking mass parameters.

\begin{figure}

\includegraphics[scale=.6]{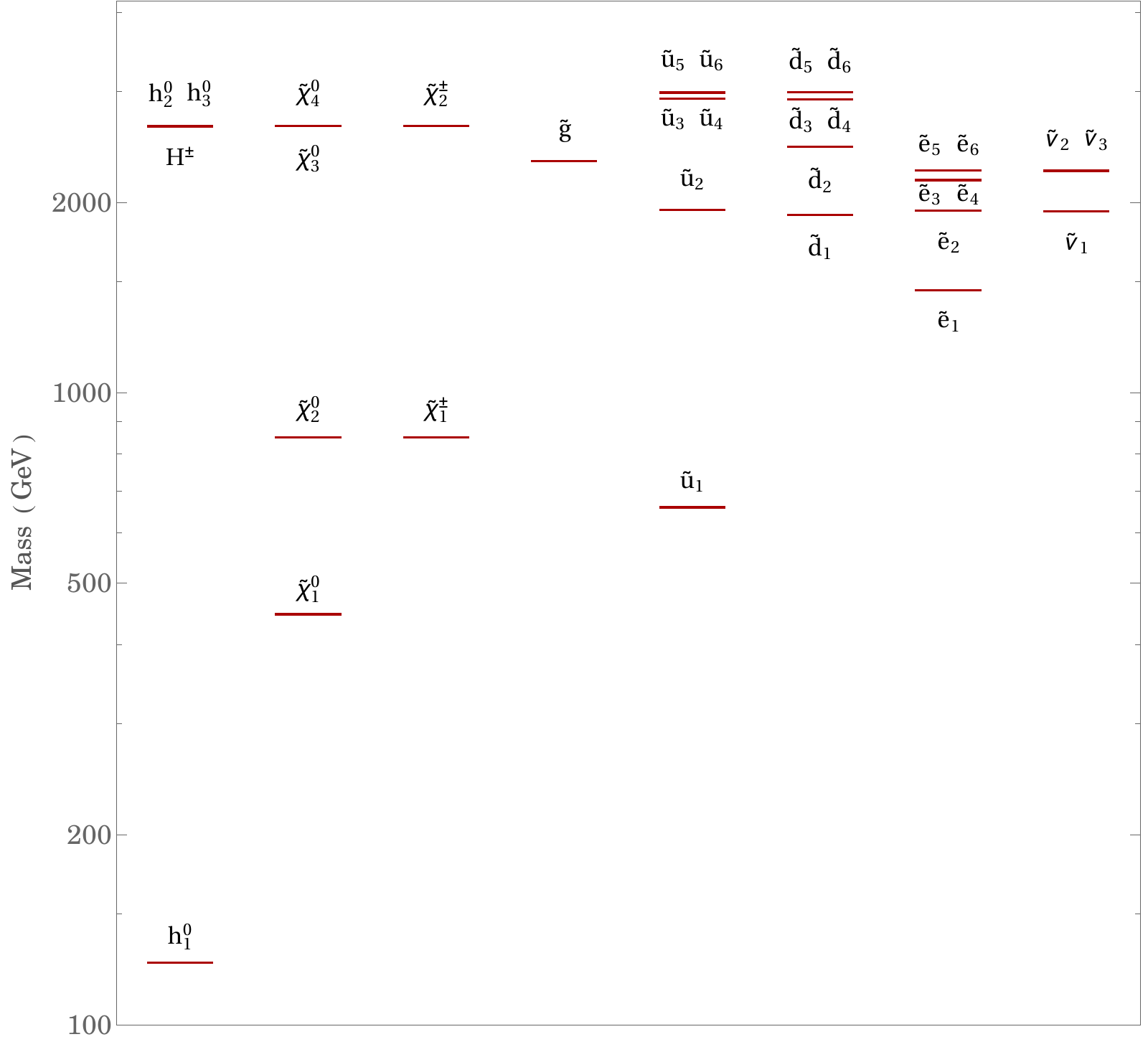}
\caption{SUSY spectrum with $SU(5)$ GUT scale boundary conditions $\frac{y_e}{y_d}=-\frac{1}{2}$, $\frac{y_\mu}{y_s}=6$, and $\frac{y_\tau}{y_b}=-\frac{3}{2}$, corresponding to our benchmark point from Section 5.}
\label{fig:spectrum}
\end{figure}

The SUSY spectrum obtained by \SusyTC~is shown in Figure \ref{fig:spectrum}.
The lightest supersymmetric particle (LSP) is a bino-like neutralino of about 445~$\GeV$. The next-to-lightest supersymmetric particle (NLSP) is a stop of about 656~$\GeV$. The SUSY scale is obtained as $Q=1128~\GeV$. The $\mu$ parameter obtained from requiring spontaneous electroweak symmetry breaking is given by $\mu = 2630~\GeV$. Note that the only experimental constraints we used were the results for quark and charged lepton masses as well as $m_H$. In particular, no bounds on the sparticle masses were applied as well as no restrictions from the neutralino relic density (which would require further assumptions on the cosmological evolution). 

Due to the large (absolute) values of the trilinear couplings, we find using the constraints from \cite{Casas:1995pd}, that the vacuum of our benchmark point is meta-stable. The scalar potential possesses charge and colour breaking (CCB) vacua, as well as one ``unbounded from below'' (UFB) field direction in parameter space. However, estimating the stability of the vacuum via the Euclidean action of the ``bounce'' solution \cite{Coleman:1977py,Callan:1977pt} (following \cite{Sarid:1998sn}) shows that the lifetime of the vacuum is many orders larger than the age of the universe.

\begin{figure}
\begin{subfigure}{0.45\textwidth}
\includegraphics[scale=.41]{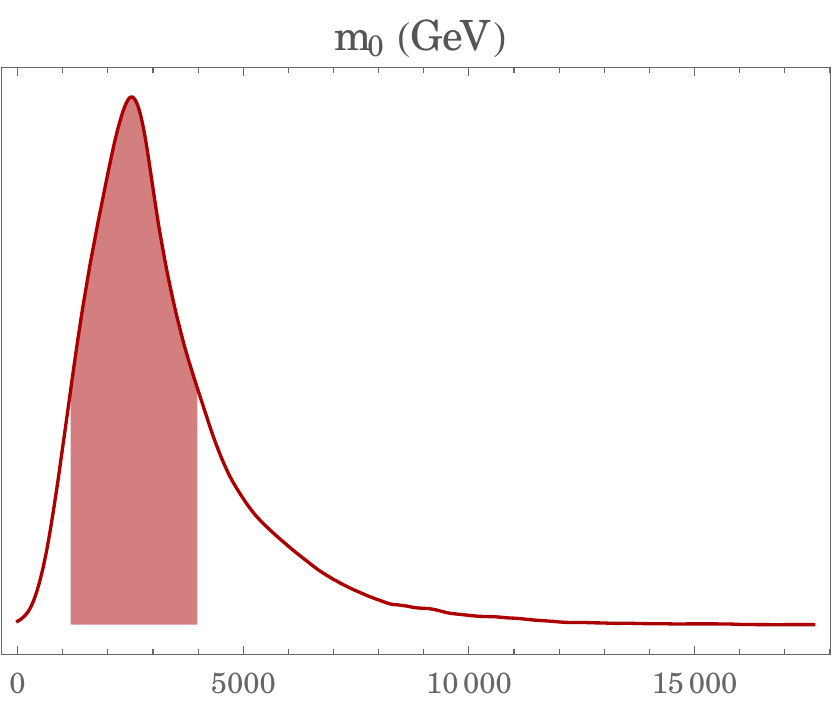}
\end{subfigure}\quad
\begin{subfigure}{0.45\textwidth}
\includegraphics[scale=.41]{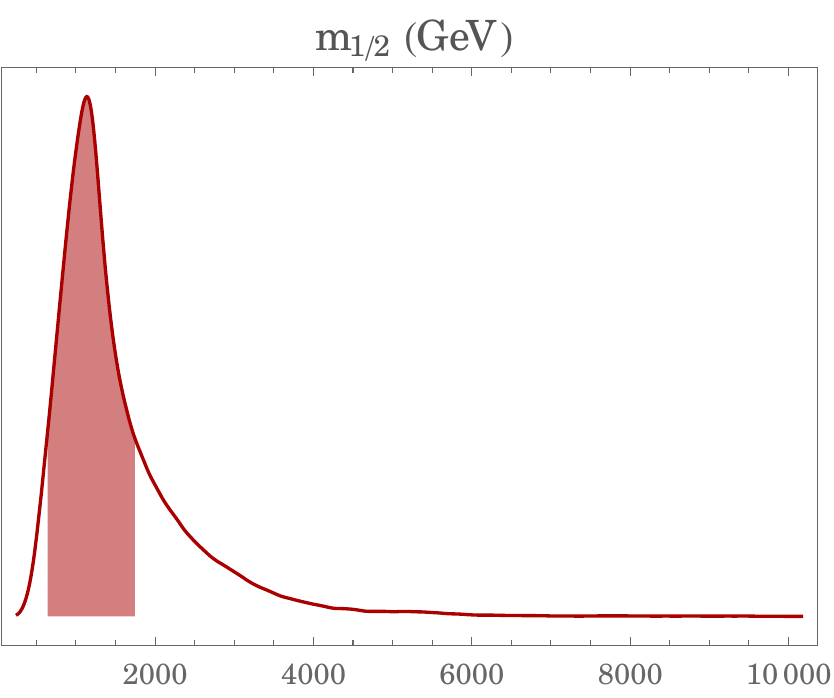}
\end{subfigure}\quad
\begin{subfigure}{0.45\textwidth}
\includegraphics[scale=.41]{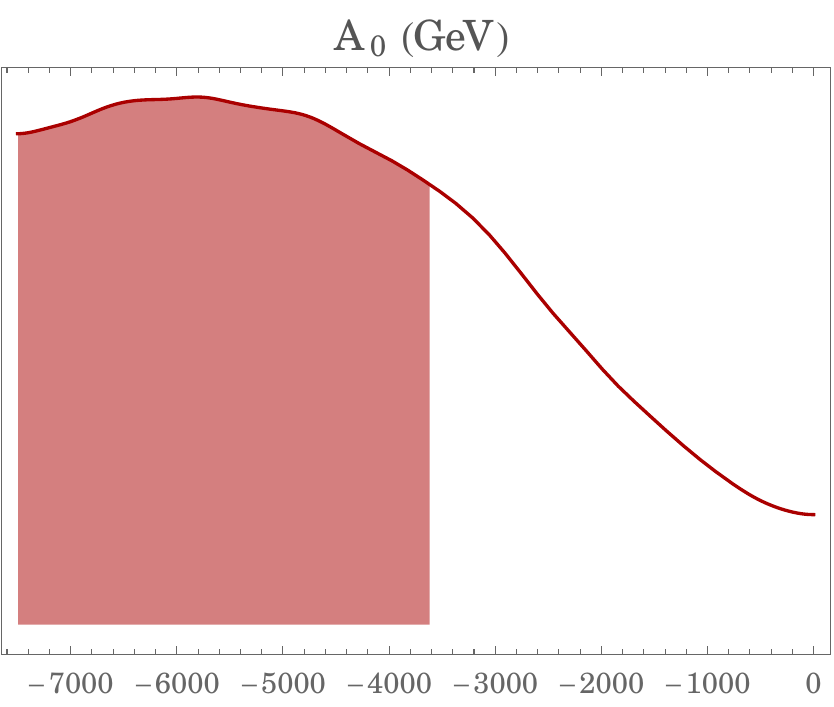}
\end{subfigure}
\caption{$1\sigma$ HPD intervals for the Constrained MSSM soft-breaking parameters.}
\label{fig:hpdcmssm}
\end{figure}

\begin{figure}
\includegraphics[scale=.6]{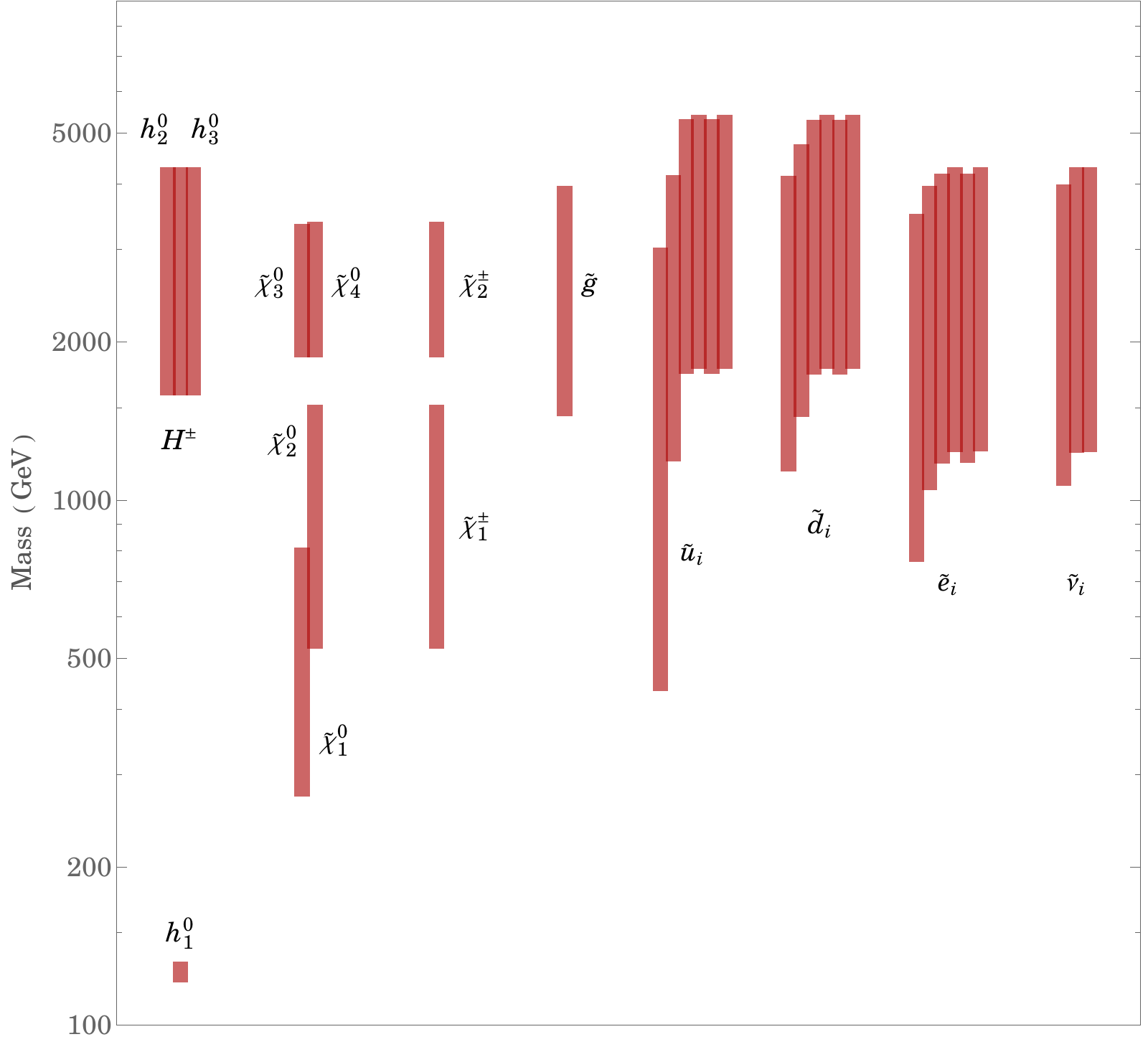}
\caption{$1\sigma$ HPD intervals for the sparticle spectrum and Higgs boson masses with $SU(5)$ GUT scale boundary conditions $\frac{y_e}{y_d}=-\frac{1}{2}$, $\frac{y_\mu}{y_s}=6$, and $\frac{y_\tau}{y_b}=-\frac{3}{2}$, corresponding to our benchmark point from Section 5. The LSP is always $\tilde \chi_1^0$ and the NLSP is always a stop.}
\label{fig:hpd}
\end{figure}

\begin{figure}
\begin{subfigure}{0.48\textwidth}
\includegraphics[scale=.38]{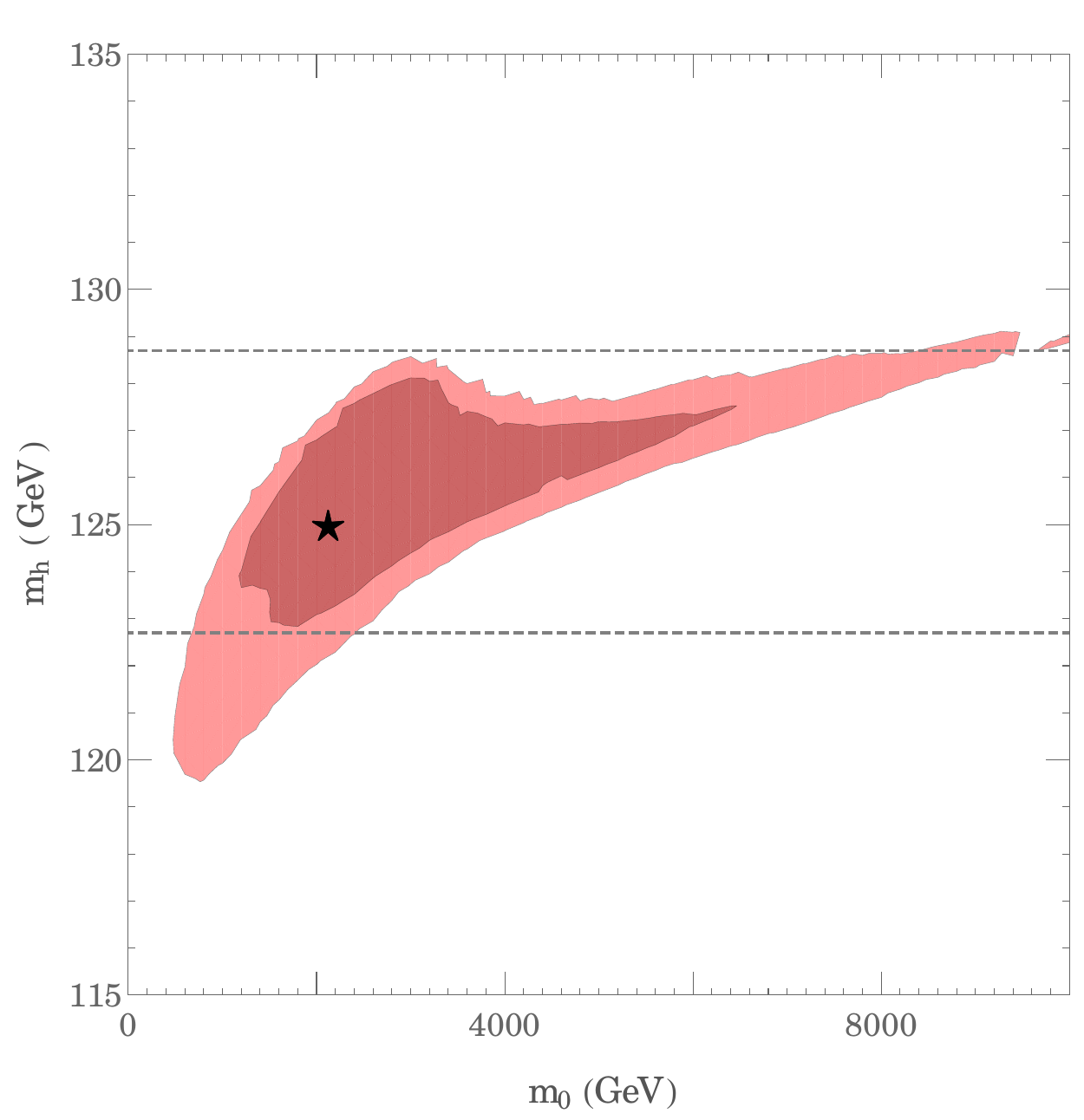}
\end{subfigure}\quad
\begin{subfigure}{0.48\textwidth}
\includegraphics[scale=.38]{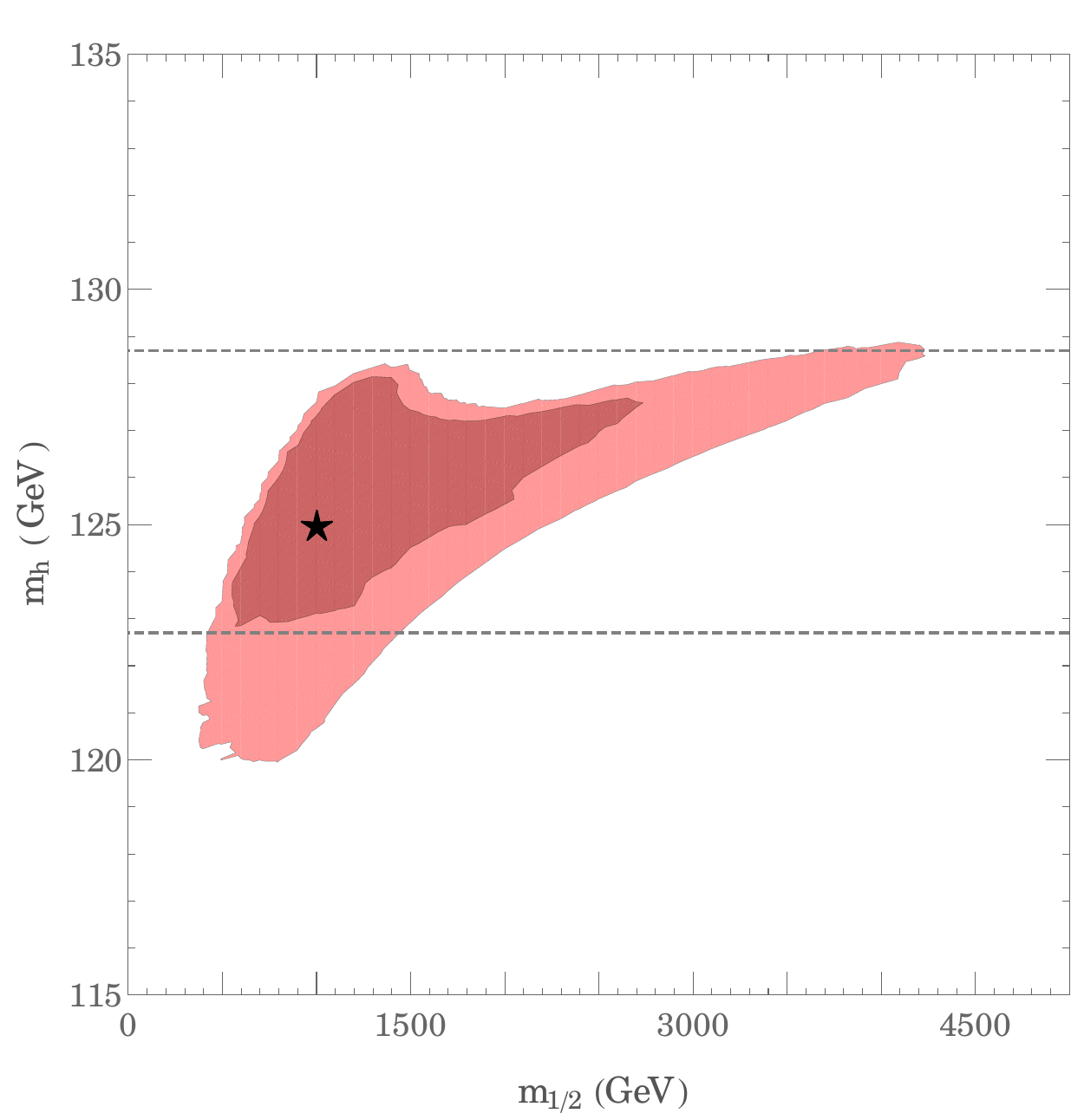}
\end{subfigure}\quad
\begin{subfigure}{0.48\textwidth}
\includegraphics[scale=.38]{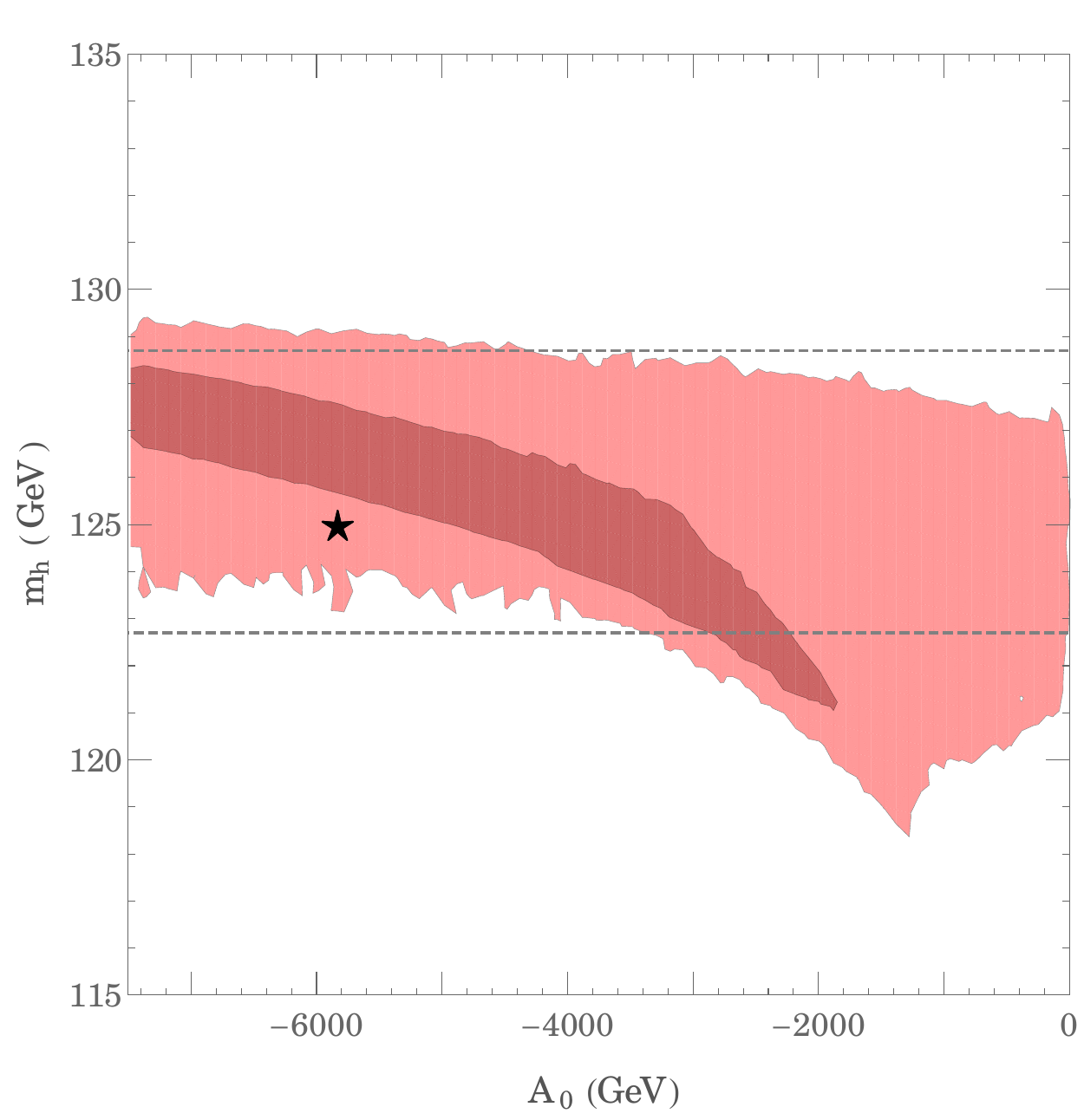}
\end{subfigure}
\caption{2D $1\sigma$ (dark) and $2\sigma$ (bright) HPD regions for Constrained MSSM soft-breaking parameters and the Higgs mass. The black star marks the benchmark point. The dashed lines correspond to the region $125.7\pm 3~\GeV$.}
\label{fig:2dhpd}
\end{figure}

\begin{figure}
\begin{subfigure}{0.48\textwidth}
\includegraphics[scale=.38]{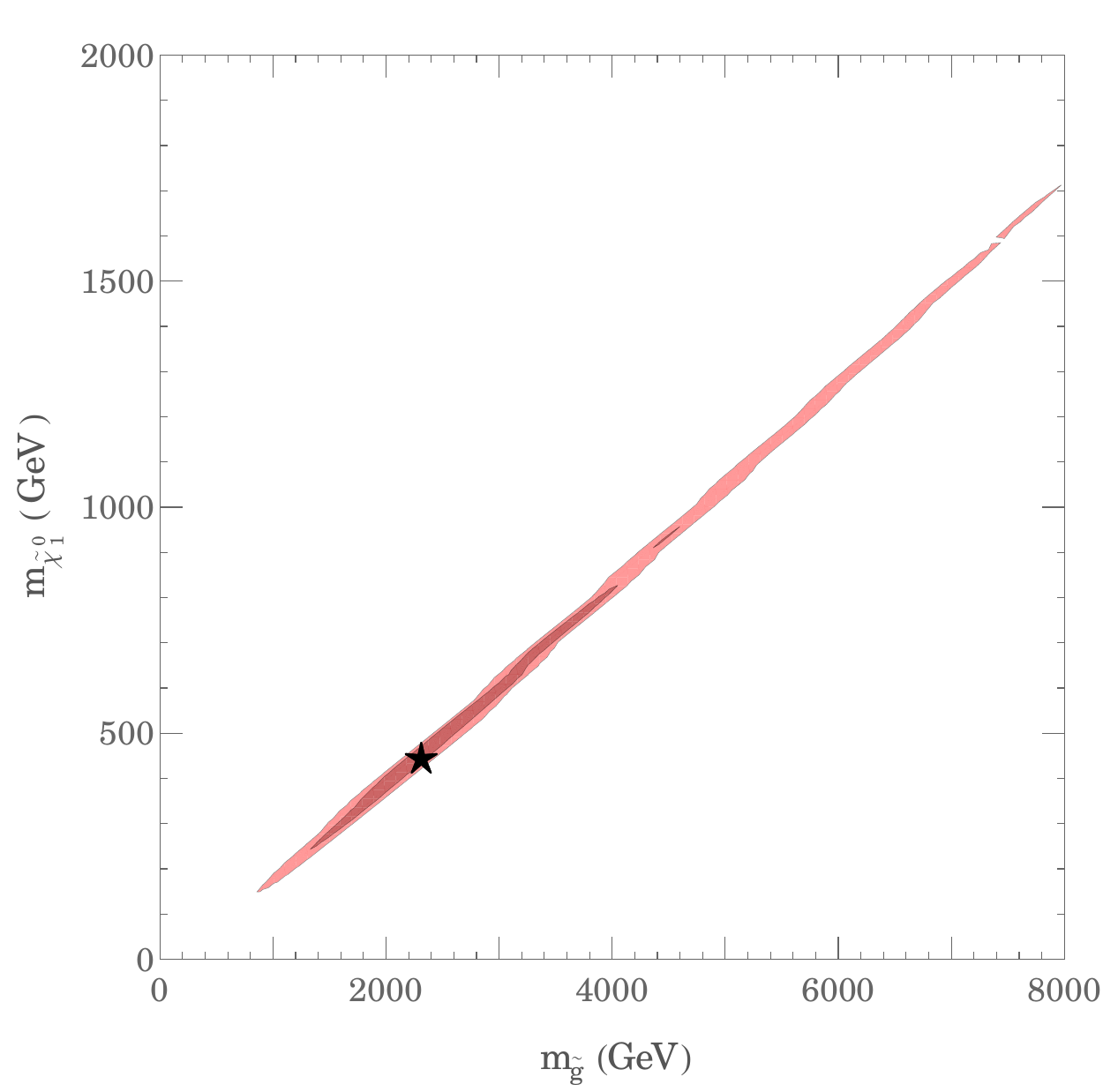}
\end{subfigure}\quad
\begin{subfigure}{0.48\textwidth}
\includegraphics[scale=.38]{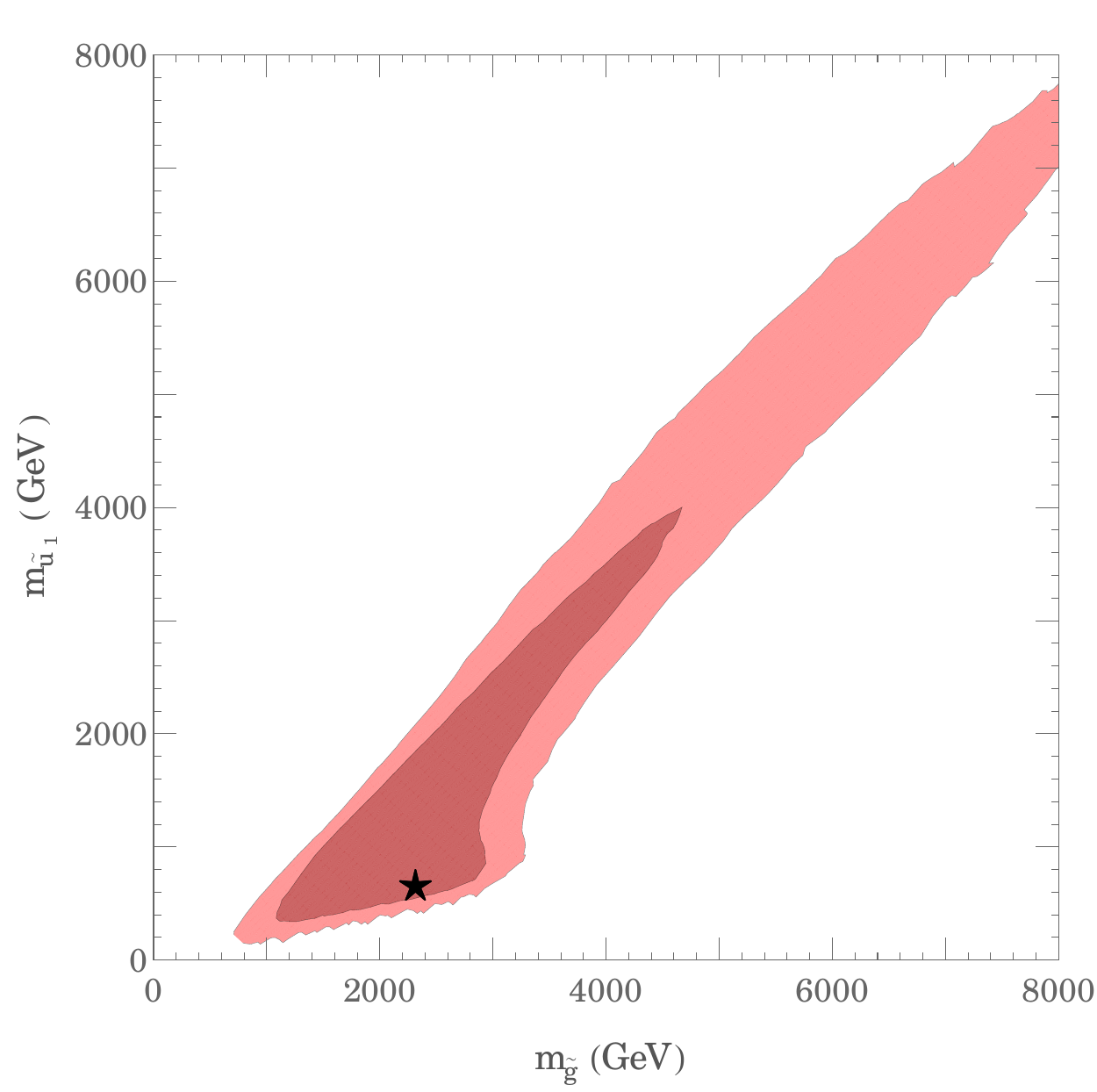}
\end{subfigure}\quad
\begin{subfigure}{0.48\textwidth}
\includegraphics[scale=.38]{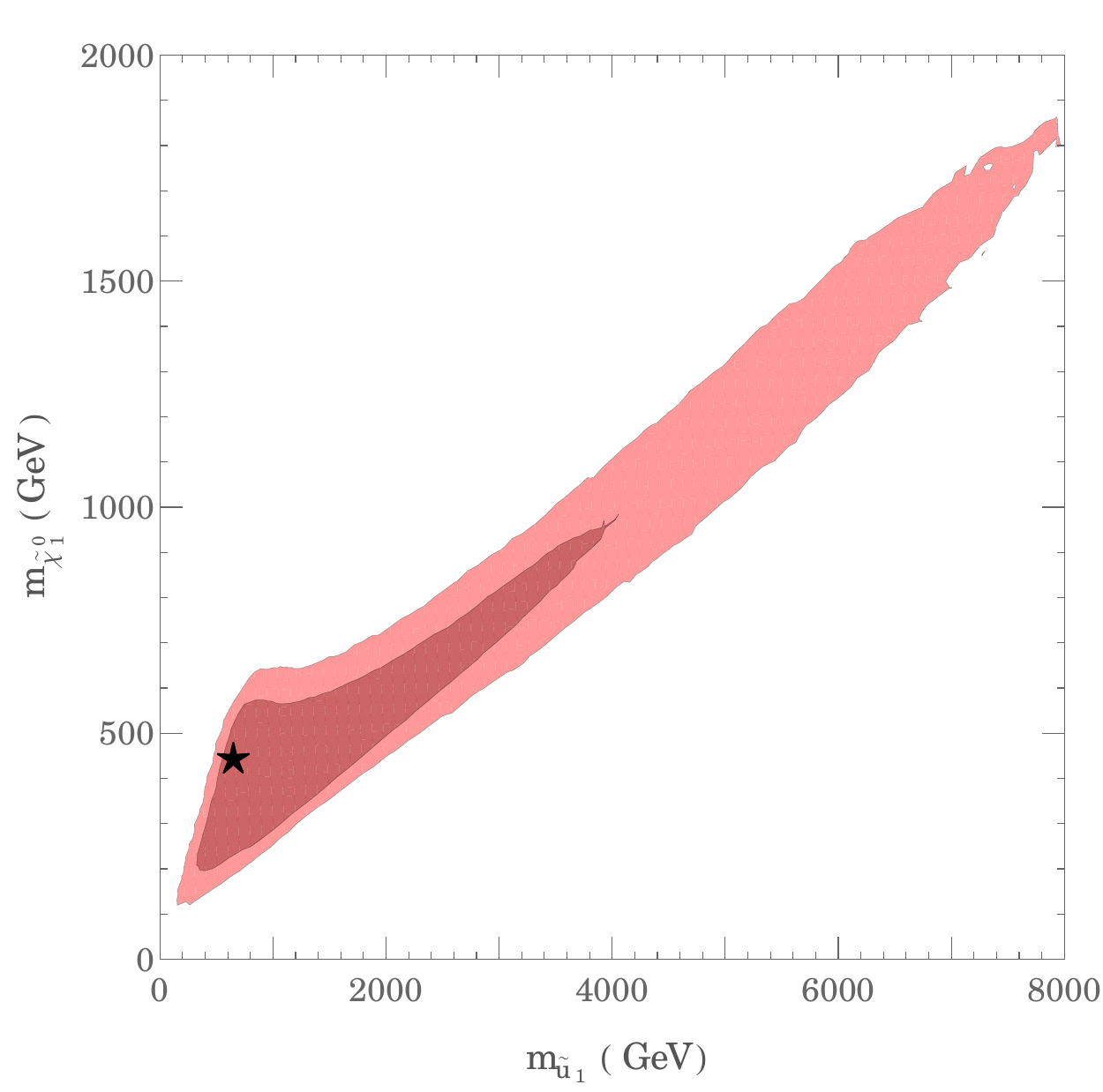}
\end{subfigure}
\caption{2D $1\sigma$ (dark) and $2\sigma$ (bright) HPD regions for the masses ofthe neutralino, gluino, and stop. The black star marks the benchmark point.}
\label{fig:corr}
\end{figure}

Confidence intervals for the sparticle masses are obtained as Bayesian ``highest posterior density'' (HPD) intervals\footnote{An $1\sigma$ HPD interval is the interval [$\theta_L$,$\theta_H$] such that $\int_{\theta_L}^{\theta_H} p(\theta)d\theta = 0.6826\dots$\ and the posterior probability density $p(\theta)$ inside the interval is higher than for any $\theta$ outside of the interval \cite{Agashe:2014kda}.} from a Markov Chain Monte Carlo sample of two million points, using a Metropolis algorithm. As additional constraint we restricted $|A_0|<7.5~\TeV$ to avoid possibly dangerous vacuum decay rates. The $1\sigma$ HPD intervals for the Constrained MSSM soft-breaking parameters are shown in Figure \ref{fig:hpdcmssm}. The $1\sigma$ HPD results of the sparticle masses are shown in Figure \ref{fig:hpd}. As for the benchmark point, for all other parameter points the LSP and NLSP are a neutralino and stop, respectively. The HPD interval for the SUSY scale is obtained as $Q_\text{HPD} = [760,3412]~\GeV$.
In Figure \ref{fig:2dhpd}, we show two-dimensional HPD regions for the correlations between Constrained MSSM soft-breaking parameters and the mass of the Higgs bosons. 
Finally, in Figure \ref{fig:corr} we show the two-dimensional HPD regions for the correlations between the masses of the lightest stop, the neutralino LSP, and the gluino.

\section{Summary and Conclusion}

In this work we discussed how predictions for the sparticle spectrum can arise from GUTs, which feature predictions for the ratios of quark and lepton Yukawa couplings at high energy. To test them by comparing with the experimental data, the RG running between high and low energy has to be performed with sufficient accuracy, including threshold corrections. In SUSY theories, the one-loop threshold corrections when matching the SUSY model to the SM are of particular importance, since they can be enhanced by $\tan \beta$ or large trilinear couplings, and thus have the potential to strongly affect the quark-lepton mass relations. Since the SUSY threshold corrections depend on the SUSY parameters, they link a given GUT flavour model to the SUSY model. In other words, via the SUSY threshold corrections, GUT models can predict properties of the sparticle spectrum from the pattern of quark-lepton mass ratios at the GUT scale. 

To accurately study such predictions, we extend and generalize various formulas in the literature which are needed for a precision analysis of SUSY flavour GUT models: For example, we extend the RGEs for the MSSM soft breaking parameters at two-loop by the additional terms in the seesaw type-I extension (cf.\ appendix \ref{app:RGE}). We generalize the one-loop calculation of $\mu$ and pole mass calculation of $m_A$ and $m_{H^+}$ to include inter-generational mixing in the self energies (cf.\ appendix \ref{app:FORM}). Furthermore, we calculate the full one-loop SUSY threshold corrections for the down-type quark, up-type quark and charged lepton Yukawa coupling matrices in the electroweak unbroken phase (cf.\ section \ref{sec:thresh}). 

We introduce the new software tool \SusyTC, a major extension to the Mathematica package \texttt{REAP}, where these formulas are implemented. In addition, \SusyTC~calculates the $\overline{\text{DR}}$ sparticle spectrum and the SUSY scale $Q$, and can provide output in SLHA ``Les Houches'' files which are the necessary input for external software, e.g.\ for performing a two-loop Higgs mass calculation. \texttt{REAP} extended by \SusyTC~accepts general GUT scale Yukawa, trilinear and soft breaking mass matrices as well as non-universal gaugino masses as input, performs the RG evolution (integrating out the right-handed neutrinos at their mass thresholds in the type I seesaw extension of the MSSM) and automatically matches the MSSM to the SM, making it a convenient tool for top-down analyses of SUSY flavour GUT models.

We applied \SusyTC~to study the predictions for the parameters of the Constrained MSSM SUSY scenario from the set of GUT-scale Yukawa relations $\frac{y_e}{y_d}=-\frac{1}{2}$, $\frac{y_\mu}{y_s}=6$, and $\frac{y_\tau}{y_b}=-\frac{3}{2}$, which has been proposed recently in the context of GUT flavour models. With a Markov Chain Monte Carlo analysis we find a ``best-fit'' benchmark point where the LSP is a bino-like neutralino with a mass of about $450~\GeV$ and the NLSP a stop with a mass of $656~\GeV$. We also find the 1$\sigma$ Bayesian confidence intervals for the sparticle masses and the correlations between the SUSY parameters. Without applying any constraints from LHC SUSY searches or dark matter, we find that the considered GUT scenario predicts a sparticle spectrum above past LHC sensitivities, but within reach of the current LHC run or a future high-luminosity upgrade.

\section*{Acknowledgements}
We would like to thank Vinzenz Maurer for help with code optimisation and Christian Hohl for testing. We also thank Eros Cazzato, Thomas Hahn, Vinzenz Maurer, Stefano Orani and Sebastian Pa\ss ehr for useful discussions. This work has been supported by the Swiss National Science Foundation.



\clearpage
\newpage

\begin{appendix}

\section{The $\beta$-functions in the seesaw type-I extension of the MSSM}
\label{app:RGE}
In this appendix we list the $\beta$-functions of the SUSY soft-breaking parameters in the MSSM extended by the additional terms in the seesaw type-I extension (obtained using the general formulas of \cite{Martin:1993zk}). Our conventions for $W$ and $\mathcal{L}_\text{soft}$ are given in \eqref{eq:W} and \eqref{eq:Lsoft}.

\subsection{One-Loop $\beta$-functions}

\begin{align}
16\pi^2 \beta_{M_1} &= \frac{66}{5} g_1^2 M_1\;, \\
16\pi^2 \beta_{M_2} &= 2 g_2^2 M_2\;, \\
16\pi^2 \beta_{M_3} &= -6 g_3^2 M_3\;,
\end{align}

\begin{align}
16\pi^2 \beta_{T_u} =\ &
     Y_u \left(2 Y_d^\dagger T_d + 4 Y_u^\dagger T_u \right)  + T_u \left(Y_d^\dagger Y_d + 5 Y_u^\dagger Y_u \right) \nonumber \\
    & + Y_u  \left( 6 \Tr(Y_u^\dagger T_u) + 2 \Tr(Y_\nu^\dagger T_\nu) + \frac{26}{15} g_1^2 M_1
+ 6 g_2^2 M_2 + \frac{32}{3} g_3^2 M_3 \right) \nonumber \\
    & + T_u  \left(3 \Tr(Y_u^\dagger Y_u) + \Tr(Y_\nu^\dagger Y_\nu) - \frac{13}{15} g_1^2 - 3 g_2^2
 - \frac{16}{3} g_3^2
    \right)\;,
\end{align}

\begin{align}
16\pi^2 \beta_{T_d} =\ &
     Y_d \left(4 Y_d^\dagger T_d + 2 Y_u^\dagger T_u \right)  + T_d \left(5 Y_d^\dagger Y_d +  Y_u^\dagger Y_u \right) \nonumber \\
    & + Y_d  \left( 6 \Tr(Y_d^\dagger T_d) + 2 \Tr(Y_e^\dagger T_e) + \frac{14}{15} g_1^2 M_1
+ 6 g_2^2 M_2 + \frac{32}{3} g_3^2 M_3 \right) \nonumber \\
    & + T_d  \left(3 \Tr(Y_d^\dagger Y_d) + \Tr(Y_e^\dagger Y_e) - \frac{7}{15} g_1^2 - 3 g_2^2
 - \frac{16}{3} g_3^2
    \right)\;,
\end{align}

\begin{align}
16\pi^2 \beta_{T_e} =\ &
     Y_e \left(4 Y_e^\dagger T_e + 2 Y_\nu^\dagger T_\nu \right)  + T_e \left(5 Y_e^\dagger Y_e +  Y_\nu^\dagger Y_\nu \right) \nonumber \\
    & + Y_e  \left( 6 \Tr(Y_d^\dagger T_d) + 2 \Tr(Y_e^\dagger T_e) + \frac{18}{5} g_1^2 M_1
+ 6 g_2^2 M_2 \right) \nonumber \\
    & + T_e  \left(3 \Tr(Y_d^\dagger Y_d) + \Tr(Y_e^\dagger Y_e) - \frac{9}{5} g_1^2 - 3 g_2^2
    \right)\;,
\end{align}

\begin{align}
16\pi^2 \beta_{T_\nu} =\ &
     Y_\nu \left(4 Y_\nu^\dagger T_\nu + 2 Y_e^\dagger T_e \right)  + T_\nu \left(5 Y_\nu^\dagger Y_\nu +  Y_e^\dagger Y_e \right) \nonumber \\
    & + Y_\nu  \left( 6 \Tr(Y_u^\dagger T_u) + 2 \Tr(Y_\nu^\dagger T_\nu) + \frac{6}{5} g_1^2 M_1
+ 6 g_2^2 M_2 \right) \nonumber \\
    & + T_\nu  \left(3 \Tr(Y_u^\dagger Y_u) + \Tr(Y_\nu^\dagger Y_\nu) - \frac{3}{5} g_1^2 - 3 g_2^2
    \right)\;,
\end{align}

\begin{align}
16\pi^2 \beta_{m^2_{\tilde L}} =\ & 2 Y_e^\dagger m^2_{\tilde e} Y_e + 2 Y_\nu^\dagger m^2_{\tilde \nu} Y_\nu \nonumber \\
&+  m^2_{\tilde L}\left( Y_e^\dagger Y_e +  Y_\nu^\dagger Y_\nu\right) 
+ \left(Y_e^\dagger Y_e + Y_\nu^\dagger Y_\nu\right) m^2_{\tilde L} \nonumber \\    
    & + 2 m^2_{H_u} Y_\nu^\dagger Y_\nu + 2 m^2_{H_d} Y_e^\dagger Y_e       
     + 2 T_e^\dagger T_e + 2 T_\nu^\dagger T_\nu \nonumber \\  
    & - \frac{6}{5} g_1^2 |M_1|^2~\mathbf{1}_3 - 6 g_2^2 |M2|^2~\mathbf{1}_3 - \frac{3}{5} g_1^2 S~\mathbf{1}_3\;,
\end{align}

\begin{align}
16\pi^2 \beta_{m^2_{\tilde Q}} =\ & 2 Y_u^\dagger m^2_{\tilde u} Y_u + 2 Y_d^\dagger m^2_{\tilde d} Y_d \nonumber \\
&+  m^2_{\tilde Q}\left( Y_u^\dagger Y_u +  Y_d^\dagger Y_d\right) 
+ \left(Y_u^\dagger Y_u + Y_d^\dagger Y_d\right) m^2_{\tilde Q} \nonumber \\    
    & + 2 m^2_{H_u} Y_u^\dagger Y_u + 2 m^2_{H_d} Y_d^\dagger Y_d       
     + 2 T_u^\dagger T_u + 2 T_d^\dagger T_d \nonumber \\  
    & - \frac{2}{15} g_1^2 |M_1|^2~\mathbf{1}_3 - 6 g_2^2 |M2|^2~\mathbf{1}_3 - \frac{32}{3} g_3^2 |M_3|^2~\mathbf{1}_3 + \frac{1}{5} g_1^2 S~\mathbf{1}_3\;,
\end{align}

\begin{align}
16\pi^2 \beta_{m^2_{\tilde u}} =\ & 4 Y_u m^2_{\tilde Q} Y_u^\dagger + 2 Y_u Y_u^\dagger m^2_{\tilde u} + 2 m^2_{\tilde u} Y_u Y_u^\dagger \nonumber \\
    & + 4 m^2_{H_u} Y_u Y_u^\dagger + 4 T_u T_u^\dagger \nonumber \\  
    & - \frac{32}{15} g_1^2 |M_1|^2~\mathbf{1}_3 - \frac{32}{3} g_3^2 |M_3|^2~\mathbf{1}_3 - \frac{4}{5} g_1^2 S~\mathbf{1}_3\;,
\end{align}

\begin{align}
16\pi^2 \beta_{m^2_{\tilde d}} =\ & 4 Y_d m^2_{\tilde Q} Y_d^\dagger + 2 Y_d Y_d^\dagger m^2_{\tilde d} + 2 m^2_{\tilde d} Y_d Y_d^\dagger \nonumber \\
    & + 4 m^2_{H_d} Y_d Y_d^\dagger + 4 T_d T_d^\dagger \nonumber \\  
    & - \frac{8}{15} g_1^2 |M_1|^2~\mathbf{1}_3 - \frac{32}{3} g_3^2 |M_3|^2~\mathbf{1}_3 + \frac{2}{5} g_1^2 S~\mathbf{1}_3\;,
\end{align}

\begin{align}
16\pi^2 \beta_{m^2_{\tilde e}} =\ & 4 Y_e m^2_{\tilde L} Y_e^\dagger + 2 Y_e Y_e^\dagger m^2_{\tilde e} + 2 m^2_{\tilde e} Y_e Y_e^\dagger \nonumber \\
    & + 4 m^2_{H_d} Y_e Y_e^\dagger + 4 T_e T_e^\dagger \nonumber \\  
    & - \frac{24}{5} g_1^2 |M_1|^2~\mathbf{1}_3 + \frac{6}{5} g_1^2 S~\mathbf{1}_3\;,
\end{align}

\begin{align}
16\pi^2 \beta_{m^2_{\tilde \nu}} =\ & 4 Y_\nu m^2_{\tilde L} Y_\nu^\dagger + 2 Y_\nu Y_\nu^\dagger m^2_{\tilde \nu} + 2 m^2_{\tilde \nu} Y_\nu Y_\nu^\dagger \nonumber \\
    & + 4 m^2_{H_u} Y_\nu Y_\nu^\dagger + 4 T_\nu T_\nu^\dagger \;,
\end{align}

\begin{align}
16\pi^2 \beta_{m^2_{H_d}} =\ &
    6 \Tr(Y_d m^2_{\tilde Q} Y_d^\dagger + Y_d^\dagger m^2_{\tilde d} Y d) \nonumber \\
    &+ 2 \Tr(Y_e m^2_{\tilde L} Y_e^\dagger + Y_e^\dagger m^2_{\tilde e} Y_e) \nonumber \\
    &+ 6 m^2_{H_d} \Tr(Y_d^\dagger Y_d)+ 2 m^2_{H_d} \Tr(Y_e^\dagger Y_e) \nonumber \\    
    &+ 6 \Tr(T_d^\dagger T_d) + 2 \Tr(T_e^\dagger T_e) \nonumber \\
    & - \frac{6}{5} g_1^2 |M_1|^2 - 6 g_2^2 |M_2|^2 - \frac{3}{5} g_1^2 S\;,
\end{align}

\begin{align}
16\pi^2 \beta_{m^2_{H_u}} =\ &
    6 \Tr(Y_u m^2_{\tilde Q} Y_u^\dagger + Y_u^\dagger m^2_{\tilde u} Y u) \nonumber \\
    &+ 2 \Tr(Y_\nu m^2_{\tilde L} Y_\nu^\dagger + Y_\nu^\dagger m^2_{\tilde \nu} Y_\nu) \nonumber \\
    &+ 6 m^2_{H_u} \Tr(Y_u^\dagger Y_u)+ 2 m^2_{H_u} \Tr(Y_\nu^\dagger Y_\nu) \nonumber \\    
    &+ 6 \Tr(T_u^\dagger T_u) + 2 \Tr(T_\nu^\dagger T_\nu) \nonumber \\
    & - \frac{6}{5} g_1^2 |M_1|^2 - 6 g_2^2 |M_2|^2 + \frac{3}{5} g_1^2 S\;,
\end{align}
with
\be
S = m^2_{H_u} + m^2_{H_d} + \Tr\left(m^2_{\tilde Q}-m^2_{\tilde L} + m^2_{\tilde d} + m^2_{\tilde e} - m^2_{\tilde u}\right)\;.
\ee

\subsection{Two-Loop $\beta$-functions}

\begin{align}
(16\pi^2)^2 \beta_{M_1}^{(2)} =\ &
    \frac{12}{5} g_1^2 \Tr(T_\nu Y_\nu^\dagger) + \frac{28}{5} g_1^2 \Tr(T_d Y_d^\dagger)
    + \frac{36}{5} g_1^2 \Tr(T_e Y_e^\dagger) + \frac{52}{5} g_1^2 \Tr(T_u Y_u^\dagger) \nonumber \\
    &- \frac{12}{5} g_1^2 M_1 \Tr(Y_\nu Y_\nu^\dagger) - \frac{28}{5} g_1^2 M_1 \Tr(Y_d Y_d^\dagger) \nonumber \\
    & - \frac{36}{5} g_1^2 M_1 \Tr(Y_e Y_e^\dagger) - \frac{52}{5} g_1^2 M_1 \Tr(Y_u Y_u^\dagger) \nonumber \\
    &+ \frac{176}{5} g_1^2 g_3^2 M_1 + \frac{176}{5} g_1^2 g_3^2 M_3 + \frac{54}{5} g_1^2 g_2^2 M_1
    + \frac{54}{5} g_1^2 g_2^2 M_2 + \frac{796}{25} g_1^4 M_1\;,
\end{align}

\begin{align}
(16\pi^2)^2 \beta_{M_2}^{(2)} =\ &
    4 g_2^2 \Tr(T_\nu Y_\nu^\dagger) + 4 g_2^2 \Tr(T_e Y_e^\dagger)
    + 12 g_2^2 \Tr(T_d Y_d^\dagger) + 12 g_2^2 \Tr(T_u Y_u^\dagger) \nonumber \\
    & - 4 g_2^2 M_2 \Tr(Y_\nu Y_\nu^\dagger) - 4 g_2^2 M_2 \Tr(Y_e Y_e^\dagger) \nonumber \\
    & - 12 g_2^2 M_2 \Tr(Y_u Y_u^\dagger) - 12 g_2^2 M_2 \Tr(Y_d Y_d^\dagger) \nonumber \\
    &+ \frac{18}{5} g_1^2 g_2^2 M_1 + \frac{18}{5} g_1^2 g_3^2 M_2 + 48 g_2^2 g_3^2 M_2
    + 48 g_2^2 g_3^2 M_3 + 100 g_2^4 M_2\;,
\end{align}

\begin{align}
(16\pi^2)^2 \beta_{M_3}^{(2)} =\ &
    8 g_3^2 \Tr(T_d Y_d^\dagger) + 8 g_3^2 \Tr(T_u Y_u^\dagger)
 - 8 g_3^2 M_3 \Tr(Y_d Y_d^\dagger) - 8 g_3^2 M_3 \Tr(Y_u Y_u^\dagger) \nonumber \\
    & + \frac{22}{5} g_1^2 g_3^2 M_1 + \frac{22}{5} g_1^2 g_3^2 M_3 + 18 g_2^2 g_3^2 M_2
    + 18 g_2^2 g_3^2 M_3 + 56 g_3^4 M_3\;,
\end{align}

\begin{align}
(16\pi^2)^2 \beta_{T_u}^{(2)} =\ & - 2 T_u\left( Y_d^\dagger Y_dY_d^\dagger Y_d + 2 Y_d^\dagger Y_d Y_u^\dagger Y_u + 3 Y_u^\dagger Y_u Y_u^\dagger Y_u\right) \nonumber \\
& - 2 Y_u\left(2 Y_d^\dagger T_d Y_d^\dagger Y_d + 2 Y_d^\dagger T_d Y_u^\dagger Y_u+ 2 Y_d^\dagger Y_d Y_d^\dagger T_d \right. \nonumber \\
& + \left.  Y_d^\dagger Y_d Y_u^\dagger T_u + 4 Y_u^\dagger T_u Y_u^\dagger Y_u + 3 Y_u^\dagger Y_u Y_u^\dagger T_u\right) \nonumber \\
& - T_u\left(Y_d^\dagger Y_d \Tr(Y_e Y_e^\dagger + 3 Y_d Y_d^\dagger) + 5 Y_u^\dagger Y_u \Tr(Y_\nu Y_\nu^\dagger + 3 Y_u Y_u^\dagger)\right) \nonumber \\
& - 2 Y_u\left( Y_d^\dagger T_d \Tr(Y_e Y_e^\dagger + 3 Y_d Y_d^\dagger) + Y_d^\dagger Y_d \Tr(T_e Y_e^\dagger + 3 T_d Y_d^\dagger) \right.\nonumber \\
& + \left. 2 Y_u^\dagger T_u \Tr(Y_\nu Y_\nu^\dagger + 3 Y_u Y_u^\dagger) + 3 Y_u^\dagger Y_u \Tr(T_\nu Y_\nu^\dagger + 3 T_u Y_u^\dagger)\vphantom{Y_d^\dagger}\right) \nonumber \\
& - T_u \left(\vphantom{Y_d^\dagger} \Tr(3 Y_\nu Y_\nu^\dagger Y_\nu Y_\nu^\dagger + Y_\nu Y_e^\dagger Y_e Y_\nu^\dagger) + 3 \Tr(Y_u Y_d^\dagger Y_d Y_u^\dagger + 3 Y_u Y_u^\dagger Y_u Y_u^\dagger )\right)\nonumber \\
& - 2 Y_u \left(\vphantom{Y_d^\dagger}\Tr(6 T_\nu Y_\nu^\dagger Y_\nu Y_\nu^\dagger + T_\nu Y_e^\dagger Y_e Y_\nu^\dagger + Y_\nu Y_e^\dagger T_e Y_\nu^\dagger ) \right. \nonumber \\
& + \left. 3 \Tr(Y_u Y_d^\dagger T_d Y_u^\dagger + T_u Y_d^\dagger Y_d Y_u^\dagger + 6 T_u Y_u^\dagger Y_u Y_u^\dagger) \right)\nonumber \\
& + \frac{2}{5} g_1^2 \left(T_u Y_d^\dagger Y_d + 2 Y_u Y_d^\dagger T_d + 3 Y_u Y_u^\dagger T_u - 2 M_1 (Y_u Y_d^\dagger Y_d + Y_u Y_u^\dagger Y_u)\right) \nonumber \\
& + 6 g_2^2 \left( Y_u Y_u^\dagger T_u + 2 T_u Y_u^\dagger Y_u - 2 M_2 Y_u Y_u^\dagger Y_u\right) \nonumber \\
& + \left(16 g_3^2 + \frac{4}{5} g_1^2 \right)\left(2 Y_u \Tr(Y_u^\dagger T_u) + T_u \Tr(Y_u^\dagger Y_u)\right) \nonumber \\
& + \frac{136}{45} g_1^2 g_3^2 T_u + \frac{15}{2} g_2^4 T_u - \frac{16}{9}g_3^4 T_u + \frac{2743}{450} g_1^4 T_u + g_1^2g_2^2 T_u + 8 g_2^2 g_3^2 T_u \nonumber \\
& - Y_u \Tr(Y_u^\dagger Y_u)\left(32 g_3^2 M_3 + \frac{8}{5} g_1^2 M_1 \right) - \frac{272}{45}g_1^2 g_3^2 Y_u (M_1+M_3)\nonumber \\
& - 2 g_1^2 g_2^2 Y_u (M_1 + M_2) - 16 g_2^2 g_3^2 Y_u (M_2+M_3)\nonumber \\
& - 30 g_2^4 M_2 Y_u - \frac{5486}{225} g_1^4 M_1 Y_u + \frac{64}{9} g_3^4 M_3 Y_u\;,
\end{align}

\begin{align}
(16\pi^2)^2 \beta_{T_d}^{(2)} =\ & - 2 T_d\left(3 Y_d^\dagger Y_dY_d^\dagger Y_d+2 Y_u^\dagger Y_u Y_d^\dagger Y_d + Y_u^\dagger Y_u Y_u^\dagger Y_u\right) \nonumber \\
& - 2 Y_d\left(4 Y_d^\dagger T_d Y_d^\dagger Y_d + 3 Y_d^\dagger Y_d Y_d^\dagger T_d + 2 Y_u^\dagger T_u Y_d^\dagger Y_d \right. \nonumber \\
& + \left.  2 Y_u^\dagger T_u Y_u^\dagger Y_u + Y_u^\dagger Y_u Y_d^\dagger T_d + 2 Y_u^\dagger Y_u Y_u^\dagger T_u\right) \nonumber \\
& - T_d\left(5 Y_d^\dagger Y_d \Tr(Y_e Y_e^\dagger + 3 Y_d Y_d^\dagger) + Y_u^\dagger Y_u \Tr(Y_\nu Y_\nu^\dagger + 3 Y_u Y_u^\dagger)\right) \nonumber \\
& - 2 Y_d\left( 2 Y_d^\dagger T_d \Tr(Y_e Y_e^\dagger + 3 Y_d Y_d^\dagger) + 3 Y_d^\dagger Y_d \Tr(T_e Y_e^\dagger + 3 T_d Y_d^\dagger) \right.\nonumber \\
& + \left. Y_u^\dagger T_u \Tr(Y_\nu Y_\nu^\dagger + 3 Y_u Y_u^\dagger) + Y_u^\dagger Y_u \Tr(T_\nu Y_\nu^\dagger + 3 T_u Y_u^\dagger)\vphantom{Y_d^\dagger}\right) \nonumber \\
& - T_d \left(\vphantom{Y_d^\dagger} \Tr(Y_e Y_\nu^\dagger Y_\nu Y_e^\dagger + 3 Y_e Y_e^\dagger Y_e Y_e^\dagger) + 3 \Tr(Y_u Y_d^\dagger Y_d Y_u^\dagger + 3 Y_d Y_d^\dagger Y_d Y_d^\dagger )\right)\nonumber \\
& - 2 Y_d \left(\vphantom{Y_d^\dagger}\Tr(Y_e Y_\nu^\dagger T_\nu Y_e^\dagger  + T_e Y_\nu^\dagger Y_\nu Y_e^\dagger + 6 T_e Y_e^\dagger Y_e Y_e^\dagger ) \right. \nonumber \\
& + \left. 3 \Tr(T_d Y_u^\dagger Y_u Y_d^\dagger + T_u Y_d^\dagger Y_d Y_u^\dagger+ 6 T_d Y_d^\dagger Y_d Y_d^\dagger ) \right)\nonumber \\
& + \frac{2}{5} g_1^2 \left(2 T_d Y_u^\dagger Y_u + 3 T_d Y_d^\dagger Y_d + 3 Y_d Y_d^\dagger T_d + 4 Y_d Y_u^\dagger T_u - 4 M_1 (Y_d Y_d^\dagger Y_d + Y_d Y_u^\dagger Y_u)\right) \nonumber \\
& + 6 g_2^2 \left( Y_d Y_d^\dagger T_d + 2 T_d Y_d^\dagger Y_d - 2 M_2 Y_d Y_d^\dagger Y_d\right) \nonumber \\
& + \left(16 g_3^2 - \frac{2}{5} g_1^2 \right)\left(2 Y_d \Tr(Y_d^\dagger T_d) + T_d \Tr(Y_d^\dagger Y_d)\right) \nonumber \\ 
& + \frac{6}{5}g_1^2\left(2 Y_d \Tr(Y_e^\dagger T_e) + T_d \Tr(Y_e^\dagger Y_e)\right) \nonumber \\
& + \frac{8}{9} g_1^2 g_3^2 T_d + \frac{15}{2} g_2^4 T_d - \frac{16}{9}g_3^4 T_d + \frac{287}{90} g_1^4 T_d + g_1^2g_2^2 T_d + 8 g_2^2 g_3^2 T_d \nonumber \\
& - Y_d \Tr(Y_d^\dagger Y_d)\left(32 g_3^2 M_3 - \frac{4}{5} g_1^2 M_1 \right) - \frac{12}{5} g_1^2 M_1 Y_d \Tr(Y_e^\dagger Y_e) \nonumber \\
& - \frac{16}{9}g_1^2 g_3^2 Y_d (M_1+M_3) - 2 g_1^2 g_2^2 Y_d (M_1 + M_2) - 16 g_2^2 g_3^2 Y_d (M_2+M_3)\nonumber \\
& - 30 g_2^4 M_2 Y_d - \frac{574}{45} g_1^4 M_1 Y_d + \frac{64}{9} g_3^4 M_3 Y_d\;,
\end{align}

\begin{align}
(16\pi^2)^2 \beta_{T_e}^{(2)} =\ & - 2 T_e\left(\vphantom{Y_d^\dagger}3 Y_e^\dagger Y_eY_e^\dagger Y_e+2 Y_\nu^\dagger Y_\nu Y_e^\dagger Y_e +  Y_\nu^\dagger Y_\nu Y_\nu^\dagger Y_\nu\right) \nonumber \\
& - 2 Y_e\left(\vphantom{Y_d^\dagger}4 Y_e^\dagger T_e Y_e^\dagger Y_e + 3 Y_e^\dagger Y_e Y_e^\dagger T_e + 2 Y_\nu^\dagger T_\nu Y_e^\dagger Y_e \right. \nonumber \\
& + \left.  2 Y_\nu^\dagger T_\nu Y_\nu^\dagger Y_\nu + Y_\nu^\dagger Y_\nu Y_e^\dagger T_e + 2 Y_\nu^\dagger Y_\nu Y_\nu^\dagger T_\nu\vphantom{Y_d^\dagger}\right) \nonumber \\
& - T_e\left(5 Y_e^\dagger Y_e \Tr(Y_e Y_e^\dagger + 3 Y_d Y_d^\dagger) + Y_\nu^\dagger Y_\nu \Tr(Y_\nu Y_\nu^\dagger + 3 Y_u Y_u^\dagger)\right) \nonumber \\
& - 2 Y_e\left( 2 Y_e^\dagger T_e \Tr(Y_e Y_e^\dagger + 3 Y_d Y_d^\dagger) + 3 Y_e^\dagger Y_e \Tr(T_e Y_e^\dagger + 3 T_d Y_d^\dagger) \right.\nonumber \\
& + \left. Y_\nu^\dagger T_\nu \Tr(Y_\nu Y_\nu^\dagger + 3 Y_u Y_u^\dagger) + Y_\nu^\dagger Y_\nu \Tr(T_\nu Y_\nu^\dagger + 3 T_u Y_u^\dagger)\vphantom{Y_d^\dagger}\right) \nonumber \\
& - T_e \left(\vphantom{Y_d^\dagger} \Tr(Y_e Y_\nu^\dagger Y_\nu Y_e^\dagger + 3 Y_e Y_e^\dagger Y_e Y_e^\dagger) + 3 \Tr(Y_d Y_u^\dagger Y_u Y_d^\dagger + 3 Y_d Y_d^\dagger Y_d Y_d^\dagger )\right)\nonumber \\
& - 2 Y_e \left(\vphantom{Y_d^\dagger}\Tr(Y_e Y_\nu^\dagger T_\nu Y_e^\dagger  + T_e Y_\nu^\dagger Y_\nu Y_e^\dagger + 6 T_e Y_e^\dagger Y_e Y_e^\dagger ) \right. \nonumber \\
& + \left. 3 \Tr(T_d Y_u^\dagger Y_u Y_d^\dagger + T_u Y_d^\dagger Y_d Y_u^\dagger+  6 T_d Y_d^\dagger Y_d Y_d^\dagger ) \right)\nonumber \\
& + \frac{6}{5} g_1^2 \left(Y_e Y_e^\dagger T_e - T_e Y_e^\dagger Y_e\right) \nonumber \\
& + 6 g_2^2 \left( Y_e Y_e^\dagger T_e + 2 T_e Y_e^\dagger Y_e - 2 M_2 Y_e Y_e^\dagger Y_e\right) \nonumber \\
& + \left(16 g_3^2 - \frac{2}{5} g_1^2 \right)\left(2 Y_e \Tr(Y_d^\dagger T_d) + T_e \Tr(Y_d^\dagger Y_d)\right) \nonumber \\ 
& + \frac{6}{5}g_1^2\left(2 Y_e \Tr(Y_e^\dagger T_e) + T_e \Tr(Y_e^\dagger Y_e)\right) \nonumber \\
& + \frac{15}{2} g_2^4 T_e + \frac{27}{2} g_1^4 T_e + \frac{9}{5} g_1^2g_2^2 T_e \nonumber \\
& - Y_e \Tr(Y_d^\dagger Y_d)\left(32 g_3^2 M_3 - \frac{4}{5} g_1^2 M_1 \right) - \frac{12}{5} g_1^2 M_1 Y_e \Tr(Y_e^\dagger Y_e) \nonumber \\
& - \frac{18}{5}g_1^2 g_2^2 Y_d (M_1+M_2) - 30 g_2^4 M_2 Y_e - 54 g_1^4 M_1 Y_e\;,
\end{align}

\begin{align}
(16\pi^2)^2 \beta_{T_\nu}^{(2)} =\ & - 2 T_\nu\left(\vphantom{Y_d^\dagger} Y_e^\dagger Y_eY_e^\dagger Y_e + 2 Y_e^\dagger Y_e Y_\nu^\dagger Y_\nu + 3 Y_\nu^\dagger Y_\nu Y_\nu^\dagger Y_\nu\right) \nonumber \\
& - 2 Y_\nu \left(\vphantom{Y_d^\dagger} 2 Y_e^\dagger T_e Y_e^\dagger Y_e + 2 Y_e^\dagger T_e Y_\nu^\dagger Y_\nu + 2 Y_e^\dagger Y_e Y_e^\dagger T_e \right. \nonumber \\
& + \left.  Y_e^\dagger Y_e Y_\nu^\dagger T_\nu + 4 Y_\nu^\dagger T_\nu Y_\nu^\dagger Y_\nu + 3 Y_\nu^\dagger Y_\nu Y_\nu^\dagger T_\nu\vphantom{Y_d^\dagger}\right) \nonumber \\
& - T_\nu\left( Y_e^\dagger Y_e \Tr(Y_e Y_e^\dagger + 3 Y_d Y_d^\dagger) + 5 Y_\nu^\dagger Y_\nu \Tr(Y_\nu Y_\nu^\dagger + 3 Y_u Y_u^\dagger)\right) \nonumber \\
& - 2 Y_\nu\left( Y_e^\dagger T_e \Tr(Y_e Y_e^\dagger + 3 Y_d Y_d^\dagger) +  Y_e^\dagger Y_e \Tr(T_e Y_e^\dagger + 3 T_d Y_d^\dagger) \right.\nonumber \\
& + \left. 2 Y_\nu^\dagger T_\nu \Tr(Y_\nu Y_\nu^\dagger + 3 Y_u Y_u^\dagger) + 3 Y_\nu^\dagger Y_\nu \Tr(T_\nu Y_\nu^\dagger + 3 T_u Y_u^\dagger)\vphantom{Y_d^\dagger}\right) \nonumber \\
& - T_\nu \left(\vphantom{Y_d^\dagger} \Tr( 3 Y_\nu Y_\nu^\dagger Y_\nu Y_\nu^\dagger +  Y_\nu Y_e^\dagger Y_e Y_\nu^\dagger) + 3 \Tr(Y_u Y_d^\dagger Y_d Y_u^\dagger + 3 Y_u Y_u^\dagger Y_u Y_u^\dagger )\right)\nonumber \\
& - 2 Y_\nu \left(\vphantom{Y_d^\dagger}\Tr(T_\nu Y_e^\dagger Y_e Y_\nu^\dagger + Y_\nu Y_e^\dagger T_e Y_\nu^\dagger + 6 T_\nu Y_\nu^\dagger Y_\nu Y_\nu^\dagger ) \right. \nonumber \\
& + \left. 3 \Tr(T_d Y_u^\dagger Y_u Y_d^\dagger + T_u Y_d^\dagger Y_d Y_u^\dagger +  6 T_u Y_u^\dagger Y_u Y_u^\dagger ) \right)\nonumber \\
& + \frac{6}{5} g_1^2 \left( 2 T_\nu Y_\nu^\dagger Y_\nu + 2 Y_\nu Y_e^\dagger T_e + Y_\nu Y_\nu^\dagger T_\nu + T_\nu Y_e^\dagger Y_e - 2 M_1 (Y_\nu Y_\nu^\dagger Y_\nu + Y_\nu Y_e^\dagger Y_e) \vphantom{Y_d^\dagger}\right) \nonumber \\
& + 6 g_2^2 \left( Y_\nu Y_\nu^\dagger T_\nu + 2 T_\nu Y_\nu^\dagger Y_\nu - 2 M_2 Y_\nu Y_\nu^\dagger Y_\nu\right) \nonumber \\
& + \left(16 g_3^2 + \frac{4}{5} g_1^2 \right)\left(2 Y_\nu \Tr(Y_u^\dagger T_u) + T_\nu \Tr(Y_u^\dagger Y_u)\right) \nonumber \\ 
& + \frac{15}{2} g_2^4 T_\nu + \frac{207}{50} g_1^4 T_\nu + \frac{9}{5} g_1^2g_2^2 T_\nu \nonumber \\
& - Y_\nu \Tr(Y_u^\dagger Y_u)\left(32 g_3^2 M_3 + \frac{8}{5} g_1^2 M_1 \right) \nonumber \\
& - \frac{18}{5}g_1^2 g_2^2 Y_\nu (M_1+M_2) - 30 g_2^4 M_2 Y_\nu - \frac{414}{25} g_1^4 M_1 Y_\nu\;,
\end{align}

\begin{align}
(16\pi^2)^2 \beta_{m^2_{\tilde L}}^{(2)} =\ & - 4 T_e^\dagger \left(\vphantom{Y_d^\dagger}T_e Y_e^\dagger Y_e + Y_e Y_e^\dagger T_e\right) - 4 T_\nu^\dagger \left(\vphantom{Y_d^\dagger}T_\nu Y_\nu^\dagger Y_\nu + Y_\nu Y_\nu^\dagger T_\nu\right)\nonumber\\
& - 2 Y_e^\dagger\left(\vphantom{Y_d^\dagger} 2 m^2_{\tilde e}Y_e Y_e^\dagger Y_e + 2 T_e T_e^\dagger Y_e + 2 Y_e T_e^\dagger T_e \right. \nonumber \\
& \left. + 2 Y_e Y_e^\dagger m^2_{\tilde e} Y_e + Y_e Y_e^\dagger Y_e m^2_{\tilde L} + 2 Y_e m^2_{\tilde L} Y_e^\dagger Y_e \vphantom{Y_d^\dagger} \right) \nonumber \\
& - 2 Y_\nu^\dagger\left(\vphantom{Y_d^\dagger} 2 m^2_{\tilde \nu}Y_\nu Y_\nu^\dagger Y_\nu + 2 T_\nu T_\nu^\dagger Y_\nu + 2 Y_\nu T_\nu^\dagger T_\nu \right. \nonumber \\
& \left. + 2 Y_\nu Y_\nu^\dagger m^2_{\tilde \nu} Y_\nu + Y_\nu Y_\nu^\dagger Y_\nu m^2_{\tilde L} + 2 Y_\nu m^2_{\tilde L} Y_\nu^\dagger Y_\nu \vphantom{Y_d^\dagger} \right)\nonumber \\
& - 2 \left(4 m^2_{H_d}+m^2_{\tilde L}\right) Y_e^\dagger Y_e Y_e^\dagger Y_e - 2\left(4 m^2_{H_u}+m^2_{\tilde L}\right) Y_\nu^\dagger Y_\nu Y_\nu^\dagger Y_\nu \nonumber \\
& - 2 T_e^\dagger \left(T_e \Tr(Y_e Y_e^\dagger +3 Y_d Y_d^\dagger)+Y_e \Tr(T_e Y_e^\dagger +3 T_d Y_d^\dagger)\right)\nonumber \\
& - 2 T_\nu^\dagger \left(\vphantom{Y_d^\dagger} T_\nu \Tr(Y_\nu Y_\nu^\dagger +3 Y_u Y_u^\dagger)+Y_\nu \Tr(T_\nu Y_\nu^\dagger +3 T_u Y_u^\dagger)\right)\nonumber \\
& - Y_e^\dagger\left(2m^2_{\tilde e}Y_e\Tr(Y_eY_e^\dagger+3Y_dY_d^\dagger)+2T_e\Tr(Y_eT_e^\dagger+3Y_dT_d^\dagger)\right.\nonumber \\
& + \left. 2 Y_e\Tr(T_eT_e^\dagger+3T_dT_d^\dagger) + Y_e m^2_{\tilde L}\Tr(Y_eY_e^\dagger+3Y_dY_d^\dagger) \right) \nonumber \\
& - Y_\nu^\dagger\left(\vphantom{Y_d^\dagger} 2m^2_{\tilde \nu}Y_\nu\Tr(Y_\nu Y_\nu^\dagger+3Y_uY_u^\dagger)+2T_\nu\Tr(Y_\nu T_\nu^\dagger+3Y_uT_u^\dagger)\right.\nonumber \\
& + \left. 2 Y_\nu\Tr(T_\nu T_\nu^\dagger+3T_uT_u^\dagger) + Y_\nu m^2_{\tilde L}\Tr(Y_\nu Y_\nu^\dagger+3Y_uY_u^\dagger) \right) \nonumber \\
& - \left(\vphantom{Y_d^\dagger}4 m^2_{H_d}+m^2_{\tilde L}\right)Y_e^\dagger Y_e \Tr(Y_eY_e^\dagger+3Y_dY_d^\dagger) \nonumber \\
& - \left(\vphantom{Y_d^\dagger}4 m^2_{H_u}+ m^2_{\tilde L}\right) Y_\nu^\dagger Y_\nu \Tr(Y_\nu Y_\nu^\dagger+3Y_uY_u^\dagger)\nonumber \\
& - 2Y_e^\dagger Y_e \Tr\left(Y_e m^2_{\tilde L}Y_e^\dagger + Y_e^\dagger m^2_{\tilde e}Y_e+3Y_d m^2_{\tilde Q}Y_d^\dagger+3Y_d^\dagger m^2_{\tilde d}Y_d\right) \nonumber \\
& - 2Y_\nu^\dagger Y_\nu \Tr\left(Y_\nu m^2_{\tilde L}Y_\nu^\dagger + Y_\nu^\dagger m^2_{\tilde \nu}Y_\nu+3Y_u m^2_{\tilde Q}Y_u^\dagger+3Y_u^\dagger m^2_{\tilde u}Y_u\right) \nonumber \\
& + \frac{6}{5} g_1^2 \left(\vphantom{Y_d^\dagger}2m^2_{H_d}Y_e^\dagger Y_e + m^2_{\tilde L}Y_e^\dagger Y_e + Y_e^\dagger Y_e m^2_{\tilde L} + 2 Y_e^\dagger m^2_{\tilde e}Y_e + 2 T_e^\dagger T_e\right)\nonumber \\
& -\frac{12}{5}g_1^2\left(M_1^*Y_e^\dagger T_e -2 |M_1|^2 Y_e^\dagger Y_e + M_1 T_e^\dagger Y_e  \right)\nonumber \\
& +\frac{621}{25}g_1^4 |M_1|^2~\mathbf{1}_3 + \frac{18}{5}g_1^2g_2^2(|M_1|^2+|M_2|^2)~\mathbf{1}_3 \nonumber \\
& + \frac{18}{5}g_1^2g_2^2Re(M_1M_2^*)~\mathbf{1}_3 + 33 g_2^4 |M_2|^2~\mathbf{1}_3 \nonumber \\
& + \frac{3}{5}g_1^2 \sigma_1~\mathbf{1}_3 + 3 g_2^2 \sigma_2~\mathbf{1}_3 - \frac{6}{5}g_1^2S'~\mathbf{1}_3\;,
\end{align}

\begin{align}
(16\pi^2)^2 \beta_{m^2_{\tilde Q}}^{(2)} =\ & - 4 T_d^\dagger \left(T_d Y_d^\dagger Y_d + Y_d Y_d^\dagger T_d\right) - 4 T_u^\dagger \left(\vphantom{Y_d^\dagger}T_u Y_u^\dagger Y_u + Y_u Y_u^\dagger T_u\right)\nonumber\\
& - 2 Y_d^\dagger\left( 2 m^2_{\tilde d}Y_d Y_d^\dagger Y_d + 2 T_d T_d^\dagger Y_d + 2 Y_d T_d^\dagger T_d \right. \nonumber \\
& \left. + 2 Y_d Y_d^\dagger m^2_{\tilde d} Y_d + Y_d Y_d^\dagger Y_d m^2_{\tilde Q} + 2 Y_d m^2_{\tilde Q} Y_d^\dagger Y_d  \right) \nonumber \\
& - 2 Y_u^\dagger\left(\vphantom{Y_d^\dagger} 2 m^2_{\tilde u}Y_u Y_u^\dagger Y_u + 2 T_u T_u^\dagger Y_u + 2 Y_u T_u^\dagger T_u \right. \nonumber \\
& \left. + 2 Y_u Y_u^\dagger m^2_{\tilde u} Y_u + Y_u Y_u^\dagger Y_u m^2_{\tilde Q} + 2 Y_u m^2_{\tilde Q} Y_u^\dagger Y_u  \right)\nonumber \\
& - 2 \left(4 m^2_{H_d}+m^2_{\tilde Q}\right) Y_d^\dagger Y_d Y_d^\dagger Y_d - 2\left(4 m^2_{H_u}+m^2_{\tilde Q}\right) Y_u^\dagger Y_u Y_u^\dagger Y_u \nonumber \\
& - 2 T_d^\dagger \left(T_d \Tr(Y_e Y_e^\dagger +3 Y_d Y_d^\dagger)+Y_d \Tr(T_e Y_e^\dagger +3 T_d Y_d^\dagger)\right)\nonumber \\
& - 2 T_u^\dagger \left(\vphantom{Y_d^\dagger} T_u \Tr(Y_\nu Y_\nu^\dagger +3 Y_u Y_u^\dagger)+Y_u \Tr(T_\nu Y_\nu^\dagger +3 T_u Y_u^\dagger)\right)\nonumber \\
& - Y_d^\dagger\left(2m^2_{\tilde d}Y_d\Tr(Y_eY_e^\dagger+3Y_dY_d^\dagger)+2T_d\Tr(Y_eT_e^\dagger+3Y_dT_d^\dagger)\right.\nonumber \\
& + \left. 2 Y_d\Tr(T_eT_e^\dagger+3T_dT_d^\dagger) + Y_d m^2_{\tilde Q}\Tr(Y_eY_e^\dagger+3Y_dY_d^\dagger) \right) \nonumber \\
& - Y_u^\dagger\left(\vphantom{Y_d^\dagger} 2m^2_{\tilde u}Y_u\Tr(Y_\nu Y_\nu^\dagger+3Y_uY_u^\dagger)+2T_u\Tr(Y_\nu T_\nu^\dagger+3Y_uT_u^\dagger)\right.\nonumber \\
& + \left. 2 Y_u\Tr(T_\nu T_\nu^\dagger+3T_uT_u^\dagger) + Y_u m^2_{\tilde Q}\Tr(Y_\nu Y_\nu^\dagger+3Y_uY_u^\dagger) \right) \nonumber \\
& - \left(4 m^2_{H_d}+m^2_{\tilde Q}\right)Y_d^\dagger Y_d \Tr(Y_eY_e^\dagger+3Y_dY_d^\dagger) \nonumber \\
& - \left(4 m^2_{H_u}+ m^2_{\tilde Q}\right) Y_u^\dagger Y_u \Tr(Y_\nu Y_\nu^\dagger+3Y_uY_u^\dagger)\nonumber \\
& - 2Y_d^\dagger Y_d \Tr\left(Y_e m^2_{\tilde L}Y_e^\dagger + Y_e^\dagger m^2_{\tilde e}Y_e+3Y_d m^2_{\tilde Q}Y_d^\dagger+3Y_d^\dagger m^2_{\tilde d}Y_d\right) \nonumber \\
& - 2Y_u^\dagger Y_u \Tr\left(Y_\nu m^2_{\tilde L}Y_\nu^\dagger + Y_\nu^\dagger m^2_{\tilde \nu}Y_\nu+3Y_u m^2_{\tilde Q}Y_u^\dagger+3Y_u^\dagger m^2_{\tilde u}Y_u\right) \nonumber \\
& + \frac{2}{5} g_1^2 \left(2m^2_{H_d}Y_d^\dagger Y_d + 4m^2_{H_u}Y_u^\dagger Y_u + m^2_{\tilde Q}Y_d^\dagger Y_d + Y_d^\dagger Y_d m^2_{\tilde Q} + 2 Y_d^\dagger m^2_{\tilde d}Y_d\right.\nonumber \\
&+\left. 2m^2_{\tilde Q}Y_u^\dagger Y_u + 2 Y_u^\dagger Y_u m^2_{\tilde Q} + 4 Y_u^\dagger m^2_{\tilde u}Y_u + 2 T_d^\dagger T_d + 4 T_u^\dagger T_u\right) \nonumber \\
& -\frac{4}{5}g_1^2\left(M_1^*Y_d^\dagger T_d+2 M_1^*Y_u^\dagger T_u-4 |M_1|^2 Y_u^\dagger Y_u\right.\nonumber \\
&+\left. M_1 T_d^\dagger Y_d + 2 M_1 T_u^\dagger Y_u -2 |M_1|^2 Y_d^\dagger Y_d \right)\nonumber \\
& - \frac{128}{3} g_3^4 |M_3|^2~\mathbf{1}_3 + \frac{2}{5}g_1^2g_2^2 Re \left(M_1M_2^*\right)~\mathbf{1}_3 + \frac{32}{45}g_1^2g_3^2 Re \left(M_1M_3^*\right)~\mathbf{1}_3\nonumber \\
& +\frac{199}{75}g_1^4 |M_1|^2~\mathbf{1}_3 + \frac{2}{5}g_1^2g_2^2(|M_1|^2+|M_2|^2)~\mathbf{1}_3 \nonumber \\
& + 32 g_2^2g_3^2(|M_2|^2 + Re(M_2M_3^*) + |M_3|^2)~\mathbf{1}_3\nonumber \\
& + \frac{32}{45}g_1^2g_3^2(|M_1|^2+|M_3|^2)~\mathbf{1}_3 + 33 g_2^4 |M_2|^2~\mathbf{1}_3 \nonumber \\
& + \frac{1}{15}g_1^2 \sigma_1~\mathbf{1}_3 + 3 g_2^2 \sigma_2~\mathbf{1}_3 + \frac{16}{3} g_3^2 \sigma_3~\mathbf{1}_3+ \frac{2}{5}g_1^2S'~\mathbf{1}_3\;,
\end{align}

\begin{align}
(16\pi^2)^2 \beta_{m^2_{\tilde u}}^{(2)} =\ & -2 \left(\vphantom{Y_d^\dagger}2m^2_{H_d}+2m^2_{H_u}+m^2_{\tilde u}\right)Y_u Y_d^\dagger Y_d Y_u^\dagger - 2 \left(\vphantom{Y_d^\dagger}4 m^2_{H_u} + m^2_{\tilde u}\right)Y_u Y_u^\dagger Y_u Y_u^\dagger \nonumber\\
& - 4 T_u \left(T_d^\dagger Y_d Y_u^\dagger + T_u^\dagger Y_u Y_u^\dagger + Y_d^\dagger Y_d T_u^\dagger + Y_u^\dagger Y_u T_u^\dagger\right)\nonumber \\
& - 4 Y_u \left(T_d^\dagger T_d Y_u^\dagger + T_u^\dagger T_u Y_u^\dagger + Y_d^\dagger T_d T_u^\dagger + Y_u^\dagger T_u T_u^\dagger\right)\nonumber \\
& - 2 Y_u \left(2 Y_d^\dagger m^2_{\tilde d}Y_d Y_u^\dagger + Y_d^\dagger Y_d Y_u^\dagger m^2_{\tilde u} + 2 Y_d^\dagger Y_d m^2_{\tilde Q}Y_u^\dagger + 2 Y_u^\dagger m^2_{\tilde u} Y_u Y_u^\dagger \right. \nonumber \\
&+ \left. Y_u^\dagger Y_u Y_u^\dagger m^2_{\tilde u} + 2 Y_u^\dagger Y_u m^2_{\tilde Q} Y_u^\dagger + 2 m^2_{\tilde Q}Y_d^\dagger Y_d Y_u^\dagger +2 m^2_{\tilde Q}Y_u^\dagger Y_u Y_u^\dagger\right)\nonumber \\
& - 2\left(\vphantom{Y_d^\dagger}4 m^2_{H_u}+m^2_{\tilde u}\right) Y_u Y_u^\dagger \Tr(Y_\nu Y_\nu^\dagger + 3 Y_u Y_u^\dagger) \nonumber \\
& - 4 T_u \left(\vphantom{Y_d^\dagger}T_u^\dagger \Tr(Y_\nu Y_\nu^\dagger +3 Y_u Y_u^\dagger)+Y_u^\dagger \Tr(Y_\nu T_\nu^\dagger +3  Y_u T_u^\dagger)\right)\nonumber \\
& - 4 Y_u \left(\vphantom{Y_d^\dagger}T_u^\dagger \Tr(T_\nu Y_\nu^\dagger +3 T_u Y_u^\dagger)+Y_u^\dagger \Tr(T_\nu T_\nu^\dagger +3  T_u T_u^\dagger)\right)\nonumber \\
& - 2 Y_u^\dagger\left(\vphantom{Y_d^\dagger} Y_u^\dagger m^2_{\tilde u}\Tr(Y_\nu Y_\nu^\dagger+3Y_uY_u^\dagger) + 2 m^2_{\tilde Q} Y_u^\dagger \Tr(Y_\nu Y_\nu^\dagger+3Y_uY_u^\dagger)\right.\nonumber \\
& + \left. 2 Y_u^\dagger \Tr(Y_\nu m^2_{\tilde L}Y_\nu^\dagger + Y_\nu^\dagger m^2_{\tilde\nu}Y_\nu + 3 Y_u m^2_{\tilde Q}Y_u^\dagger + 3 Y_u^\dagger m^2_{\tilde u}Y_u)\right)\nonumber \\
& + \left(6 g_2^2 - \frac{2}{5}g_1^2\right)\left(m^2_{\tilde u}Y_u Y_u^\dagger + Y_u Y_u^\dagger m^2_{\tilde u} + 2 m^2_{H_u}Y_u Y_u^\dagger + 2 Y_u m^2_{\tilde Q}Y_u^\dagger + 2 T_u T_u^\dagger\right)\nonumber\\
& - 12 g_2^2\left(\vphantom{Y_d^\dagger}M_2^* T_u Y_u^\dagger - 2 |M_2|^2 Y_u Y_u^\dagger + M_2 Y_u T_u^\dagger\right)\nonumber \\
& + \frac{4}{5}g_1^2\left(\vphantom{Y_d^\dagger}M_1^* T_u Y_u^\dagger - 2 |M_1|^2 Y_u Y_u^\dagger + M_1 Y_u T_u^\dagger\right)\nonumber \\
& - \frac{128}{3}g_3^4 |M_3|^2~\mathbf{1}_3 + \frac{512}{45}g_1^2 g_3^2 \left(|M_1|^2+ Re(M_1M_3^*)+|M_3|^2\right)~\mathbf{1}_3 \nonumber \\
& + \frac{3424}{75}g_1^4 |M_1|^2~\mathbf{1}_3 + \frac{16}{15}g_1^2\sigma_1~\mathbf{1}_3 + \frac{16}{3}g_3^2\sigma_3~\mathbf{1}_3 - \frac{8}{5}g_1^2S'~\mathbf{1}_3\;,
\end{align}

\begin{align}
(16\pi^2)^2 \beta_{m^2_{\tilde d}}^{(2)} =\ & -2 \left(\vphantom{Y_d^\dagger}2m^2_{H_d}+2m^2_{H_u}+m^2_{\tilde d}\right)Y_d Y_u^\dagger Y_u Y_d^\dagger - 2 \left(\vphantom{Y_d^\dagger}4 m^2_{H_d} + m^2_{\tilde d}\right)Y_d Y_d^\dagger Y_d Y_d^\dagger \nonumber\\
& - 4 T_d \left(T_d^\dagger Y_d Y_d^\dagger + T_u^\dagger Y_u Y_d^\dagger + Y_d^\dagger Y_d T_d^\dagger + Y_u^\dagger Y_u T_d^\dagger\right)\nonumber \\
& - 4 Y_d \left(T_d^\dagger T_d Y_d^\dagger + T_u^\dagger T_u Y_d^\dagger + Y_d^\dagger T_d T_d^\dagger + Y_u^\dagger T_u T_d^\dagger\right)\nonumber \\
& - 2 Y_d \left(2 Y_d^\dagger m^2_{\tilde d}Y_d Y_d^\dagger + Y_d^\dagger Y_d Y_d^\dagger m^2_{\tilde d} + 2 Y_d^\dagger Y_d m^2_{\tilde Q}Y_d^\dagger + 2 Y_u^\dagger m^2_{\tilde u} Y_u Y_d^\dagger \right. \nonumber \\
&+ \left. Y_u^\dagger Y_u Y_d^\dagger m^2_{\tilde d} + 2 Y_u^\dagger Y_u m^2_{\tilde Q} Y_d^\dagger + 2 m^2_{\tilde Q}Y_d^\dagger Y_d Y_d^\dagger +2 m^2_{\tilde Q}Y_u^\dagger Y_u Y_d^\dagger\right)\nonumber \\
& - 2\left(\vphantom{Y_d^\dagger}4 m^2_{H_d}+m^2_{\tilde d}\right) Y_d Y_d^\dagger \Tr(Y_e Y_e^\dagger + 3 Y_d Y_d^\dagger) \nonumber \\
& - 4 T_d \left(\vphantom{Y_d^\dagger}T_d^\dagger \Tr(Y_e Y_e^\dagger +3 Y_d Y_d^\dagger)+Y_d^\dagger \Tr(Y_e T_e^\dagger +3  Y_d T_d^\dagger)\right)\nonumber \\
& - 4 Y_d \left(\vphantom{Y_d^\dagger}T_d^\dagger \Tr(T_e Y_e^\dagger +3 T_d Y_d^\dagger)+Y_d^\dagger \Tr(T_e T_e^\dagger +3  T_d T_d^\dagger)\right)\nonumber \\
& - 2 Y_d^\dagger\left(Y_d^\dagger m^2_{\tilde d}\Tr(Y_e Y_e^\dagger+3Y_dY_d^\dagger) + 2 m^2_{\tilde Q} Y_d^\dagger \Tr(Y_e Y_e^\dagger+3Y_dY_d^\dagger)\right.\nonumber \\
& + \left. 2 Y_d^\dagger \Tr(Y_e m^2_{\tilde L}Y_e^\dagger + Y_e^\dagger m^2_{\tilde e}Y_e + 3 Y_d m^2_{\tilde Q}Y_d^\dagger + 3 Y_d^\dagger m^2_{\tilde d}Y_d)\right)\nonumber \\
& + \left(6 g_2^2 + \frac{2}{5}g_1^2\right)\left(m^2_{\tilde d}Y_d Y_d^\dagger + Y_d Y_d^\dagger m^2_{\tilde d} + 2 m^2_{H_d}Y_d Y_d^\dagger + 2 Y_d m^2_{\tilde Q}Y_d^\dagger + 2 T_d T_d^\dagger\right)\nonumber\\
& - 12 g_2^2\left(M_2^* T_d Y_d^\dagger - 2 |M_2|^2 Y_d Y_d^\dagger + M_2 Y_d T_d^\dagger\right)\nonumber \\
& - \frac{4}{5}g_1^2\left(M_1^* T_d Y_d^\dagger - 2 |M_1|^2 Y_d Y_d^\dagger + M_1 Y_d T_d^\dagger\right)\nonumber \\
& - \frac{128}{3}g_3^4 |M_3|^2~\mathbf{1}_3 + \frac{128}{45}g_1^2 g_3^2 \left(|M_1|^2+ Re(M_1M_3^*)+|M_3|^2\right)~\mathbf{1}_3 \nonumber \\
& + \frac{808}{75}g_1^4 |M_1|^2~\mathbf{1}_3 + \frac{4}{15}g_1^2\sigma_1~\mathbf{1}_3 + \frac{16}{3}g_3^2\sigma_3~\mathbf{1}_3 + \frac{4}{5}g_1^2S'~\mathbf{1}_3\;,
\end{align}

\begin{align}
(16\pi^2)^2 \beta_{m^2_{\tilde e}}^{(2)} =\ & -2 \left(\vphantom{Y_d^\dagger}2m^2_{H_d}+2m^2_{H_u}+m^2_{\tilde e}\right)Y_e Y_\nu^\dagger Y_\nu Y_e^\dagger - 2 \left(\vphantom{Y_d^\dagger}4 m^2_{H_d} + m^2_{\tilde e}\right)Y_e Y_e^\dagger Y_e Y_e^\dagger \nonumber\\
& - 4 T_e \left(\vphantom{Y_d^\dagger}T_e^\dagger Y_e Y_e^\dagger + T_\nu^\dagger Y_\nu Y_e^\dagger + Y_e^\dagger Y_e T_e^\dagger + Y_\nu^\dagger Y_\nu T_e^\dagger\right)\nonumber \\
& - 4 Y_e \left(\vphantom{Y_d^\dagger}T_e^\dagger T_e Y_e^\dagger + T_\nu^\dagger T_\nu Y_e^\dagger + Y_e^\dagger T_e T_e^\dagger + Y_\nu^\dagger T_\nu T_e^\dagger\right)\nonumber \\
& - 2 Y_e \left(\vphantom{Y_d^\dagger}2 Y_e^\dagger m^2_{\tilde e}Y_e Y_e^\dagger + Y_e^\dagger Y_e Y_e^\dagger m^2_{\tilde e} + 2 Y_e^\dagger Y_e m^2_{\tilde L}Y_e^\dagger + 2 Y_\nu^\dagger m^2_{\tilde \nu} Y_\nu Y_e^\dagger \right. \nonumber \\
&+ \left. Y_\nu^\dagger Y_\nu Y_e^\dagger m^2_{\tilde e} + 2 Y_\nu^\dagger Y_\nu m^2_{\tilde L} Y_e^\dagger + 2 m^2_{\tilde L}Y_e^\dagger Y_e Y_e^\dagger +2 m^2_{\tilde L}Y_\nu^\dagger Y_\nu Y_e^\dagger\right)\nonumber \\
& - 2\left(\vphantom{Y_d^\dagger}4 m^2_{H_d}+m^2_{\tilde e}\right) Y_e Y_e^\dagger \Tr(Y_e Y_e^\dagger + 3 Y_d Y_d^\dagger) \nonumber \\
& - 4 T_e \left(\vphantom{Y_d^\dagger}T_e^\dagger \Tr(Y_e Y_e^\dagger +3 Y_d Y_d^\dagger)+Y_e^\dagger \Tr(Y_e T_e^\dagger +3  Y_d T_d^\dagger)\right)\nonumber \\
& - 4 Y_e \left(\vphantom{Y_d^\dagger}T_e^\dagger \Tr(T_e Y_e^\dagger + 3T_d Y_d^\dagger)+Y_e^\dagger \Tr(T_e T_e^\dagger +3  T_d T_d^\dagger)\right)\nonumber \\
& - 2 Y_e^\dagger\left(Y_e^\dagger m^2_{\tilde e}\Tr(Y_e Y_e^\dagger+3Y_dY_d^\dagger) + 2 m^2_{\tilde L} Y_e^\dagger \Tr(Y_e Y_e^\dagger+3Y_dY_d^\dagger)\right.\nonumber \\
& + \left. 2 Y_e^\dagger \Tr(Y_e m^2_{\tilde L}Y_e^\dagger + Y_e^\dagger m^2_{\tilde e}Y_e + 3 Y_d m^2_{\tilde Q}Y_d^\dagger + 3 Y_d^\dagger m^2_{\tilde d}Y_d)\right)\nonumber \\
& + \left(6 g_2^2 - \frac{6}{5}g_1^2\right)\left(\vphantom{Y_d^\dagger}m^2_{\tilde e}Y_e Y_e^\dagger + Y_e Y_e^\dagger m^2_{\tilde e} + 2 m^2_{H_d}Y_e Y_e^\dagger + 2 Y_e m^2_{\tilde L}Y_e^\dagger + 2 T_e T_e^\dagger\right)\nonumber\\
& - 12 g_2^2\left(\vphantom{Y_d^\dagger}M_2^* T_e Y_e^\dagger - 2 |M_2|^2 Y_e Y_e^\dagger + M_2 Y_e T_e^\dagger\right)\nonumber \\
& + \frac{12}{5}g_1^2\left(\vphantom{Y_d^\dagger}M_1^* T_e Y_e^\dagger - 2 |M_1|^2 Y_e Y_e^\dagger + M_1 Y_e T_e^\dagger\right)\nonumber \\
& + \frac{2808}{25}g_1^4 |M_1|^2~\mathbf{1}_3 + \frac{12}{5}g_1^2\sigma_1~\mathbf{1}_3 + \frac{12}{5}g_1^2S'~\mathbf{1}_3\;,
\end{align}

\begin{align}
(16\pi^2)^2 \beta_{m^2_{\tilde \nu}}^{(2)} =\ & -2 \left(\vphantom{Y_d^\dagger}2m^2_{H_d}+2m^2_{H_u}+m^2_{\tilde \nu}\right)Y_\nu Y_e^\dagger Y_e Y_\nu^\dagger - 2 \left(\vphantom{Y_d^\dagger}4 m^2_{H_u} + m^2_{\tilde \nu}\right)Y_\nu Y_\nu^\dagger Y_\nu Y_\nu^\dagger \nonumber\\
& - 4 T_\nu \left(\vphantom{Y_d^\dagger}T_e^\dagger Y_e Y_\nu^\dagger + T_\nu^\dagger Y_\nu Y_\nu^\dagger + Y_e^\dagger Y_e T_\nu^\dagger + Y_\nu^\dagger Y_\nu T_\nu^\dagger\right)\nonumber \\
& - 4 Y_\nu \left(\vphantom{Y_d^\dagger}T_e^\dagger T_e Y_\nu^\dagger + T_\nu^\dagger T_\nu Y_\nu^\dagger + Y_e^\dagger T_e T_\nu^\dagger + Y_\nu^\dagger T_\nu T_\nu^\dagger\right)\nonumber \\
& - 2 Y_\nu \left(\vphantom{Y_d^\dagger}2 Y_e^\dagger m^2_{\tilde e}Y_e Y_\nu^\dagger + Y_e^\dagger Y_e Y_\nu^\dagger m^2_{\tilde \nu} + 2 Y_e^\dagger Y_e m^2_{\tilde L}Y_\nu^\dagger + 2 Y_\nu^\dagger m^2_{\tilde \nu} Y_\nu Y_\nu^\dagger \right. \nonumber \\
&+ \left. Y_\nu^\dagger Y_\nu Y_\nu^\dagger m^2_{\tilde \nu} + 2 Y_\nu^\dagger Y_\nu m^2_{\tilde L} Y_\nu^\dagger + 2 m^2_{\tilde L}Y_e^\dagger Y_e Y_\nu^\dagger +2 m^2_{\tilde L}Y_\nu^\dagger Y_\nu Y_\nu^\dagger\vphantom{Y_d^\dagger}\right)\nonumber \\
& - 2\left(\vphantom{Y_d^\dagger}4 m^2_{H_u}+m^2_{\tilde \nu}\right) Y_\nu Y_\nu^\dagger \Tr(Y_\nu Y_\nu^\dagger + 3 Y_u Y_u^\dagger) \nonumber \\
& - 4 T_\nu \left(\vphantom{Y_d^\dagger}T_\nu^\dagger \Tr(Y_\nu Y_\nu^\dagger +3 Y_u Y_u^\dagger)+Y_\nu^\dagger \Tr(Y_\nu T_\nu^\dagger +3 Y_u T_u^\dagger)\right)\nonumber \\
& - 4 Y_\nu \left(\vphantom{Y_d^\dagger}T_\nu^\dagger \Tr(T_\nu Y_\nu^\dagger +3 T_u Y_u^\dagger)+Y_\nu^\dagger \Tr(T_\nu T_\nu^\dagger +3 T_u T_u^\dagger)\right)\nonumber \\
& - 2 Y_\nu^\dagger\left(\vphantom{Y_d^\dagger}Y_\nu^\dagger m^2_{\tilde \nu}\Tr(Y_\nu Y_\nu^\dagger+3Y_uY_u^\dagger) + 2 m^2_{\tilde L} Y_\nu^\dagger \Tr(Y_\nu Y_\nu^\dagger+3Y_uY_u^\dagger)\right.\nonumber \\
& + \left. 2 Y_\nu^\dagger \Tr(Y_\nu m^2_{\tilde L}Y_\nu^\dagger + Y_\nu^\dagger m^2_{\tilde \nu}Y_\nu + 3 Y_u m^2_{\tilde Q}Y_u^\dagger + 3 Y_u^\dagger m^2_{\tilde u}Y_u)\vphantom{Y_d^\dagger}\right)\nonumber \\
& + \left(6 g_2^2 + \frac{6}{5}g_1^2\right)\left(\vphantom{Y_d^\dagger}m^2_{\tilde \nu}Y_\nu Y_\nu^\dagger + Y_\nu Y_\nu^\dagger m^2_{\tilde \nu} + 2 m^2_{H_u}Y_\nu Y_\nu^\dagger + 2 Y_\nu m^2_{\tilde L}Y_\nu^\dagger + 2 T_\nu T_\nu^\dagger\right)\nonumber\\
& - 12 g_2^2\left(\vphantom{Y_d^\dagger}M_2^* T_\nu Y_\nu^\dagger - 2 |M_2|^2 Y_\nu Y_\nu^\dagger + M_2 Y_\nu T_\nu^\dagger\right)\nonumber \\
& - \frac{12}{5}g_1^2\left(\vphantom{Y_d^\dagger}M_1^* T_\nu Y_\nu^\dagger - 2 |M_1|^2 Y_\nu Y_\nu^\dagger + M_1 Y_\nu T_\nu^\dagger\right)\;,
\end{align}

\begin{align}
(16\pi^2)^2 \beta_{m^2_{H_d}}^{(2)} =\ & -2\left(\vphantom{Y_d^\dagger}m^2_{H_d}+m^2_{H_u}\right)\Tr(Y_e Y_\nu^\dagger Y_\nu Y_e^\dagger + 3 Y_d Y_u^\dagger Y_u Y_d^\dagger)\nonumber \\
& - 12 m^2_{H_d}\Tr(Y_e Y_e^\dagger Y_e Y_e^\dagger + 3 Y_d Y_d^\dagger Y_d Y_d^\dagger) \nonumber\\
& -12 \Tr(T_e^\dagger T_e Y_e^\dagger Y_e + 3 T_d^\dagger T_d Y_d^\dagger Y_d + T_e^\dagger Y_e Y_e^\dagger T_e +3 T_d^\dagger Y_d Y_d^\dagger T_d) \nonumber \\
& - 2\Tr(T_e T_\nu^\dagger Y_\nu Y_e^\dagger + 3 T_d T_u^\dagger Y_u Y_d^\dagger + T_e Y_\nu^\dagger Y_\nu T_e^\dagger +3  T_d Y_u^\dagger Y_u T_d^\dagger) \nonumber \\
& - 2\Tr(Y_e T_\nu^\dagger T_\nu Y_e^\dagger + 3 Y_d T_u^\dagger T_u Y_d^\dagger + Y_e Y_\nu^\dagger T_\nu T_e^\dagger + 3  Y_d Y_u^\dagger T_u T_d^\dagger)\nonumber \\
& - 36 \Tr(Y_d Y_d^\dagger m^2_{\tilde d}Y_d Y_d^\dagger + Y_d^\dagger Y_d m^2_{\tilde Q}Y_d^\dagger Y_d) \nonumber \\
& -12 \Tr(Y_e Y_e^\dagger m^2_{\tilde e} Y_e Y_e^\dagger + Y_e^\dagger Y_e m^2_{\tilde L}Y_e^\dagger Y_e)\nonumber \\
& - 6 \Tr(Y_u Y_d^\dagger m^2_{\tilde d} Y_d Y_u^\dagger + Y_d Y_u^\dagger m^2_{\tilde u} Y_u Y_d^\dagger + Y_u^\dagger Y_u m^2_{\tilde Q} Y_d^\dagger Y_d + Y_d^\dagger Y_d m^2_{\tilde Q} Y_u^\dagger Y_u )\nonumber \\
& - 2 \Tr(Y_\nu Y_e^\dagger m^2_{\tilde e} Y_e Y_\nu^\dagger + Y_e Y_\nu^\dagger m^2_{\tilde \nu} Y_\nu Y_e^\dagger + Y_\nu^\dagger Y_\nu m^2_{\tilde L} Y_e^\dagger Y_e + Y_e^\dagger Y_e m^2_{\tilde L} Y_\nu^\dagger Y_\nu )\nonumber \\
& + \left(32 g_3^2 - \frac{4}{5}g_1^2\right)\Tr(T_d^\dagger T_d + m^2_{H_d}Y_d^\dagger Y_d+Y_d m^2_{\tilde Q}Y_d^\dagger + Y_d^\dagger m^2_{\tilde d}Y_d)\nonumber \\
& + \frac{12}{5}g_1^2 \Tr(T_e^\dagger T_e + m^2_{H_d}Y_e^\dagger Y_e + Y_e m^2_{\tilde L}Y_e^\dagger + Y_e^\dagger m^2_{\tilde e}Y_e)\nonumber \\
& - \frac{12}{5}g_1^2\Tr(M_1^* Y_e^\dagger T_e + M_1 T_e^\dagger Y_e - 2 |M_1|^2 Y_e^\dagger Y_e)\nonumber \\
& + \frac{4}{5}g_1^2\Tr(M_1^* Y_d^\dagger T_d + M_1 T_d^\dagger Y_d - 2 |M_1|^2 Y_d^\dagger Y_d)\nonumber\\
& - 32g_3^2\Tr(M_3^* Y_d^\dagger T_d + M_3 T_d^\dagger Y_d - 2 |M_3|^2 Y_d^\dagger Y_d)\nonumber\\
& + \frac{18}{5}g_1^2 g_2^2\left(\vphantom{Y_d^\dagger}|M_1|^2 + Re(M_1 M_2^*)+|M_2|^2\right)\nonumber\\
& + \frac{621}{25}g_1^4 |M_1|^2 + 33 g_2^4 |M_2|^2 + \frac{3}{5}g_1^2\sigma_1 + 3 g_2^2\sigma_2 -\frac{6}{5} g_1^2S'\;,
\end{align}

\begin{align}
(16\pi^2)^2 \beta_{m^2_{H_u}}^{(2)} =\ & -2\left(\vphantom{Y_d^\dagger}m^2_{H_d}+m^2_{H_u}\right)\Tr(Y_e Y_\nu^\dagger Y_\nu Y_e^\dagger + 3 Y_d Y_u^\dagger Y_u Y_d^\dagger)\nonumber \\
& - 12 m^2_{H_u}\Tr(Y_\nu Y_\nu^\dagger Y_\nu Y_\nu^\dagger + 3 Y_u Y_u^\dagger Y_u Y_u^\dagger) \nonumber\\
& -12 \Tr(T_\nu^\dagger T_\nu Y_\nu^\dagger Y_\nu + 3 T_u^\dagger T_u Y_u^\dagger Y_u + T_\nu^\dagger Y_\nu Y_\nu^\dagger T_\nu +3 T_u^\dagger Y_u Y_u^\dagger T_u) \nonumber \\
& - 2\Tr(T_e T_\nu^\dagger Y_\nu Y_e^\dagger + 3 T_d T_u^\dagger Y_u Y_d^\dagger + T_e Y_\nu^\dagger Y_\nu T_e^\dagger +3  T_d Y_u^\dagger Y_u T_d^\dagger) \nonumber \\
& - 2\Tr(Y_e T_\nu^\dagger T_\nu Y_e^\dagger + 3 Y_d T_u^\dagger T_u Y_d^\dagger + Y_e Y_\nu^\dagger T_\nu T_e^\dagger + 3  Y_d Y_u^\dagger T_u T_d^\dagger)\nonumber \\
& - 36 \Tr(Y_u Y_u^\dagger m^2_{\tilde u}Y_u Y_u^\dagger + Y_u^\dagger Y_u m^2_{\tilde Q}Y_u^\dagger Y_u) \nonumber \\
& -12 \Tr(Y_\nu Y_\nu^\dagger m^2_{\tilde \nu} Y_\nu Y_\nu^\dagger + Y_\nu^\dagger Y_\nu m^2_{\tilde L}Y_\nu^\dagger Y_\nu)\nonumber \\
& - 6 \Tr(Y_u Y_d^\dagger m^2_{\tilde d} Y_d Y_u^\dagger + Y_d Y_u^\dagger m^2_{\tilde u} Y_u Y_d^\dagger + Y_u^\dagger Y_u m^2_{\tilde Q} Y_d^\dagger Y_d + Y_d^\dagger Y_d m^2_{\tilde Q} Y_u^\dagger Y_u )\nonumber \\
& - 2 \Tr(Y_\nu Y_e^\dagger m^2_{\tilde e} Y_e Y_\nu^\dagger + Y_e Y_\nu^\dagger m^2_{\tilde \nu} Y_\nu Y_e^\dagger + Y_\nu^\dagger Y_\nu m^2_{\tilde L} Y_e^\dagger Y_e + Y_e^\dagger Y_e m^2_{\tilde L} Y_\nu^\dagger Y_\nu )\nonumber \\
& + \left(32 g_3^2 + \frac{8}{5}g_1^2\right)\Tr(T_u^\dagger T_u + m^2_{H_u}Y_u^\dagger Y_u+Y_u m^2_{\tilde Q}Y_u^\dagger + Y_u^\dagger m^2_{\tilde u}Y_u)\nonumber \\
& - \frac{8}{5}g_1^2\Tr(M_1^* Y_u^\dagger T_u + M_1 T_u^\dagger Y_u - 2 |M_1|^2 Y_u^\dagger Y_u)\nonumber\\
& - 32g_3^2\Tr(M_3^* Y_u^\dagger T_u + M_3 T_u^\dagger Y_u - 2 |M_3|^2 Y_u^\dagger Y_u)\nonumber\\
& + \frac{18}{5}g_1^2 g_2^2\left(\vphantom{Y_d^\dagger}|M_1|^2 + Re(M_1 M_2^*)+|M_2|^2\right)\nonumber\\
& + \frac{621}{25}g_1^4 |M_1|^2 + 33 g_2^4 |M_2|^2 + \frac{3}{5}g_1^2\sigma_1 + 3 g_2^2\sigma_2 +\frac{6}{5} g_1^2S'\;,
\end{align}
with
\begin{align}
\sigma_1 & = \frac{1}{5}g_1^2\left(3 m^2_{H_u} + 3 m^2_{H_d} + \Tr\left(m^2_{\tilde Q}+ 3 m^2_{\tilde L} + 2 m^2_{\tilde d} + 6 m^2_{\tilde e} + 8 m^2_{\tilde u}\right)\right)\;,\\
\sigma_2^2 & = g_2^2 \left(m^2_{H_u} + m^2_{H_d} + \Tr\left(3 m^2_{\tilde Q}+ m^2_{\tilde L}\right)\right)\;,\\
\sigma_3 & = g_3^2 \Tr\left(2 m^2_{\tilde Q}+ m^2_{\tilde d} +  m^2_{\tilde u}\right)\;,
\end{align}
and
\begin{align}
S'  = &\ m^2_{H_d}\Tr\left(Y_e Y_e^\dagger + 3 Y_d Y_d^\dagger \right) - m^2_{H_u}\Tr\left(Y_\nu Y_\nu^\dagger + 3 Y_u Y_u^\dagger\vphantom{Y_d^\dagger}\right)\nonumber \\
& +\Tr\left(\vphantom{Y_d^\dagger}m^2_{\tilde L}Y_e^\dagger Y_e+ m^2_{\tilde L}Y_\nu^\dagger Y_\nu\right) -\Tr\left(m^2_{\tilde Q}Y_d^\dagger Y_d+ m^2_{\tilde Q}Y_u^\dagger Y_u\right)\nonumber\\
& - 2 \Tr\left(Y_d Y_d^\dagger m^2_{\tilde d} + Y_e Y_e^\dagger m^2_{\tilde e}-2Y_u Y_u^\dagger m^2_{\tilde u}\right)\nonumber \\
& + \frac{1}{30}g_1^2\left(9 m^2_{H_u} - 9 m^2_{H_d} + \Tr\left(m^2_{\tilde Q}- 9 m^2_{\tilde L} + 4 m^2_{\tilde d} + 36 m^2_{\tilde e} -32  m^2_{\tilde u}\right)\right)\nonumber \\
& + \frac{3}{2}g_2^2\left(m^2_{H_u} - m^2_{H_d} + \Tr\left(m^2_{\tilde Q}- m^2_{\tilde L}\right)\right)\nonumber \\
& + \frac{8}{3}g_3^2 \Tr\left(m^2_{\tilde Q}+ m^2_{\tilde d} - 2 m^2_{\tilde u}\right)\;.
\end{align}

\section{Self-energies and one-loop tadpoles including inter-generational mixing}
\label{app:FORM}
Here we present the used formulas for the self-energies $\Pi_{ZZ}^T$, $\Pi_{H^+H^-}$, $\Pi_{AA}$, and the one-loop tadpoles $t_u$, $t_d$, which are based on \cite{Pierce:1996zz} but generalized to include inter-generational mixing. In this appendix we employ SLHA 2 conventions \cite{Allanach:2008qq} in the Super-CKM and Super-PMNS basis, to agree with the convention of \cite{Pierce:1996zz}. The soft-breaking mass matrices in the Super-CKM/Super-PMNS basis are obtained from our flavour basis conventions \eqref{eq:W} and \eqref{eq:Lsoft} by
\begin{align}
y_f & \equiv Y_f^\text{diag} = U_f^\dagger Y_f V_f \;, \nonumber \\
\hat T_f &\equiv U_f^\dagger T_f V_f\;, \nonumber \\
\hat m_{\tilde Q}^2 &\equiv V_d^\dagger m_{\tilde Q}^2 V_d\;,\nonumber \\
\hat m_{\tilde L}^2 &\equiv V_e^\dagger m_{\tilde L}^2 V_e\;,\nonumber \\
\hat m_{\tilde u}^2 &\equiv U_u^\dagger m_{\tilde u}^2 U_u\;,\nonumber \\
\hat m_{\tilde d}^2 &\equiv U_d^\dagger m_{\tilde d}^2 U_d\;,\nonumber \\
\hat m_{\tilde e}^2 &\equiv U_e^\dagger m_{\tilde e}^2 U_e\;.
\end{align}

Let us briefly review our generalization to the sfermion mass matrices of \cite{Pierce:1996zz}: We define the sfermion mixing matrices by\footnote{The SLHA 2 convention sfermion mixing matrices $R_{\tilde f}$ can be obtained via $R_{\tilde f}=W_{\tilde f}^\dagger$.}
\be
M_{\tilde f}^2 = W_{\tilde f} M_{\tilde f}^{2 \text{ diag}} W_{\tilde f}^\dagger\;,
\ee
with the sfermion mass matrices in the Super-CKM/Super-PMNS basis
\begin{align}
M_{\tilde u}^2 = & \begin{pmatrix}
V_\text{CKM}\hat m_{\tilde Q}^2V_\text{CKM}^\dagger + \frac{v^2}{2} y_u^2 \sin^2\beta + D_{u,L} & \frac{v}{\sqrt{2}} \hat T_{ u}^\dagger\sin\beta - \mu \frac{v}{\sqrt{2}} y_u \cos\beta \\
      \frac{v}{\sqrt{2}} \hat T_{ u} \sin\beta - \mu^* \frac{v}{\sqrt{2}} y_u \cos\beta &
        \hat m_{\tilde u}^2 + \frac{v^2}{2} y_u^2 \sin^2\beta + D_{u,R}
\end{pmatrix}\;, \nonumber\\
M_{\tilde d}^2 = & \begin{pmatrix}
\hat m_{\tilde Q}^2 + \frac{v^2}{2} y_d^2 \cos^2\beta + D_{d,L} & \frac{v}{\sqrt{2}} \hat T_{ d}^\dagger\cos\beta - \mu \frac{v}{\sqrt{2}} y_d \sin\beta \\
      \frac{v}{\sqrt{2}} \hat T_{ d} \cos\beta - \mu^* \frac{v}{\sqrt{2}} y_d \sin\beta &
        \hat m_{\tilde d}^2 + \frac{v^2}{2} y_d^2 \cos^2\beta + D_{d,R}
\end{pmatrix}\;, \nonumber\\
M_{\tilde e}^2 = & \begin{pmatrix}
\hat m_{\tilde L}^2 + \frac{v^2}{2} y_e^2 \cos^2\beta + D_{e,L} & \frac{v}{\sqrt{2}} \hat T_{ e}^\dagger\cos\beta - \mu \frac{v}{\sqrt{2}} y_e \sin\beta \\
      \frac{v}{\sqrt{2}} \hat T_{ e} \cos\beta - \mu^* \frac{v}{\sqrt{2}} y_e \sin\beta &
        \hat m_{\tilde e}^2 + \frac{v^2}{2} y_e^2 \cos^2\beta + D_{e,R}
\end{pmatrix}\;, \nonumber\\
M_{\tilde \nu}^2 = & V_\text{PMNS}^\dagger \hat m_{\tilde L}^2 V_\text{PMNS} + D_{\nu,L}\;.
\end{align}
The D-terms are given by
\begin{align}
D_{f,L} & = \hat M_Z^2 (I_3 - Q_e \sin^2\theta_W) \cos(2 \beta) ~\mathbf{1}_3\;, \nonumber \\
D_{f,R} & = \hat M_Z^2 Q_e \cos(2 \beta) \sin^2\theta_W ~\mathbf{1}_3\;,
\end{align}
where $I_3$ denotes the $SU(2)_L$ isospin and $Q_e$ the electric charge of the flavour $f$, and $\theta_W$ denotes the weak mixing angle. Note that our convention for $\mu$ differs by a sign from the convention in \cite{Pierce:1996zz}. 

For the sake of completeness we also list the conventions for neutralino and chargino mass matrices and mixing matrices: The neutralino mixing matrix is defined by
\be
M_{\psi^0} = N^T M_{\psi^0}^\text{diag} N\;,
\ee
with
\be
M_{\psi^0} = \begin{pmatrix}
 M_1 & 0 & -M_Z \cos\beta \sin\theta_W & M_Z \sin\beta \sin\theta_W \\ 
 0 & M_2 & M_Z \cos\beta \cos\theta_W & -M_Z \sin\beta \cos\theta_W \\ 
 -M_Z \cos\beta \sin\theta_W & M_Z \cos\beta \cos\theta_W & 0 & -\mu \\ 
 M_Z \sin\beta \sin\theta_W & -M_Z \sin\beta \cos\theta_W & -\mu & 0
\end{pmatrix}\;.
\ee
The chargino mixing matrix is defined by 
\be
M_{\psi^+} = U^T M_{\psi^+}^\text{diag} V\;,
\ee
with
\be
M_{\psi^+} = \begin{pmatrix}
M_2 & \sqrt{2} M_W \sin\beta \\
\sqrt{2} M_W \cos\beta & \mu
\end{pmatrix}\;.
\ee

We now present the generalization of $\Pi_{ZZ}^T$, $\Pi_{H^+H^-}$, $\Pi_{AA}$, $t_u$, and $t_d$ of \cite{Pierce:1996zz} to include inter-generational mixing. For all we have checked that our equations reduce to the corresponding equations in \cite{Pierce:1996zz} when 
\begin{align}
W_{\tilde f_{i\,i}} & =  W_{\tilde f_{i+3\,i+3}} = \cos\theta_{\tilde f_i} \nonumber \\
W_{\tilde f_{i\,i+3}} & =  - W_{\tilde f_{i+3\,i}} = - \sin\theta_{\tilde f_i}\;,
\end{align}
where $i=1\dots 3$.

We keep the abbreviations of \cite{Pierce:1996zz}:
\begin{align}
s_{\alpha\beta} &\equiv \sin(\alpha - \beta)\;,\\
c_{\alpha\beta} &\equiv \cos(\alpha - \beta)\;, \\
g_{f_L} &\equiv I_3 - Q_e \sin^2\theta_W\;,\\
g_{f_R}&\equiv Q_e \sin^2\theta_W\;,\\
e & \equiv g_2 \sin\theta_W\;, \\
N_C &\equiv \begin{cases}
    3       & \quad \text{for (s)quarks}\\
    1  & \quad \text{for (s)fermions}\\
  \end{cases}\;.
\end{align}
The conventions for the one-loop scalar functions $A_0$, $B_{22}$, $\tilde B_{22}$, $H$, $G$, and $F$ \cite{'tHooft:1978xw} are adopted from appendix B of \cite{Pierce:1996zz}. Summations $\sum_f$ are over all fermions, whereas summations $\sum_{f_u}$, $\sum_{f_d}$ are restricted to up-type and down-type fermions, respectively. Summations $\sum_Q$, $\sum_{\tilde Q}$ are over $SU(2)$ (s)quark doublets, and analogously for (s)leptons. In summations over sfermions the indices $i$, $j$, $s$, and $t$ run from 1 to 6 for $\tilde u$, $\tilde d$, and $\tilde e$ and from 1 to 3 for $\tilde \nu$. In summations of neutralinos (charginos) the indices $i$, $j$ run from 1 to 4 (2). The summations $\sum_{h^0}$ runs over all neutral Higgs- and Goldstone bosons, the summation $\sum_{h^+}$ over the charged ones.

\begin{align}
16\pi^2\frac{\cos^2\theta_W}{g_2^2} \Pi_{ZZ}^T(p^2)  = & -s^2_{\alpha\beta} \left(\tilde B_{22}(m_A,m_H)+\tilde B_{22}(M_Z,m_h) - M_Z^2 B_0(M_Z,m_h)\right)\nonumber \\
& -c^2_{\alpha\beta} \left(\tilde B_{22}(M_Z,m_H)+\tilde B_{22}(m_A,m_h) - M_Z^2 B_0(M_Z,m_h)\right)\nonumber \\
& - 2\cos^4\theta_W \left(2 p^2 + M_W^2 - M_Z^2 \frac{\sin^4\theta_W}{\cos^2\theta_W} \right)B_0(M_W,M_W) \nonumber \\
& - \left(8 \cos^4\theta_W + \cos^2(2 \theta_W)\right)\tilde B_{22}(M_W,M_W)\nonumber \\
& - \cos^2(2 \theta_W)\tilde B_{22}(m_{H^+}, m_{H^+})\nonumber \\
& - \sum_{\tilde f} N_C \sum_{s,t} \left|I_3 \sum_i W_{\tilde f_{is}}^* W_{\tilde f_{it}} - Q_e \sin^2\theta_W \delta_{st}\right|^2 4 B_{22}(m_{\tilde f_s},m_{\tilde f_t}) \nonumber \\
& + \frac{1}{2}\sum_{\tilde f} N_C \sum_s\left((1-8 I_3 Q_e \sin^2\theta_W)\sum_i W_{\tilde f_{is}}^*W_{\tilde f_{is}} \right. \nonumber \\
& +\left.  4 Q_e^2 \sin^4\theta_W\vphantom{\sum_i}\right) A_0(m_{\tilde f_s})\nonumber \\
& + \sum_f N_C \left( \left(g_{f_L}^2 + g_{f_R}^2\right)H(m_f,m_f)-4 g_{f_L}g_{f_R} m_f^2 B_0(m_{f},m_{f})\right)\nonumber \\
& + \frac{\cos^2\theta_W}{2 g_2^2} \sum_{i,j} f_{ijZ}^0H(m_{\tilde \chi_i^0},m_{\tilde \chi_j^0}) + 2g_{ijZ}^0 m_{\tilde \chi^0_i}m_{\tilde\chi^0_j}B_0(m_{\tilde \chi_i^0},m_{\tilde \chi_j^0})\nonumber \\
& + \frac{\cos^2\theta_W}{g_2^2} \sum_{i,j} f_{ijZ}^+H(m_{\tilde \chi_i^+},m_{\tilde \chi_j^+}) + 2g_{ijZ}^+ m_{\tilde \chi^+_i}m_{\tilde\chi^+_j} B_0(m_{\tilde \chi_i^+},m_{\tilde \chi_j^+})\;.
\end{align}
The couplings $f_Z^0$, $f_Z^+$, $g_Z^0$, and $g_Z^+$ are given in Eqs.~(A.7) and (D.5) of \cite{Pierce:1996zz}.

\begin{align}
16\pi^2 \Pi_{H^+H^-}(p^2)  = & \sum_{Q} N_C \left(\vphantom{\sum_i}\left(\cos^2\beta y_{u}^2+\sin^2\beta y_d^2\right)G(m_u,m_d) \right. \nonumber \\
& -\left. 2 \sin(2\beta)y_u y_d m_u m_d B_0(m_u,m_d)\vphantom{\sum_i}\right)\nonumber \\
& + \sum_{L} \vphantom{\sum_i}\sin^2\beta y_e^2 G(0,m_e) \nonumber \\
& + \sum_{\tilde Q} N_C \sum_{i,j}\left(\lambda_{H^+\tilde Q}\right)_{ij}^2B_0(m_{\tilde u_i},m_{\tilde d_j})\nonumber \\
& + \sum_{\tilde L} \sum_{i,j}\left(\lambda_{H^+\tilde L}\right)_{ij}^2B_0(m_{\tilde \nu_i},m_{\tilde e_j})\nonumber \\
& + \sum_{\tilde f} \sum_i \left(\lambda_{H^+H^-\tilde f}\right)_{ii}A_0(m_{\tilde f_i})\nonumber \\
& + \frac{g_2^2}{4}\left(\vphantom{\frac{\cos^2(2\theta_W)}{\cos^2\theta_W}} s_{\alpha\beta}^2 F(m_H,M_W) + c_{\alpha\beta}^2 F(m_h,M_W)+F(m_A,M_W)\right.\nonumber\\
&+\left.\frac{\cos^2(2\theta_W)}{\cos^2\theta_W}F(m_{H^+},M_Z)\right)\nonumber \\
& + e^2 F(m_{H^+},0) + 2g_2^2 A_0(M_W) + g_2^2 \cos^2(2\theta_W) A_0(M_Z)\nonumber \\
& + \sum_{h^+} \left(\sum_{h^0} \left(\lambda_{H^+h^0h^-}\right)^2 B_0(m_{h^0},m_{h^+}) +\lambda_{H^+H^-h^+h^-} A_0(m_{h^+}) \right)\nonumber \\
& + g_2^2\frac{M_W^2}{4} B_0(M_W,m_A) \nonumber \\
& + \frac{1}{2}\sum_{h^0}\lambda_{H^+H^-h^0h^0}A_0(m_{h^0})\nonumber \\
& +  \sum_{j,i} f_{ijH^+} G(m_{\tilde \chi_j^+},m_{\tilde \chi_i^0}) - 2 m_{\tilde \chi_j^+}m_{\tilde \chi_i^0}g_{ijH^+} B_0(m_{\tilde \chi_j^+},m_{\tilde \chi_i^0})\;.
\end{align}
The couplings $f_{H^+}$ and $g_{H^+}$ are given in Eq.~(D.70) and Eqs.~(D.39-D.42) of \cite{Pierce:1996zz}. The couplings $\lambda_{H^+H^-h^-}$, $\lambda_{H^+H^-h^+h^-}$, and $\lambda_{H^+H^-h^0h^0}$ are defined in Eqs.~(D.63-D.65) and Eq.~(D.67) of \cite{Pierce:1996zz}. The couplings to sfermions in the case of inter-generational mixing are given by
\begin{align}
\lambda_{H^+\tilde Q} &= W_{\tilde u}
\begin{pmatrix}
\frac{g2}{\sqrt{2}} \hat M_W \sin(2 \beta)~\mathbf{1}_3 - \left(y_u^2+y_d^2\right) \cos\beta \frac{v \sin\beta}{\sqrt{2}} & - \hat T_d^\dagger \sin\beta - \mu y_d \cos\beta \\
-\hat T_u \cos\beta - \mu^* y_u \sin\beta & -y_uy_d \frac{v}{\sqrt{2}}
\end{pmatrix}
W_{\tilde d}\;,\\
\lambda_{H^+\tilde L} &= W_{\tilde \nu}
\begin{pmatrix}
      \frac{g_2}{\sqrt{2}} \hat M_W \sin(2 \beta)~\mathbf{1}_3 - y_e^2 \cos\beta \frac{v \sin\beta}{\sqrt{2}} \quad&\quad - \hat T_e^\dagger \sin\beta - \mu y_e \cos\beta
\end{pmatrix}
W_{\tilde e}\;,\\
\lambda_{H^+H^-\tilde f} &= -W_{\tilde f}^\dagger
\begin{pmatrix}
        -\frac{g_2^2}{2} \cos(2 \beta) \left(\frac{\cos(2 \theta_W)}{\cos^2\theta_W} I_3 + \tan^2\theta_W Q_e\right) ~\mathbf{1}_3 & 0 \\
0 &  \frac{g_2^2}{2} \cos(2 \beta) \tan^2\theta_W Q_e ~\mathbf{1}_3 
\end{pmatrix}W_{\tilde f} \nonumber\\
& + 
\begin{cases}
W_{\tilde u}^\dagger 
\begin{pmatrix}
         y_d^2 \sin^2\beta & 0 \\
0 &    y_u^2  \cos^2\beta
\end{pmatrix}W_{\tilde u}\qquad \tilde f = \tilde u \\[5mm]
W_{\tilde d}^\dagger 
\begin{pmatrix}
        y_u^2  \cos^2\beta & 0 \\
0 &   y_d^2 \sin^2\beta
\end{pmatrix}W_{\tilde d}\qquad \tilde f = \tilde d \\[5mm]
W_{\tilde e}^\dagger 
\begin{pmatrix}
        0 & 0 \\
0 &    y_e^2 \sin^2\beta
\end{pmatrix}W_{\tilde e}\qquad \tilde f = \tilde e
\end{cases}\;,\\
\lambda_{H^+H^-\tilde \nu} &= -W_{\tilde \nu}^\dagger
\left(
-\frac{g_2^2}{2} \cos(2 \beta) \left(\frac{\cos(2 \theta_W)}{\cos^2\theta_W} I_3 + \tan^2\theta_W Q_e\right) ~\mathbf{1}_3 -  y_e^2\sin^2\beta\right)W_{\tilde\nu}
\end{align}

\begin{align}
16\pi^2 \Pi_{AA}(p^2)  = & \cos^2\beta \sum_{f_u} N_C y_{f_u}^2 \left(p^2 B_0(m_{f_u},m_{f_u})-2 A_0(m_{f_u})\right)\nonumber \\
& + \sin^2\beta \sum_{f_d} N_C y_{f_d}^2 \left(p^2 B_0(m_{f_d},m_{f_d})-2 A_0(m_{f_d})\right)\nonumber \\
& + \sum_{\tilde f}\sum_i N_C \left(\left(\lambda_{AA\tilde f}\right)_{ii}A_0(m_{\tilde f_i})+\sum_j \left(\lambda_A\tilde f\right)^2_{ij} B_0(m_{\tilde f_i},m_{\tilde f_j})\right)\nonumber \\
& + \frac{g_2^2}{4}\left(2 F(m_{H^+},M_W)+ \frac{s_{\alpha\beta}^2}{\cos^2\theta_W} F(m_H,M_Z) + \frac{c_{\alpha\beta}^2}{\cos^2\theta_W} F(m_h,M_Z) \right)\nonumber \\
& + \frac{1}{2}\sum_{h^0_n}\left(\sum_{h^0_m}\lambda_{Ah^0_nh^0_m}B_0(m_{h^0_n},m_{h^0_m})
+\lambda_{AAh^0_nh^0_n}A_0(m_{h^0_n})\right)\nonumber \\
& + g_2^2\left( \frac{M_W^2}{2}B_0(M_W,m_{H_p}) + 2 A_0(M_W) + \frac{1}{\cos^2\theta_W}A_0(M_Z)\right)\nonumber \\
& + \sum_{h^+}\lambda_{AAh^+h^+} A_0(m_{h^+}) \nonumber \\
& + \frac{1}{2} \sum_{i,j}f_{ijA}^0 G(m_{\tilde \chi^0_i},m_{\tilde\chi^0_j})- 2 g_{ijA}^0 m_{\tilde \chi^0_i}m_{\tilde\chi^0_j}B_0(m_{\tilde \chi^0_i},m_{\tilde\chi^0_j}) \nonumber \\
& + \sum_{i,j}f_{ijA}^+ G(m_{\tilde \chi^+_i},m_{\tilde\chi^+_j})- 2 g_{ijA}^+ m_{\tilde \chi^+_i}m_{\tilde\chi^+_j}B_0(m_{\tilde \chi^+_i},m_{\tilde\chi^+_j})\;.
\end{align}
The couplings $f_{A}^0$, $g_{A}^0$, $f_{A}^+$, and $g_{A}^+$ are given in Eq.~(D.70) and Eqs.~(D.34-D.38) of \cite{Pierce:1996zz}. The couplings $\lambda_{Ah^0h^0}$, $\lambda_{AAh^0h^0}$, and $\lambda_{AAh^+h^+}$ are defined in Eqs.~(D.63-D.65) and Eq.~(D.67) of \cite{Pierce:1996zz}. The couplings to sfermions in the case of inter-generational mixing are given by
\begin{align}
\lambda_{A\tilde u} & = W_{\tilde u}^\dagger
\begin{pmatrix}
0 & -\frac{1}{\sqrt{2}}\left(\hat T_{ u}^\dagger \cos\beta +\mu y_u \sin\beta\right) \\
\frac{1}{\sqrt{2}}\left(\hat T_{ u} \cos\beta +\mu^* y_u \sin\beta\right)
\end{pmatrix}W_{\tilde u}\;,\\
\lambda_{A\tilde e(\tilde d)} & = W_{\tilde e(\tilde d)}^\dagger
\begin{pmatrix}
0 & -\frac{1}{\sqrt{2}}\left(\hat T_{ e( d)}^\dagger \sin\beta +\mu y_{e(d)} \cos\beta\right) \\
\frac{1}{\sqrt{2}}\left(\hat T_{ e( d)} \sin\beta +\mu^* y_{e(d)} \cos\beta\right)
\end{pmatrix}W_{\tilde e(\tilde d)}\;,\\
\lambda_{A\tilde \nu} & = 0\;,\\
\lambda_{AA\tilde f} & = W_{\tilde f}^\dagger
\begin{pmatrix}
        -\frac{g_2^2}{2} \cos(2 \beta) \left(\frac{1}{\cos^2\theta_W} I_3 - \tan^2\theta_W Q_e\right) ~\mathbf{1}_3 & 0 \\
0 &  -\frac{g_2^2}{2} \cos(2 \beta) \tan^2\theta_W Q_e ~\mathbf{1}_3 
\end{pmatrix}W_{\tilde f} \nonumber\\
& + 
\begin{cases}
W_{\tilde u}^\dagger 
\begin{pmatrix}
         y_u^2 \cos^2\beta & 0 \\
0 &    y_u^2 \cos^2\beta
\end{pmatrix}W_{\tilde u}\qquad \tilde f = \tilde u \\[5mm]
W_{\tilde d}^\dagger 
\begin{pmatrix}
         y_d^2 \sin^2\beta & 0 \\
0 &    y_d^2 \sin^2\beta
\end{pmatrix}W_{\tilde d}\qquad \tilde f = \tilde d \\[5mm]
W_{\tilde e}^\dagger 
\begin{pmatrix}
        y_e^2 \sin^2\beta & 0 \\
0 &   y_e^2 \sin^2\beta
\end{pmatrix}W_{\tilde e}\qquad \tilde f = \tilde e
\end{cases}\;,\\
\lambda_{AA\tilde \nu} & = -\frac{g_2^2}{2} \cos(2 \beta) \left(\frac{1}{\cos^2\theta_W} I_3 - \tan^2\theta_W Q_e\right)~\mathbf{1}_3
\end{align}

\begin{align}
 16\pi^2 t_d & = -2\sum_{f_d} N_C y_{f_d}^2 A_0(m_{f_d}) \nonumber \\
 & + \sum_{\tilde f} N_C \sum_i \frac{g_2^2}{2 M_W \cos\beta} \left(\lambda_{d\tilde f}\right)_{ii} A_0(m_{\tilde f_i}) \nonumber \\
 & - g_2^2 \frac{\cos(2\beta)}{8\cos^2\theta_W}\left(A_0(m_A)+ 2 A_0(m_{H^+})\right) + \frac{g_2^2}{2}A_0(m_{H_p})\nonumber \\
 & + \frac{g_2^2}{8\cos^2\theta_W}\left(3 \sin^2\alpha-\cos^2\alpha+\sin(2\alpha)\tan\beta\right)A_0(m_h) \nonumber \\
  & + \frac{g_2^2}{8\cos^2\theta_W}\left(3 \cos^2\alpha-\sin^2\alpha-\sin(2\alpha)\tan\beta\right)A_0(m_H) \nonumber \\
  & - g_2^2 \sum_i  \frac{m_{\tilde \chi_i^0}}{M_W\cos\beta}Re\left(N_{i3}(N_{i2}-N_{i,1}\tan\theta_W)\right) A_0(m_{\tilde\chi^0_i})\nonumber \\
  & - \sqrt{2}g_2^2\sum_i \frac{m_{\tilde \chi_i^+}}{M_W\cos\beta}Re\left(V_{i1}U_{i2}\right) A_0(m_{\tilde\chi^+_i})\nonumber \\
  & + \frac{3}{4}g_2^2\left(2 A_0(M_W)+ \frac{A_0(M_Z)}{\cos^2\theta_W}\right)\nonumber \\
  & + g_2^2 \frac{\cos(2\beta)}{8\cos^2\theta_W}\left(2 A_0(M_W)+  A_0(M_Z) \right)\;.
\end{align}
The couplings to sfermion in the case of inter-generational mixing are given by
\begin{align}
\lambda_{d\tilde u} & = W_{\tilde u}^\dagger
\begin{pmatrix}
g_2 \frac{M_Z}{\cos\theta_W} g_{u_L} \cos\beta  ~\mathbf{1}_3 & - y_u \frac{\mu}{\sqrt{2}} \\
- y_u \frac{\mu^*}{\sqrt{2}} & g_2 \frac{M_Z}{\cos\theta_W} g_{u_R} \cos\beta  ~\mathbf{1}_3
\end{pmatrix} W_{\tilde u}\;,\\
\lambda_{d\tilde f} & = W_{\tilde f}^\dagger
\begin{pmatrix}
g_2 \frac{M_Z}{\cos\theta_W} g_{f_L} \cos\beta  ~\mathbf{1}_3 & \hat T_{ f}^\dagger \frac{1}{\sqrt{2}} \\
\hat T_{ f} \frac{1}{\sqrt{2}} & g_2 \frac{M_Z}{\cos\theta_W} g_{f_R} \cos\beta  ~\mathbf{1}_3
\end{pmatrix}\nonumber \\
& + \begin{cases}
W_{\tilde d}^\dagger 
\begin{pmatrix}
        Y_d^2 v \cos\beta & 0 \\
0 &   Y_d^2 v \cos\beta 
\end{pmatrix}W_{\tilde d}\qquad \tilde f = \tilde d \\[5mm]
W_{\tilde e}^\dagger 
\begin{pmatrix}
        Y_e^2 v \cos\beta & 0 \\
0 &   Y_e^2 v \cos\beta 
\end{pmatrix}W_{\tilde e}\qquad \tilde f = \tilde e
\end{cases}\;,\\
\lambda_{d\tilde \nu} & = g_2 \frac{M_Z}{\cos\theta_W} g_{\nu_L} \cos\beta  ~\mathbf{1}_3
\end{align}

\begin{align}
 16\pi^2 t_u & = -2\sum_{f_u} N_C y_{f_u}^2 A_0(m_{f_u}) \nonumber \\
 & + \sum_{\tilde f} N_C \sum_i \frac{g_2^2}{2 M_W \sin\beta} \left(\lambda_{u\tilde f}\right)_{ii} A_0(m_{\tilde f_i}) \nonumber \\
 & + g_2^2 \frac{\cos(2\beta)}{8\cos^2\theta_W}\left(A_0(m_A)+ 2 A_0(m_{H^+})\right) + \frac{g_2^2}{2}A_0(m_{H_p})\nonumber \\
  & + \frac{g_2^2}{8\cos^2\theta_W}\left(3 \cos^2\alpha-\sin^2\alpha+\sin(2\alpha)\cot\beta\right)A_0(m_h) \nonumber \\
   & + \frac{g_2^2}{8\cos^2\theta_W}\left(3 \sin^2\alpha-\cos^2\alpha-\sin(2\alpha)\cot\beta\right)A_0(m_H) \nonumber \\
  & - g_2^2 \sum_i  \frac{m_{\tilde \chi_i^0}}{M_W\sin\beta}Re\left(N_{i4}(N_{i2}-N_{i,1}\tan\theta_W)\right) A_0(m_{\tilde\chi^0_i})\nonumber \\
  & - \sqrt{2}g_2^2\sum_i \frac{m_{\tilde \chi_i^+}}{M_W\sin\beta}Re\left(V_{i2}U_{i1}\right) A_0(m_{\tilde\chi^+_i})\nonumber \\
  & + \frac{3}{4}g_2^2\left(2 A_0(M_W)+ \frac{A_0(M_Z)}{\cos^2\theta_W}\right)\nonumber \\
  & - g_2^2 \frac{\cos(2\beta)}{8\cos^2\theta_W}\left(2 A_0(M_W)+  A_0(M_Z) \right)\;.
\end{align}
The couplings to sfermion in the case of inter-generational mixing are given by
\begin{align}
\lambda_{u\tilde u} & = W_{\tilde u}^\dagger
\begin{pmatrix}
-g_2 \frac{M_Z}{\cos\theta_W} g_{u_L} \sin\beta  ~\mathbf{1}_3 + y_u^2 v \sin\beta &  \hat T_u^\dagger \frac{1}{\sqrt{2}} \\
\hat T_u \frac{1}{\sqrt{2}} & -g_2 \frac{M_Z}{\cos\theta_W} g_{u_R} \sin\beta  ~\mathbf{1}_3 + y_u^2 v \sin\beta
\end{pmatrix} W_{\tilde u}\;,\\
\lambda_{u\tilde f} & = W_{\tilde f}^\dagger
\begin{pmatrix}
-g_2 \frac{M_Z}{\cos\theta_W} g_{f_L} \sin\beta  ~\mathbf{1}_3 & -y_{\tilde f} \frac{\mu}{\sqrt{2}} \\
-y_{\tilde f} \frac{\mu^*}{\sqrt{2}} & -g_2 \frac{M_Z}{\cos\theta_W} g_{f_R} \sin\beta  ~\mathbf{1}_3
\end{pmatrix}\qquad\tilde f = \tilde e\;,\tilde d\;,\nonumber \\
\lambda_{u\tilde \nu} & = - g_2 \frac{M_Z}{\cos\theta_W} g_{\nu_L} \sin\beta  ~\mathbf{1}_3\;.
\end{align}

\section{\SusyTC~documentation}
\label{app:DOC}
Here we present a documentation of the \texttt{REAP} extension \SusyTC. To get started, please follow first the steps described in Section 4.

We now describe that additional features of \SusyTC:
In addition to the features of \texttt{REAP} package \texttt{RGEMSSMsoftbroken.m} (described in the \texttt{REAP} documentation), \SusyTC~adds the following options to the command \texttt{RGEAdd}:
\begin{itemize}
\item \texttt{STCsign$\mu$} is the general factor $e^{i\phi_\mu}$ in front of $\mu$ in \eqref{eq:mu}. (default: $+1$)
\item \texttt{STCcMSSM} is a switch to change between the CP-violating (complex) MSSM and CP-conserving (real) MSSM. (default: \texttt{True})
\item \texttt{STCSusyScale} sets the SUSY scale $Q$ (in $\GeV$), where the MSSM is matched to the SM. If set to \texttt{"Automatic"}, \SusyTC~determines $Q$ automatically from the sparticle spectrum. (default: \texttt{"Automatic"})
\item \texttt{STCSusyScaleFromStops} is a switch to choose whether \SusyTC~calculates the SUSY scale $Q$ as geometric mean of the stop masses $Q = \sqrt{m_{\tilde t_1}m_{\tilde t_2}}$,
where the stop masses are defined by the up-type squark mass eigenstates $\tilde u_i$ with the largest mixing to $\tilde t_1$ and $\tilde t_2$, or as geometric mean of the lightest and heaviest up-type squarks $Q=\sqrt{m_{\tilde u_1}m_{\tilde u_6}}$. Without effect if \texttt{STCSusyScale} is not set to \texttt{"Automatic"}. (default: \texttt{True})
\item \texttt{STCSearchSMTransition} is a switch to enable or disable the matching to the SM and the calculation of supersymmetric threshold corrections and sparticle spectrum. (default: \texttt{True})
\item \texttt{STCCCBConstraints} is a switch to enable or disable a warning message for potentially dangerous charge and colour breaking vacua, if large trilinear couplings violate the constraints of \cite{Casas:1996de} at the SUSY scale $Q$
\begin{equation}
|T_{ij}|^2 \leq \left(m_{R_{ii}}^2 + m_{L_{jj}}^2+m_{H_f}^2 + |\mu|^2 \right) \begin{cases}
(y_i^2+y_j^2) \quad i \neq j\\[5mm]
3 y_i^2 \quad i = j
\end{cases}\;,
\end{equation}
where $m_L$, $m_R$ and $m_{H_f}$ denote the soft-breaking mass parameters of the scalar fields associated with the trilinear coupling $T$ in the basis of diagonal Yukawa matrices. (default: \texttt{True})
\item \texttt{STCUFBConstraints} is a switch to enable or disable a warning message for possibly dangerous ``unbounded from below'' directions in the scalar potential, if the constraints of \cite{Casas:1995pd} are violated at the SUSY scale $Q$
\begin{align}
m_{H_u}^2 + |\mu|^2 + m^2_{\tilde L_i} - \frac{|m_3|^4}{m_{H_d}^2+|\mu|^2-m^2_{\tilde L_i}} > 0 \qquad \text{UFB-2}\;, \\
m_{H_u}^2 + m^2_{\tilde L_i} > 0 \qquad \text{UFB-3}\;,
\end{align}
evaluated in the basis of \eqref{eq:rotY}. Note that the UFB-I constraint is automatically satisfied, since \SusyTC~calculates $m_3^2$ from $m^2_{H_u}$, $m^2_{H_d}$, $M_Z$, and $\tan\beta$ by requiring the existence of electroweak symmetry breaking. (default: \texttt{True})
\end{itemize}
Thus, a typical call of \texttt{RGEAdd[]} might look like\\

\hspace{1cm}\textbf{	RGEAdd[}\texttt{"MSSMsoftbroken",RGEtan$\beta$->20,STCcMSSM->False,STCsign$\mu$->-1}\textbf{];}\\

In addition to the parameters known from the \texttt{MSSM} \texttt{REAP} model, the following soft-breaking parameters are available as input for \texttt{RGESetInitial}:
\begin{itemize}
\item \texttt{RGETu}, \texttt{RGETd}, \texttt{RGETe}, and \texttt{RGET$\nu$} are the soft-breaking trilinear coupling matrices.  If given, the Constrained MSSM parameter \texttt{RGEA0} for the corresponding trilinear coupling is overwritten. (default: Constrained MSSM)
\item \texttt{RGEM1}, \texttt{RGEM2}, and \texttt{RGEM3} are the soft-breaking gaugino mass parameters.  If given, the Constrained MSSM parameter \texttt{RGEM12} for the corresponding gaugino is overwritten. (default: Constrained MSSM)
\item \texttt{RGEm2Q}, \texttt{RGEm2L}, \texttt{RGEm2u}, \texttt{RGEm2d}, \texttt{RGEm2e}, \texttt{RGEm2$\nu$} are the soft-breaking squared mass matrices $m^2_{\tilde f}$ for the sfermions.  If given, the Constrained MSSM parameter \texttt{RGEm0} for the corresponding scalar masses is overwritten. (default: Constrained MSSM)
\item \texttt{RGEm2Hd} and \texttt{RGEm2Hu} are the soft-breaking squared masses for $H_d$ and $H_u$, respectively.  If given, the Constrained MSSM parameter \texttt{RGEm0} for the corresponding scalar mass is overwritten. (default: Constrained MSSM)
\item \texttt{RGEM12} is the Constrained MSSM parameter for gaugino mass parameters in GeV. (default: 750)
\item \texttt{RGEm0} is the Constrained MSSM parameter for all soft-breaking masses of scalars in GeV. (default: 1500)
\item \texttt{RGEA0} is the Constrained MSSM parameter $A_0$ for trilinear couplings, e.g.\ $T_{ f} = A_0 Y_f$. (default: -500)
\end{itemize}
\noindent
An example for an input at the GUT scale would be\\

\hspace{1cm}\textbf{RGESetInitial[}\texttt{2$\cdot$10\textasciicircum 16,}\textbf{RGEM1}\texttt{->863,}\textbf{RGEM2}\texttt{->131,}

\hspace{1.5cm}\textbf{RGEM3}\texttt{->-392,}\textbf{RGESuggestion}\texttt{->"GUT"}\textbf{];}\\

The solution at a lower energy scale such as $M_Z$ can now be obtained by the \texttt{REAP} command \texttt{RGESolve}:\\

\hspace{1cm}\textbf{RGESolve[}\texttt{91.19,2$\cdot$10\textasciicircum 16}\textbf{];}\\

\noindent Some parameter points might lead to tachyonic sparticle masses. In such instances the evaluation of \SusyTC~is stopped and an error message is returned using the Mathematica command \texttt{Throw}. In order to properly catch such error messages, we therefore recommend to use instead\\

\hspace{1cm}\textbf{Catch[RGESolve[}\texttt{91.19,2$\cdot$10\textasciicircum 16}\textbf{]}\texttt{,TachyonicMass}\textbf{];}\\

In addition to the parameters known from the \texttt{MSSM} \texttt{REAP} model, the following soft-breaking parameters are available for \texttt{RGEGetSolution} at all energy scales higher than the SUSY scale $Q$:
\begin{itemize}
\item \texttt{RGETu}, \texttt{RGETd}, \texttt{RGETe}, and \texttt{RGET$\nu$} are used to get the soft-breaking trilinear coupling matrices.
\item \texttt{RawT$\nu$} is used to get the raw (internal representation) of the soft-breaking trilinear matrix for sneutrinos.
\item \texttt{RGEM1}, \texttt{RGEM2}, and \texttt{RGEM3} are used to get the soft-breaking gaugino mass parameters.
\item \texttt{RGEm2Q}, \texttt{RGEm2L}, \texttt{RGEm2u}, \texttt{RGEm2d}, \texttt{RGEm2e}, \texttt{RGEm2$\nu$} are used to get the soft-breaking squared mass matrices $m^2_{\tilde f}$ for the sfermions.
\item \texttt{RGEm2Hd} and \texttt{RGEm2Hu} are used to get the soft-breaking squared masses for $H_d$ and $H_u$, respectively.
\end{itemize}
To obtain the running $\overline{\text{DR}}$ gluino mass at a scale of two TeV for example, one uses\\

\hspace{1cm}\textbf{RGEGetSolution[}\texttt{2000,RGEM3}\textbf{];}\\

With \SusyTC~the $\overline{\text{DR}}$ sparticle spectrum is automatically calculated. The following functions are included in \SusyTC:
\begin{itemize}
\item \texttt{STCGetSUSYScale[]} returns the SUSY scale $Q$.

\item \texttt{STCGetSUSYSpectrum[]} returns a list of replacement rules for the SUSY scale $Q$, the $\overline{\text{DR}}$ tree-level values of $\mu$ and $m_3$, and the $\overline{\text{DR}}$ sparticle masses and (tree-level) mixing matrices at the SUSY scale. In detail it contains
\begin{itemize}
\item \texttt{"Q"} the SUSY scale $Q$.
\item \texttt{"$\mu$","$m_3$"} the values of $\mu$ and $m_3$.
\item \texttt{"M1","M2","M3"} are the gaugino mass parameters.
\item \texttt{"Mh","MH","MA","MHp"} the (tree-level) masses of the MSSM Higgs bosons.\footnote{Note that there is no CP-violation in the MSSM Higgs sector on tree-level.}
\item \texttt{"m$\chi$0"} a list of the four neutralino masses.
\item \texttt{"m$\chi$p"} a list of the two chargino masses.
\item \texttt{"msude"} a 3$\times$6 array of the six up-type quarks, down-type squarks and charged slepton  masses, respectively.
\item \texttt{"ms$\nu$"} a list of the three light sneutrino masses.
\item \texttt{"$\theta$W"} the weak mixing angle.
\item \texttt{"tan$\alpha$"} the mixing angle of the CP-even Higgs bosons.
\item \texttt{"N"} the mixing matrix of neutralinos.
\item \texttt{"U","V"} the mixing matrices for charginos.
\item \texttt{"Wude"} a list of the three sparticle mixing matrices for up-type squarks, down-type squarks and charged sleptons.
\item \texttt{"W$\nu$"} the mixing matrix of the three light sneutrinos.
\end{itemize} 
To obtain for example the SUSY scale and the tree-level masses of the charginos call

\hspace{1cm}\texttt{"Q"/.}\textbf{STCGetSUSYSpectrum[];}

\hspace{1cm}\texttt{"m$\chi$p"/.}\textbf{STCGetSUSYSpectrum[];}

The squark masses and charged slepton masses are contained in a joint list as \texttt{\{$m_{\tilde u}$,$m_{\tilde d}$,$m_{\tilde e}$\}}, and analogously for the sfermion mixing matrices. To obtain for example the up-type squark masses, the charged slepton mixing matrix, and the sneutrino masses type

\hspace{1cm}\texttt{("msude"/.}\textbf{STCGetSUSYSpectrum[]}\texttt{)[[1]];}

\hspace{1cm}\texttt{("Wude"/.}\textbf{STCGetSUSYSpectrum[]}\texttt{)[[3]];}

\hspace{1cm}\texttt{"ms$\nu$"/.}\textbf{STCGetSUSYSpectrum[];}

\item \texttt{STCGetOneLoopValues[]} returns a list of replacement rules containing 
\begin{itemize}
\item \texttt{"$\mu$","$m_3$"} the one-loop corrected $\overline{\text{DR}}$ $\mu$-parameter and $m_3$ as in \eqref{eq:mu} and \eqref{eq:m3} at the SUSY scale $Q$.
\item \texttt{"vev"} the one-loop $\overline{\text{DR}}$ vev $\hat v$ as in \eqref{eq:vev} at the SUSY scale $Q$.
\item \texttt{"MHp" ("MA")} the pole-mass $m_{H^+}$ ($m_A$) of the charged (CP-odd) Higgs boson for \texttt{STCcMSSM} = \texttt{True} (\texttt{False}).
\end{itemize}

 The value of $\mu$ can for example be obtained from

\hspace{1cm}\texttt{"$\mu$"/.}\textbf{STCGetOneLoopValues[];}

\item \texttt{STCGetSCKMValues[]} returns a list of replacement rules with the soft-breaking mass squared and trilinear coupling matrices in the SCKM basis, where sparticles are rotated analogously with their corresponding superpartners\footnote{We use the term ``SCKM'' for the super CKM and super PMNS basis.}. Since they are used for the self-energies calculation as described in the previous appendix, they are returned in SLHA2 convention \cite{Allanach:2008qq}! In detail, there are
\begin{itemize}
\item \texttt{"VCKM"} the CKM mixing matrix.
\item \texttt{"VPMNS"} the PMNS mixing matrix.
\item \texttt{SCKMBasis["m2Q"], SCKMBasis["m2u"], SCKMBasis["m2d"]} the squark soft-breaking mass squared matrices in the super CKM basis with SLHA2 convetions.
\item \texttt{SCKMBasis["m2L"], SCKMBasis["m2e"]} the slepton soft-breaking mass squared matrices in the super PMNS basis with SLHA2 conventions. 
\item \texttt{SCKMBasis["T"]} a list of the three trilinear coupling matrices for up-type squarks, down-type squarks and charged sleptons in the SCKM basis with SLHA2 conventions.
\item \texttt{SCKMBasis["Y"]} a 3$\times$3 array of the Yukawa coupling singular values for up-type squarks, down-type squarks and charged sleptons.
\end{itemize}

To obtain the down-type trilinear coupling matrix and the mass squared matrix of the left-handed up-type squarks in the SLHA basis for example, type

\hspace{1cm}\texttt{(SCKMBasis["T"]/.}\textbf{STCGetSCKMValues[]}\texttt{)[[2]];}

\hspace{1cm}\texttt{SCKMBasis["mQ2u"]/.}\textbf{STCGetSCKMValues[];}

\item \texttt{STCGetInternalValues[]} returns everything that is internally used for the calculation of the threshold corrections and sparticle spectrum, i.e.\ the results from \texttt{STCGetSCKMValues[]} and \texttt{STCGetSUSYSpectrum[]} with the one-loop corrected parameters from \texttt{STCGetOneLoopValues[]} replacing tree-level ones if available. We recommend to the user to use those separate functions instead.
\end{itemize}

As additional feature, \SusyTC~optionally supports input and output as SLHA ``Les Houches'' files. These files follow SLHA conventions \cite{Skands:2003cj,Allanach:2008qq}:

\begin{itemize}
\item \texttt{STCSLHA2Input[}\textit{``Path''}\texttt{]} loads an ``Les Houches'' input file stored in \textit{``Path''} and executes \texttt{REAP} and \SusyTC. If no path is given, the default path is assumed as \textit{``SusyTC.in''} in the Mathematica Notebook Directory. An important difference to other spectrum calculators is the pure ``top-down'' approach by \SusyTC, i.e.\ there is no attempt of fitting SM inputs at a low scale or calculating a GUT-scale from gauge couplings unification. Instead, all input is given at a user-defined high energy scale, which is then evolved to lower scales. The input should be given in the flavour basis in SLHA 2 convention \cite{Allanach:2008qq}, with analogous convention for $Y_\nu$ and convention for $M_n$ as in (\ref{eq:W}). The relation between the \SusyTC~conventions and the SLHA 2 conventions is given in Section 3. In the following, we list all SLHA 2 input blocks, which are available in \SusyTC:
\begin{itemize}
\item \texttt{Block MODSEL}: The only available switch is:\\
5\quad:~(Default = 2) CP violation (\texttt{STCcMSSM})
\item \texttt{Block SusyTCInput}: Switches (1=\texttt{True}, 0=\texttt{False}) are defined for \texttt{RGEAdd[]}:\\
1\quad:~(Default = 1) \texttt{STCSusyScaleFromStops}\\
2\quad:~(Default = 1) \texttt{STCSearchSMTransition}\\
3\quad:~(Default = 1) \texttt{STCCCBConstraints}\\
4\quad:~(Default = 1) \texttt{STCUFBConstraints}\\
5\quad:~(Default = 1) Print a Warning in case of Tachyonic masses\\
6\quad:~(Default = 1) One or Two Loop RGEs
\item \texttt{Block MINPAR}: Constrained MSSM parameters as defined in \cite{Skands:2003cj,Allanach:2008qq}. Note however, that the input value of $\tan\beta$ is interpreted to be given at the SUSY scale.
\item \texttt{Block IMMINPAR}: Reads the $\sin\phi_\mu$ in case of the complex MSSM:\\
4\quad:~ $\sin\phi_\mu$
\item \texttt{Block EXTPAR}:\\
0\quad:~(Default = $2\cdot 10^{16}$): $M_{input}$ Input scale\\
Note that with \SusyTC~an automatic calculation of the GUT scale is not possible. The remainder of the block works as usual, e.g.\ optionally one can overwrite common Constrained MSSM gaugino or Higgs soft-breaking parameters:\\
1\quad:~$M_1(M_{input})$ bino mass (real part)\\ 
2\quad:~$M_2(M_{input})$ wino mass (real part)\\
3\quad:~$M_3(M_{input})$ gluino mass (real part)\\
21\quad:~$m_{H_d}^2(M_{input})$\\
22\quad:~$m_{H_u}^2(M_{input})$\\
Imaginary components for the gaugino masses can be given in \texttt{Block IMEXTPAR}.
\item \texttt{Block IMEXTPAR}: as defined in \cite{Allanach:2008qq}.
\item \texttt{Block QEXTPAR}: low energy input:\\
0\quad:~(Default = $91.1876$): The low energy scale to which \texttt{REAP} evolves the SM RGEs.\\
23\quad:~``SUSY scale'' $Q$, where the MSSM is matched to the SM. If this entry is set, it overwrites the automatically calculated SUSY scale.
\item \texttt{Block GAUGE}: the $\overline{\text{DR}}$ gauge couplings at the input scale\\
1\quad:~$g_1(M_{input})$  $U(1)$ gauge coupling\\ 
2\quad:~$g_2(M_{input})$  $SU(2)$ gauge coupling\\ 
3\quad:~$g_3(M_{input})$ $SU(3)$ gauge coupling
\item \texttt{Block YU, Block YD, Block YE, Block YN}: The real parts of the Yukawa matrices $Y_u$, $Y_d$, $Y_e$, and $Y_\nu$ in the flavour basis  \cite{Allanach:2008qq}. They should be given in the \texttt{FORTRAN} format \texttt{(1x,I2,1x,I2,3x,1P,E16.8,0P,3x,`\#',1x,A)}, where the first two integers correspond to the indices and the double precision number to $Re(Y_{ij})$.
\item \texttt{Block IMYU, Block IMYD, Block IMYE, Block IMYN}: The imaginary parts of the Yukawa matrices $Y_u$, $Y_d$, $Y_e$, and $Y_\nu$ in the flavour basis \cite{Allanach:2008qq}. They are given in the same format as the real parts.
\item \texttt{Block MN}: The real part of the symmetric Majorana mass matrix $M_n$ of the right-handed neutrinos in the flavour basis (\ref{eq:W}). Only the ``upper-triangle'' entries should be given, the input format is as for the Yukawa matrices.
\item \texttt{Block IMMN}: The imaginary part of the symmetric Majorana mass matrix $M_n$ of the right-handed neutrinos in the flavour basis (\ref{eq:W}). Only the ``upper-triangle'' entries should be given, the input format is as for the Yukawa matrices.
\end{itemize}
The remaining blocks can be given optionally to overwrite Constrained MSSM input boundary conditions:
\begin{itemize}
\item \texttt{Block TU, Block TD, Block TE, Block TN}: The real parts of the trilinear soft-breaking matrices $T_u$, $T_d$, $T_e$, and $T_\nu$ in the flavour basis \cite{Allanach:2008qq}. They should be given in the same format as the Yukawa matrices.
\item \texttt{Block IMTU, Block IMTD, Block IMTE, Block IMTN}: The imaginary parts of the trilinear soft-breaking matrices $T_u$, $T_d$, $T_e$, and $T_\nu$ in the flavour basis \cite{Allanach:2008qq}. They should be given in the same format as the Yukawa matrices.
\item \texttt{Block MSQ2, Block MSU2, Block MSD2, Block MSL2, Block MSE2,}\\\texttt{Block MSN2}: The real parts of the soft-breaking mass squared matrices $m^2_{\tilde Q}$, $m^2_{\tilde u}$, $m^2_{\tilde d}$, $m^2_{\tilde L}$, $m^2_{\tilde e}$, and $m^2_{\tilde \nu}$ in the flavour basis \cite{Allanach:2008qq}. Only the ``upper-triangle'' entries should be given, the input format is as for the Yukawa matrices.
\item \texttt{Block IMMSQ2, Block IMMSU2, Block IMMSD2,Block IMMSL2,}\\\texttt{Block IMMSE2, Block IMMSN2}: The imaginary parts of the soft-breaking mass squared matrices $m^2_{\tilde Q}$, $m^2_{\tilde u}$, $m^2_{\tilde d}$, $m^2_{\tilde L}$, $m^2_{\tilde e}$, and $m^2_{\tilde \nu}$ in the flavour basis \cite{Allanach:2008qq}. Only the ``upper-triangle'' entries should be given, the input format is as for the Yukawa matrices.
\end{itemize}

\item \texttt{STCWriteSLHA2Output[}\textit{``Path''}\texttt{]} writes an ``Les Houches'' \cite{Skands:2003cj,Allanach:2008qq} output file to \textit{``Path''}. If no path is given, the output is saved in the Mathematica Notebook directory as \textit{``SusyTC.out''}. The output follows SLHA conventions, with the following exceptions:
\begin{itemize}
\item \texttt{Block MASS}: The mass spectrum is given as $\overline{\text{DR}}$ masses at the SUSY scale. The only exception is the pole mass $M_{H^+}$ ($M_A$) for CP violation turned on (off).
\item \texttt{Block ALPHA}: the tree-level Higgs mixing angle $\alpha_\text{tree}$.
\item \texttt{Block HMIX}: Instead of $M_A$ we give\\
101\quad:~$m_3$
\end{itemize}
The other blocks follow the SLHA2 output conventions, e.g.\ $\overline{\text{DR}}$ values at the SUSY scale in the Super-CKM/Super-PMNS basis. To avoid confusion, the blocks \texttt{Block DSQMIX, Block USQMIX, Block SELMIX, Block SNUMIX} and the corresponding blocks for the imaginary entries, return the sfermion mixing matrices $R_{\tilde f}$ in SLHA 2 convention.
\end{itemize}

\end{appendix}


\end{document}